\documentclass[11pt]{article}
\usepackage{amsmath,amsthm,latexsym,amssymb,amsfonts,epsfig}

\makeatletter
\@addtoreset{equation}{section}
\makeatother

\newtheorem{Theorem}{Theorem}[section]

\def\be{\begin{equation}}
\def\ee{\end{equation}}
\def\ba{\begin{eqnarray}}
\def\ea{\end{eqnarray}}
\def\Nl{{\mathchoice
{\setbox0=\hbox{$\displaystyle\rm N$}\hbox{\hbox to0pt
{\kern0.4\wd0\vrule height0.9\ht0\hss}\box0}}
{\setbox0=\hbox{$\textstyle\rm N$}\hbox{\hbox to0pt
{\kern0.4\wd0\vrule height0.9\ht0\hss}\box0}}
{\setbox0=\hbox{$\scriptstyle\rm N$}\hbox{\hbox to0pt
{\kern0.4\wd0\vrule height0.9\ht0\hss}\box0}}
{\setbox0=\hbox{$\scriptscriptstyle\rm N$}\hbox{\hbox to0pt
{\kern0.4\wd0\vrule height0.9\ht0\hss}\box0}}}}
\def\Zl{{\mathchoice
{\setbox0=\hbox{$\displaystyle\rm Z$}\hbox{\hbox to0pt
{\kern0.4\wd0\vrule height0.9\ht0\hss}\box0}}
{\setbox0=\hbox{$\textstyle\rm Z$}\hbox{\hbox to0pt
{\kern0.4\wd0\vrule height0.9\ht0\hss}\box0}}
{\setbox0=\hbox{$\scriptstyle\rm Z$}\hbox{\hbox to0pt
{\kern0.4\wd0\vrule height0.9\ht0\hss}\box0}}
{\setbox0=\hbox{$\scriptscriptstyle\rm Z$}\hbox{\hbox to0pt
{\kern0.4\wd0\vrule height0.9\ht0\hss}\box0}}}}
\def\Ql{{\mathchoice
{\setbox0=\hbox{$\displaystyle\rm Q$}\hbox{\hbox to0pt
{\kern0.4\wd0\vrule height0.9\ht0\hss}\box0}}
{\setbox0=\hbox{$\textstyle\rm Q$}\hbox{\hbox to0pt
{\kern0.4\wd0\vrule height0.9\ht0\hss}\box0}}
{\setbox0=\hbox{$\scriptstyle\rm Q$}\hbox{\hbox to0pt
{\kern0.4\wd0\vrule height0.9\ht0\hss}\box0}}
{\setbox0=\hbox{$\scriptscriptstyle\rm Q$}\hbox{\hbox to0pt
{\kern0.4\wd0\vrule height0.9\ht0\hss}\box0}}}}
\def\Rl{{\mathchoice
{\setbox0=\hbox{$\displaystyle\rm R$}\hbox{\hbox to0pt
{\kern0.4\wd0\vrule height0.9\ht0\hss}\box0}}
{\setbox0=\hbox{$\textstyle\rm R$}\hbox{\hbox to0pt
{\kern0.4\wd0\vrule height0.9\ht0\hss}\box0}}
{\setbox0=\hbox{$\scriptstyle\rm R$}\hbox{\hbox to0pt
{\kern0.4\wd0\vrule height0.9\ht0\hss}\box0}}
{\setbox0=\hbox{$\scriptscriptstyle\rm R$}\hbox{\hbox to0pt
{\kern0.4\wd0\vrule height0.9\ht0\hss}\box0}}}}
\def\Cl{{\mathchoice
{\setbox0=\hbox{$\displaystyle\rm C$}\hbox{\hbox to0pt
{\kern0.4\wd0\vrule height0.9\ht0\hss}\box0}}
{\setbox0=\hbox{$\textstyle\rm C$}\hbox{\hbox to0pt
{\kern0.4\wd0\vrule height0.9\ht0\hss}\box0}}
{\setbox0=\hbox{$\scriptstyle\rm C$}\hbox{\hbox to0pt
{\kern0.4\wd0\vrule height0.9\ht0\hss}\box0}}
{\setbox0=\hbox{$\scriptscriptstyle\rm C$}\hbox{\hbox to0pt
{\kern0.4\wd0\vrule height0.9\ht0\hss}\box0}}}}
\def\Hl{{\mathchoice
{\setbox0=\hbox{$\displaystyle\rm H$}\hbox{\hbox to0pt
{\kern0.4\wd0\vrule height0.9\ht0\hss}\box0}}
{\setbox0=\hbox{$\textstyle\rm H$}\hbox{\hbox to0pt
{\kern0.4\wd0\vrule height0.9\ht0\hss}\box0}}
{\setbox0=\hbox{$\scriptstyle\rm H$}\hbox{\hbox to0pt
{\kern0.4\wd0\vrule height0.9\ht0\hss}\box0}}
{\setbox0=\hbox{$\scriptscriptstyle\rm H$}\hbox{\hbox to0pt
{\kern0.4\wd0\vrule height0.9\ht0\hss}\box0}}}}
\def\Ol{{\mathchoice
{\setbox0=\hbox{$\displaystyle\rm O$}\hbox{\hbox to0pt
{\kern0.4\wd0\vrule height0.9\ht0\hss}\box0}}
{\setbox0=\hbox{$\textstyle\rm O$}\hbox{\hbox to0pt
{\kern0.4\wd0\vrule height0.9\ht0\hss}\box0}}
{\setbox0=\hbox{$\scriptstyle\rm O$}\hbox{\hbox to0pt
{\kern0.4\wd0\vrule height0.9\ht0\hss}\box0}}
{\setbox0=\hbox{$\scriptscriptstyle\rm O$}\hbox{\hbox to0pt
{\kern0.4\wd0\vrule height0.9\ht0\hss}\box0}}}}

\title{{\sf Loop Quantum Gravity:}\\ {\sf An Inside View}}
\author{{\sf T. 
Thiemann}\thanks{{\sf 
thiemann@aei.mpg.de,tthiemann@perimeterinstitute.ca}}\\
\\
{\sf MPI f. Gravitationsphysik, Albert-Einstein-Institut,} \\
{\sf Am M\"uhlenberg 1, 14476 Potsdam, Germany}\\
\\
{\sf and}\\
\\
{\sf Perimeter Institute for Theoretical Physics,}\\ 
{\sf 31 Caroline Street N, Waterloo, ON N2L 2Y5, Canada}}
\date{{\small\sf Preprint AEI-2006-066}}

\begin{document}

\maketitle

\begin{abstract}
{\sf
This is a (relatively) non -- technical summary of the status of the 
quantum dynamics 
in Loop Quantum Gravity (LQG). We explain in detail the historical 
evolution of the subject and why the results obtained so far are non -- 
trivial. The present text can be viewed in part as a response to an 
article by Nicolai, Peeters and Zamaklar. We also explain why certain 
no go conclusions drawn from a mathematically correct calculation in a 
recent paper by Helling et al are physically incorrect.
}
\end{abstract}

\newpage

\tableofcontents

\newpage

\section{Introduction}
\label{s1}

Loop Quantum Gravity (LQG) \cite{books,TB,reviews} has become a serious 
competitor to string theory as a candidate theory of quantum gravity.
Its popularity is steadily growing because it has transpired that the 
major obstacle to be solved in combing the principles of General 
Relativity and Quantum Mechanics is to preserve a key feature of 
Einstein's theory, namely {\it background independence}. LQG, in contrast 
to the present formulation of string theory, has background independence 
built in by construction. 

Loosely speaking, background independence means 
that the spacetime metric is not an external structure on which matter 
fields and gravitational perturbations propagate. Rather, the metric is a 
dynamical entity which becomes a fluctuating quantum operator. These 
fluctuations will be huge in extreme astrophysical (centre of black 
holes) and cosmological (big bang singularity) situations and the notion 
of a (smooth) background metric disappears, the framework of quantum field 
theory 
on (curved) background metrics \cite{Wald,Fredenhagen} becomes 
meaningless. Since quantum gravity is supposed to take over as a more 
complete theory precisely in those situations when there is no meaningful 
concept of a (smooth) metric {\it at all} available, background 
independence is 
{\it a necessary feature} of a successful quantum gravity theory. 

Indeed, the modern formulation of ordinary QFT on background 
spacetimes uses the algebraic approach \cite{Haag} and fundamental for 
this framework is the locality axiom: Two (scalar) field operators 
$\widehat{\phi(f)},\;\widehat{\phi(f')}$ which  
are smeared with test functions $f,\;f'$ whose supports are {\it spacelike 
separated} are axiomatically required to commute. In other words, the 
causality structure of the external background metric {\it defines the 
algebra of field operators} $\mathfrak{A}$. One then looks for Hilbert 
space representations of $\mathfrak{A}$. It follows, that without a 
background metric one {\it cannot even define} quantum fields in the usual 
setting. 

Notice that background independence implies that a non -- perturbative 
formulation must be found. For, if we split the metric as $g=g_0+h$ where
$g_0$ is a given background metric and treat the deviation $h$ (graviton) 
as a quantum field propagating on $g_0$, then we break background 
independence because we single out $g_0$. We also break the symmetry group 
of Einstein's theory which is the group of diffeomorphisms of the given 
spacetime manifold $M$. Moreover, it is well -- known 
that this perturbative formulation leads to a non -- renormalisable 
theory,
with \cite{Sagnotti} or without \cite{Goroff} supersymmetry. String theory 
\cite{Polchinsky} in its current formulation is also background dependent 
because one has to fix a target space background metric (mostly Minkowski 
space or maybe AdS) and let strings propagate on it. Some excitations of 
the string are interpreted as gravitons and one often hears that string 
theory is perturbatively finite to all orders, in contrast to perturbative 
quantum gravity. However, this has been established only to second 
order and only for the heterotic string \cite{DP} which is better but 
still far from a perturbatively finite theory. In fact, even perturbative 
finiteness is not the real issue because one can formulate perturbation 
theory in such a way that UV divergences never arise \cite{EP}. The issue 
is whether 1. only a finite number of free renormalisation 
constants need to be fixed (predictability) and 2. whether the 
perturbation series converges. Namely, in a fundamental theory as string 
theory claims to be, there is no room for singularities such as a a 
divergent perturbation series. This is different from QED which is 
believed to be only an effective theory. Hence, before one does not prove 
convergence of the perturbation series, string theory has not been shown 
to be a fundemantal theory. All these issues are obviously avoided in a 
manifestly non -- perturbative formulation.

One of the reasons why LQG is gaining in its degree of popularity as 
compared to string theory is that LQG has ``put its cards on the table''.
LQG has a clear conceptual setup which follows from physical 
considerations and is based on a rigorous mathematical formulation.
The ``rules of the game'' have been written and are not tinkered with.
This makes it possible even for outsiders of the field \cite{NPZ,NPZ1} 
to get 
a relatively good understanding of the physical and mathematical details.
There is no room for extra, unobserved structures, the approach is 
intendedly minimalistic. In LQG one just tries to make quantum gravity and 
general relativity work together harmonically. However, in order to do so 
one must be ready to go beyond some of the mathematical structures that we 
got used to from ordinary QFT as we have explained. Much of the criticism 
against LQG of which some can be found in \cite{NPZ} has to do with the 
fact that physicists equipped with a particle physics background feel 
uneasy when one explains to them that in LQG we cannot use perturbation 
theory, Fock spaces, background metrics etc. This is not the fault of LQG.
It will be a common feature of all quantum gravity theories which preserve 
background independence. In such theories, the task is to construct a new 
type of QFT, namely a QFT on a {\it differential manifold $M$} rather than 
a QFT on a background spacetime $(M,g_0)$. Since such a theory ``quantises 
all backgrounds at once'' in a coherent fashion, the additional task is 
then to show that for any background metric $g_0$ the theory contains a 
semiclassical sector which looks like ordinary QFT on $(M,g_0)$. This is 
what LQG is designed to do, not more and not less.

Another criticism which is raised against LQG is that it is not a unified 
theory of all interactions in the sense that string theory claims to be.
Indeed, in LQG one quantises geometry together with the fields of the 
(supersymmetric extension of the) standard model. At present there seem
to be no constraints on the particle content or the dimensionality of the 
world. In fact, this is not quite true because the size of 
the physical Hilbert space of the theory may very well depend on the 
particle content and moreover, the concrete algebra $\mathfrak{A}$ which 
one quantises in LQG is available only in 3+1 dimensions. But apart from
that it is certainly true that LQG cannot give a prediction of the matter 
content. The fact that all matter can be treated may however be an 
advantage because, given the fact that in the past 100 years we 
continuously 
discovered substructures of particles up to the subnuclear scale makes
it likely that we find even more structure until the Planck scale which 
is some sixteen orders of magnitude smaller than what the LHC can resolve.
Hence, LQG is supposed to deliver a universal framework for combining 
geometry and matter, however, it does not uniquely predict which 
matter and does not want to. Notice that while theorists would find a 
``unique'' theory aesthetical, there is no logical reason why a 
fundamental theory should be mathematically unique. 

In this context we would like to point out the following:
One often hears that string theory is mathematically unique, predicting 
supersymmetry, the dimensionality (ten) of the world and the particle 
content. What one means by that is that a consistent quantum string theory 
based on the Polyakov action on the Minkowski target spacetime exists
only if one is in ten dimensions and only if the theory is 
supersymmetric and there are only five such theories.
However, this is not enough in order to have a unique theory because 
string theory must be decompactified from ten to four dimensions with 
supersymmetry being broken at sufficiently high energies in order 
to be phenomenologically acceptable. Recent findings \cite{Taylor} show that
for Minkowski space there are an at least 
countably infinite number of physically different, phenomenologically 
acceptable ways to compactify
string theory from ten to four dimensions. These possibilities are 
labelled by flux vacua and the resulting collection of quantum string 
theories is called {\it the landscape}. In this sense, {\it string theory 
is far from being mathematically unique.} The presence of an infinite 
number of possibilities could be interpreted  
as the loss of predictability of string theory and the use of the 
anthropic principle was proposed \cite{Susskind}. The question, 
whether a physical theory that needs the anthropic principle still can 
be called a fundamental theory, was discussed in \cite{SmolinAnthropic}.

Our interpretation of the landscape is the following which is
in agreement with \cite{SmolinBackground}:\\ 
The anthropic principle should be avoided by all means in a fundamental 
theory, hence a new idea is needed and it is here where background 
independence could help. We notice that each landscape vacuum is based 
on a different background structure (flux charges, moduli). In addition,
a landscape will exist for each of the uncountably infinite number 
of target space background spacetimes\footnote{   
Currently string theory can be 
quantised only on a hand full of target space background spacetimes,
mostly only on those on which the theory becomes a free field theory.
For instance, on AdS$_5\times$S$^5$, which is much discussed in the 
context of the famous AdS/CFT correspondence \cite{Maldacena}, string  
theory is interacting and to the best knowledge of the author currently no 
quantum string theory on this spacetime was constructed.} Thus, the full 
string theory landscape is presumably labelled not only by flux vacua
on a given spacetime  
but also by the background spacetimes. Suppose one could rigorously 
quantise 
string theory on all of these background structures. Then, if one knew 
how to combine all of these distinct quantum theories into a single one,
thus achieving {\it background independence}, then the landscape would 
have disappeared. The understanding of the present author is 
that this is what M -- theory is supposed to achieve but currently, to the 
best knowledge of the author, there 
seems to be no convincing idea for how to do that.\\
\\
In the present article we summarise the status of the quantum dynamics 
of LQG which has been the focal point of criticism in \cite{NPZ}. 
Our intention is to give a self -- contained inside point of view of the 
subject
which is complementary to \cite{NPZ} in the sense that we explain in some 
detail why the constructions used in LQG are physically well motivated
and sometimes even mathematically unique, hence less ambiguous than 
described in \cite{NPZ}. We exactly define what is meant 
by canonical quantisation of general relativity, indicating explicitly the 
freedom that one has at various stages of the quantisation programme. 
We will see that the theory has much less freedom than \cite{NPZ} 
makes it look like. In particular, we evaluate ``what has been gained in 
LQG as compared to the old geometrodynamics approach'' and we will see 
that the amount of progress is non -- trivial. We also include more recent 
results such as the Master Constraint Programme \cite{MCP} which tightens 
the implementations of the quantum dynamics and enables to systematically 
construct the physical inner product, which was not possible as of three
years ago. Furthermore, in order to show that there is not only 
mathematical progress in LQG, we also fill the gaps that \cite{NPZ} 
did not report on such as the semiclassical sector of the theory, 
matter coupling, quantum black hole physics, quantum cosmology and LQG 
phenomenology. Finally 
we also show that the key technique that was used to
make the Wheeler -- DeWitt operator well -- defined \cite{QSDI} is also 
the underlying reason for the success of Loop Quantum Cosmology (LQC)
which is the usual cosmological minisuperspace toy model quantised with 
the methods of LQG. 

Here we only sketch these results since we want to reach a rather general 
audience. All the technical details can be found in the comprehensive
and self -- contained monograph \cite{TB}.\\
\\
%
%
%
{\bf Remark}\\
Since no theoretical Ansatz concerning Quantum Gravity has been yet
brought to completion each of these Ans\"atze is understandably subject to
criticisms. This is clearly the case also for LQG, and in particular such
criticisms can be found e.g., in \cite{NPZ,NPZ1}.
We want to stress that the purpose of this article, while in part a 
response to \cite{NPZ}, is {\it not} to 
criticise \cite{NPZ}. In fact, the considerable effort of the non expert
authors of \cite{NPZ} to present LQG in as much technical detail as they 
did from a 
particle physicist's perspective is highly appreciated. Rather, what we 
have in mind is to draw a more 
optimistic picture than \cite{NPZ} did, to hopefully resolve confusions 
that may have arisen from gaps in \cite{NPZ} and to give a more complete 
picture 
of all the research being done in LQG than \cite{NPZ} did. The discussion 
will be kept objectively, problems with the present formulation of LQG 
will 
not be swept under the rug but rather discussed in great detail together
with their possible solutions.

\newpage

\section{Classical preliminaries}
\label{s2}

The starting point is a Lagrangean formulation of the classical field 
theory, say the Einstein Hilbert Lagrangean for General Relativity. 
Hence one has an action 
\be \label{2.1}
S=\int_M \; d^{n+1}X\; L(\Phi,\partial \Phi)
\ee
where $L$ is the Lagrangean density, that is, a scalar density of weight 
one constructed in a covariant fashion from the fields $\Phi$ and their 
first partial derivatives\footnote{Higher derivative theories can also be 
treated canonically \cite{Tyutin}, however, they are generically  
pathological, that is, unstable \cite{Woodard}.} which is sufficient for 
gravity and all known matter. Here $\Phi$ stands for a collection of 
fields including the metric and all known matter. $M$ is an $(n+1)$ -- 
dimensional, 
smooth differential manifold. 

If one wants to have a well -- posed initial value formulation, then 
the metric fields $g$ that live on $M$ are such that $(M,g)$ is globally 
hyperbolic which implies \cite{Geroch} that $M$ is diffeomorphic to the 
direct product 
$\Rl\times \sigma$ where $\sigma$ is an $n-$dimensional smooth 
manifold\footnote{Unless otherwise stated we take $\sigma$ to be compact 
without boundary in order to avoid a lengthy discussion of boundary 
terms.}. Since the action (\ref{2.1}) is invariant under Diff$(M)$, 
the diffeomorphisms $Y:\;\Rl\times\sigma\to M;\;(t,x)\mapsto Y_t(x)$ 
are a symmetry of the action. For each $Y$ we obtain a foliation 
of $M$ into a one parameter family of spacelike hypersurfaces 
$\Sigma_t=Y_t(\sigma)$. One now pulls all fields back by $Y$
and obtains an action on $\Rl\times \sigma$. Due to the arbitrariness of 
$Y$, this action contains $n+1$ fields, usually called lapse and shift 
fields, which appear without time derivatives, they are Lagrange 
multipliers. The Legendre transformation 
is therefore singular and leads to constraints on the resulting phase 
space \cite{Dirac, Henneaux}. They can be obtained by
extremisation of the action with respect to the Lagrange multipliers.

Hence, after the Legendre transformation we obtain a phase space 
${\cal M}$ of canonical fields $\phi$ which are the pull -- backs to 
$\sigma$ of the spacetime fields $\Phi$ together with their canonically 
conjugate momenta $\pi$. The symplectic manifold ${\cal M}$ with 
coordinates $(\phi(x),\pi(x))_{x\in \sigma}$ equipped with the 
corresponding canonical bracket\footnote{We suppress spatial tensor 
and internal Lie algebra indices.}
$\{\phi(x),\pi(x')\}=\delta^{(n)}(x,x')$ is a time $t$ independent 
object. 

As one can show by purely geometric arguments \cite{Hojman}, these 
constraints 
are automatically of first class in the terminology of Dirac, that is, 
they close under 
their mutual Poisson brackets, irrespective of the matter content of the 
theory. As we will need them in some detail later on, let us display this so 
called Dirac algebra $\mathfrak{D}$ in more detail
\ba \label{2.2}
\{D(\vec{N}),D(\vec{N}')\} 
&=& 8\pi G_{{\rm Newton}} \;\; D({\cal L}_{\vec{N}}\vec{N}')
\nonumber\\    
\{D(\vec{N}),H(N')\} 
&=& 8\pi G_{{\rm Newton}} \;\; H({\cal L}_{\vec{N}}N')
\nonumber\\    
\{H(N),H(N')\} 
&=& 8\pi G_{{\rm Newton}} \;\; D(q^{-1}(N dN'- N' dN)
\ea
The notation is as follows: We distinguish between the so -- called 
spatial diffeomorphism constraints $D_a(x),\;a=1,..,n;\; x\in\sigma$
and the Hamiltonian constraints $H(x),\; x\in\sigma$. Notice that these 
are 
infinitely many constraints, $n+1$ per $x\in \sigma$. We smear them with 
test functions $N^a,N$, specifically
$D(\vec{N})=\int_\sigma\; d^3x \; N^a D_a$ and 
$H(N)=\int_\sigma\; d^3x \; N H$. Finally, $q_{ab}$ is the pull -- back
to $\sigma$ of the spacetime metric with inverse $q^{ab}$ and $\cal L$
denotes the Lie derivative.

The algebra $\mathfrak{D}$ is {\it universal} and underlies the 
canonical 
formulation of any field theory based on an action which is Diff$(M)$
invariant and contains General Relativity in $n+1$ dimensions as for 
instance the closed bosonic string\footnote{A closed bosonic string
is an embedding of a circle $\sigma:=S^1$ into a $D+1$ dimensional 
target 
space background manifold, mostly $D+1$ dimensional Minkowski space.
The spacetime manifold of the string is therefore $M=\Rl\times S^1$,
i.e. a 2D cylinder. In 2D gravity is topological, hence gravity is also
trivially contained in string theory.}   
\cite{String}. 

As we can read off from 
(\ref{2.2}), it has the following structure: The first line in 
(\ref{2.2}) says that elements of the form 
$D(\vec{N})$ generate a subalgebra which can be identified with the Lie 
algebra diff$(\sigma)$ of the spatial diffeomorphism group 
Diff$(\sigma)$ of $\sigma$. This is why the $D(\vec{N})$, where the
$\vec{N}$ are arbitrary smooth vector fields on $\sigma$ of rapid 
decrease, are called spatial diffeomorphism constraints. The second line 
in (\ref{2.2}) says that diff$(\sigma)$ is not an ideal of $\mathfrak{D}$ 
because the Hamiltonian constraints $H(N)$, where the $N$ are 
arbitrary smooth functions on $\sigma$ of rapid decrease, are not 
diff$(\sigma)$ invariant. The name Hamiltonian constraint stems 
from the fact that the Hamiltonian flow of this constraint on the 
phase space generates gauge motions which, when the equations of motion 
hold, can be identified with 
spacetime diffeomorphisms generated by vector fields orthogonal 
to the hypersurfaces $\Sigma_t$. Finally the third line in (\ref{2.2}) 
says 
that (\ref{2.2}) is not a Lie algebra in the strict sense of the 
word because, while the right hand side of the Poisson bracket between 
two Hamiltonian constraints is a linear combination of spatial 
diffeomorphism constraints, the coefficients in that linear combination 
have non -- trivial phase space dependence through the 
tensor $q^{ab}(x)$. 

A pecularity happens in the case $n=1$, such as the closed bosonic 
string: In $n=1$ dimensions, $p-$times contravariant and $q-$times 
covariant tensors are the same thing as scalar densities of weight
$q-p$. For this reason, in contrast to $n>1$ dimensions, in $n=1$ 
dimensions the constraints 
come with a natural density weight of two rather than one while the 
smearing functions acquire density weight $-1$ rather than $0$. One can 
think of this as if the actual constraints had been multiplied by a 
factor of $\sqrt{q}$ while the smearing functions had been multiplied 
by a factor of $\sqrt{q}^{-1}$ which however does not change the Poisson 
bracket because gravity is not dynamical in 2D. For this reason the 
factor $q^{-1}$ must be absent in the third relation in (\ref{2.2})
in order to match the density weights on both sides of the equations.
This is why {\it only in 2D} the Dirac algebra $\mathfrak{D}$ is a true 
Lie algebra. In fact, using the combinations $V_\pm:=H\pm D$ one 
realises that the Dirac algebra acquires the structure a direct sum 
of loop algebras (Lie algebra of the diffeomorphism group of the circle) 
$\mathfrak{D}\cong {\rm diff}(S^1)\oplus {\rm diff}(S^1)$, see 
\cite{String} for all the details.
Thus, {\it in 2D the Dirac algebra trivialises}. It does not even 
faintly display the complications that come with the non -- Lie 
algebra structure of $\mathfrak{D}$ in realistic field 
theories, i.e. $D=4$. Therefore, any comparisons made between structures 
in 2D and 4D which hide this important difference 
are void of any lesson.

Proceeding with the general classical theory, what we are given is a 
phase space $\cal M$ subject to a collection of constraints 
$C_I,\;I\in {\cal I}$ where in our case the labelling set 
comprises the $N,\vec{N}$. These constraints force us 
to consider the constraint hypersurface 
$\overline{{\cal M}}:=\{m\in {\cal M};\;C_I(m)=0\;\forall\;I\in {\cal 
I}\}$. The closure of $\mathfrak{D}$ means that the Hamiltonian flow
of the $C_I$ preserves $\overline{{\cal M}}$. Since the $C_I$ generate 
gauge transformations (namely spacetime diffeomorphisms), all the points
contained in the gauge orbit $[m]$ through $m\in \overline{{\cal M}}$ 
must be identified as physically equivalent. As one can show in 
general \cite{Woodhouse}, the set of orbits 
$\widehat{{\cal M}}:=\{[m];\;m\in \overline{{\cal M}}\}$ is again a 
symplectic manifold and known as the 
reduced phase space. 

It is mathematically more 
convenient to consider functions on all of $\cal M$ which are invariant
under gauge transformations, called Dirac observables. Their 
restrictions to 
$m\in\overline{{\cal M}}$ are completely determined by $[m]$. The 
physical idea to construct such functions is due to Rovelli \cite{Rovelli}
and its mathematical implementation has been much improved recently 
in \cite{Bianca} (see also \cite{TR}). For a particularly simple 
realisation of this so -- called relational Ansatz in terms of suitable 
matter see \cite{Phantom}. We consider functions $T_I$ on $\cal M$
which have the property that the matrix with entries 
$A_{IJ}:=\{C_I,T_J\}$ is invertible (at least locally). Let $X_I$ be the 
Hamiltonian vector field of the constraint 
$C'_I:=\sum_J (A^{-1})_{IJ} C_J$. The set of constraints $C'_I$ is 
equivalent to the set of the $C_I$ but the $C'_I$ have the advantage 
that the vector fields $X_I$ are weakly (that is, on 
$\overline{{\cal M}}$) mutually commuting. Now given any smooth function 
$f$ on $\cal M$ and any real numbers $\tau_I$ in the range of 
the $T_I$ consider 
\be \label{2.3}
O_f(\tau):=[\alpha_t(f)]_{t=\tau-T},\;\; \alpha_t(f):=[\exp(\sum_I t_I 
X_I) 
\cdot f]
\ee 
Notice that one is supposed to first evaluate $\alpha_t(f)$ with 
$t_I$ 
considered as real numbers and then evaluates the result at the phase 
space dependent point $t_I=\tau_I-T_I$. It is not difficult to show 
that (\ref{2.3}) is a weak Dirac observable, that is
$\{C_I,O_f(\tau)\}_{|\overline{{\cal M}}}=0$. It has the physical 
interpretation of displaying the value of $f$ in the gauge $T_I=\tau_I$.
Equivalently, it is the gauge invariant extension of $f$ off the gauge 
cut $T=\tau$ and in fact can be expanded in a power series in $\tau-T$ 
by expanding the exponential function in (\ref{2.3}). 

The relational Ansatz solves the problem of time of canonical quantum 
gravity: By this one means that in generally covariant systems there is 
no Hamiltonian, there are only Hamiltonian constraints. Since the 
observables of the theory are the gauge invariant functions on phase 
space, that is, the Dirac observables, ``nothing moves in canonical 
quantum gravity'' because the Poisson brackets between the Hamiltonian 
constraints and the Observables vanishes (weakly) by construction. The 
missing evolution of the Dirac observables 
$O_f(\tau)$ is now supplied as follows: Using the fact that the map
$\alpha_t$ in (\ref{2.3}) is actually a Poisson automorphism (i.e. a 
canonical transformation) one can show that if 1. the phase space 
coordinates can be subdivided into canonical pairs $(T_I,\pi_I)$ and 
$(q^a,p_a)$ and 2. if $f$ is a function of only\footnote{Nothing is lost 
by this assumption because $T_I$ is pure gauge and the constraints can 
be solved for $\pi_I$ in terms of $q^a, p_a, T_I$.} $q^a,p_a$ then 
the evolution in $\tau_I$ has a Hamiltonian generator \cite{TR}. That 
is, there exist Dirac observables $H_I(\tau)$ such that 
$\partial O_f(\tau)/\partial\tau_I=\{O_f(\tau),H_I(\tau)\}$.

The task left is then to single out a one parameter family 
$s\mapsto \tau_I(s)$ such that the corresponding Hamiltonian 
\be \label{2.4}
H(s)=\sum_I \dot{\tau}_I(s) H_I(\tau(s))
\ee
is positive, $s-$independent 
and reduces to the Hamiltonian of the standard model on flat space.
This has been achieved recently in \cite{Phantom} using suitable
matter which supplies the clocks $T_I$ with the required properties.
It follows that the gauge invariant functions $O_f(s)$ then evolve
according to the physical Hamiltonian $H$. Moreover, they satisfy the 
algebra $\{O_f(s),O_{f'}(s)\}=O_{\{f,f'\}}(s)$ because the $s$ 
evolution has the canonical generator $H$.\\
\\
To summarise:\\
Classical canonical gravity has a clear conceptual and technical 
formulation with no mysteries or unsolved conceptual problems. Certainly 
classical General Relativity is not an integrable system and thus not 
everything is technically solvable (for instance not all solutions to 
the field equations are known) but one exactly knows what to do in order 
to try to solve a given problem. The canonical formulation that we have 
used here for a generally covariant field theory is widely used in 
numerical General Relativity with great success. General covariance 
is manifestly built into the framework and is faithfully represented 
in terms of the Dirac algebra $\mathfrak{D}$ (\ref{2.2}) which is the 
key object to construct the invariants (\ref{2.3}) of the theory and 
the physical Hamiltonian $H$ (\ref{2.4}) according to which they evolve. 
At no 
point in those constructions did one use a background metric or did one 
violate spacetime diffeomorphism invariance. This is because, while one 
did use a split of spacetime into space and time, one did consider all 
splits simultaneously which is reflected in the constraints that in turn 
enforce spacetime diffeomorphism invariance.

For clarity we mention that diffeomorphism 
invariance should not be confused with Poincar\'e invariance. Poincar\'e 
invariance is an invariance of a special solution to Einstein's vacuum 
equations. It is not a symmetry or a gauge invariance of the theory.
The gauge group is Diff$(M)$ which is a background metric independent 
object because it only refers to the differential manifold $M$ but to 
no
metric. In fact, if $\sigma$ is compact as appropriate for certain 
cosmological models, then the Poincar\'e group 
$\cal P$ has no place in the theory. If $M$ is equipped with 
asymptotically flat boundary conditions then in fact one can in addition 
define Poincar\'e generators of $\cal P$ as functions on phase space,
called ADM charges \cite{Beig}. These are particular Dirac observables.
Notice that $\cal P$ is not contained in Diff$(M)$ because 
diffeomorphisms are of rapid decrease at spatial infinity (at least they 
vanish there). This must be because $\cal P$ is a symmetry and not a 
local gauge invariance like Diff$(M)$.

\section{Canonical quantisation programme}
\label{s3}

The programme of canonical quantisation is a mathematical formalism which 
seeks to provide a quantum field theory from a given classical field 
theory. There are several choices to be made within the formalism and 
the outcome depends on it. This applies to ordinary field theories such as 
free scalar fields on Minkowski space as well as to more complicated 
situations. In the presence of constraints such as in General Relativity
one would ideally solve the constraints classically before quantising 
the theory. That is, one studies the representation theory of the algebra 
of 
invariants such as (\ref{2.3}). Unfortunately, this is generically to 
difficult because the algebra of invariants is complictated and thus 
usually prevents one from using standard representations for simple 
algebras such as a Fock representation for usal CCR (canonical 
commutation relation) or CAR (canonical anticommuttion relation) algebras.

Thus, in order to start the quantisation process one follows Dirac 
\cite{Dirac} and starts with a redundant set of functions on phase space 
which generate a sufficiently simple Poisson algebra so that suitable 
representations thereof can be found. These functions are not gauge 
invariant but provide a system of coordinates for $\cal M$. Then, in a 
second step, provided 
that the constraints themselves can be represented on the chosen,
so called kinematical, Hilbert 
space as closable\footnote{That is, the adjoint is also densely 
defined.} and densely defined operators, one looks for the generalised 
joint kernel of the constraint operators. Here generalised refers to the 
fact that the joint kernel typically has trivial intersection with the 
Hilbert 
space, that is, the non zero solutions of the constraints are not 
normalisable. Rather, they are elements of the physical Hilbert space 
which is not a subspace of the kinematical Hilbert space. The physical 
Hilbert space is induced from the kinematical Hilbert space by applying
standard spectral theory to the constraint operators. Once the physical 
Hilbert space is known, at least in principle, it automatically carries 
a self -- adjoint representation of the algebra of strong observables, 
that is, those operators that commute with all quantum constraints and 
for which (\ref{2.3}) and (\ref{2.4}) are the classical counterparts. 

All of this is of course difficult, if not impossible, to carry out 
exactly and in full completeness for General Relativity because, after 
all, one is dealing with a rather non -- linear and highly interacting 
QFT. Hence, in praxis one will have to develop and rely on approximation 
schemes. However, these are only technical difficulties coming from the 
complexity of the theory. There are no in principle obstacles, the 
programme of canonical quantisation follows a clear sequence of steps
at each of which one knows exactly what one has to do and sometimes one 
has a certain freedom which one will exploit using physical intuition.

After the above sketch of the programme, we will now become somewhat 
more detailed and pin down explicitly the freedom that one has and the 
choices that one has to make.\\
\\
The starting point is a then a symplectic manifold $\cal M$ subject to 
real valued, first class constraints $C_I,\;I\in {\cal I}$. 
That is, we have $\{C_I,C_J\}=f_{IJ}\;^K \;C_K$ for some, possibly phase 
space dependent functions $f_{IJ}\;^K$, called structure functions.
We will assume for simplicity, as it is 
the case in 
General Relativity, that we are dealing with a completely parametrised 
system, that is, there is no apriori gauge invariant Hamiltonian. 
In order to simplify the discussion for the purposes of this short 
review, we display here for concreteness a recently proposed strategy
\cite{MCP,Test1} to deal with those constraints:
Consider instead of the individual constraints $C_I$ the single
Master Constraint $M:=\sum_I C_I K_{IJ} C_J$. Here $K=(K_{IJ})$ is a 
positive 
definite matrix valued function on phase space. The Master Constraint
contains the same information about the gauge redundancy of the system 
as the individual $C_I$ since $M=0$ is equivalent with $C_I=0$ for all 
$I$ and the equation $\{O,\{O,M\}\}_{M=0}$ is equivalent with 
$\{O,C_I\}_{M=0}$ for all $I$. Hence the Master Constraint selects the 
the same reduced phase space as the original set of constraints. The 
reason for the using the matrix $K$ is that we can and often must use 
the associated freedom to regularise the square of the constraints:
namely, typically the $C_I$ become operator valued distributions and 
their square is therefore ill -- defined. By a judicious choice of $K$ 
(which also becomes an operator) one can remove the corresponding UV 
singularity. See e.g. \cite{Test2} for examples.

Given this set up, 
the programme of canonical quantisation consists of the 
following\footnote{What follows is still a simplified version. See 
\cite{TB} for a complete discussion.}:
\begin{itemize}
\item[I.] {\it Algebra of elementary functions $\mathfrak{E}$}\\
Select a Poisson sub$^\ast-$algebra $\mathfrak{E}$ of
$C^\infty({\cal M})$, called elementary functions, which separates the 
points 
of $\cal M$. That is, $\mathfrak{E}$ should be closed under taking 
Poisson barckets and complex conjugation and for any  
$m\not=m'$ there exists $e\in \mathfrak{E}$ such that $e(m)\not=e(m')$.
The latter property implies that any $f\in C^\infty({\cal M})$ can be 
thought of as a function of the elements of $\mathfrak{E}$ so that 
$\mathfrak{E}$ is a sytstem of coordinates for $\cal M$ (which maybe 
redundant). The choice of $\mathfrak{E}$ will be guided by mathematical 
convenience and physical intuition: One will try to use bounded 
functions, such as the Weyl elements used in free field theories, in 
order to deal with bounded operators later on which avoids domain 
questions. Of course, the algebra $\mathfrak{E}$ should be sufficiently 
simple in order that one can manage to find representations of the 
corresponding quantum algebra at all. Furthermore, one will choose 
$\mathfrak{E}$ in such a way that its elements transform in a simple way 
under the gauge group of the system in question.
\item[II.] {\it Quantum $^\ast-$algebra $\mathfrak{A}$}\\
One now constructs a $^\ast-$algebra $\mathfrak{A}$ using 
the following well known procedure: We consider the free algebra 
$\mathfrak{F}$ of finite linear combinations of formal words. A word 
is a formal finite sequence of elements $w=(e_1.. e_N)$. Multiplication 
of words consists in combining sequences, e.g. 
$w\cdot w'=(e_1..e_N)\cdot (e'_1..e'_{N'}):=   
(e_1..e_N e'_1..e'_{N'})$. The involutive structure is defined by 
$w^\ast:=(e_N^\ast..e_1^\ast)$. We now consider the two sided ideal 
$\mathfrak{I}$ generated by elements of the form 
1. $ ee'-e' e-i\hbar\{e,e'\}$ and 2. $e^\ast-\overline{e}$. The quantum 
algebra 
is the quotient $\mathfrak{A}:=\mathfrak{F}/\mathfrak{I}$.
\item[III.] {\it Kinematical Hilbert space}\\
Next we study the representation theory of 
$\mathfrak{A}$. As is well known, for field theories such as General 
Relativity the number of unitarily inequvalent representations is 
usually uncountably infinite. For instance, all Fock representations of 
a free massive scalar field with different masses are unitarily 
inequivalent. This follows by a simple application of Haag's 
theorem \cite{Haag}. Hence, in order to select from this multitude of 
possibilities one must use dynamical input, such as the mass in the 
scalar field example. In the case of the presence of the constraints,
dynamical input into the representation problem is provided for instance 
by asking that (parts of) the gauge group be represented unitarily on 
the corresponding Hilbert space or that the constraints be represented 
at all as closable and densely defined operators, possibly subject to 
some choice of factor ordering and maybe after some sort of 
regularisation and renormalisation. For the purpose of this discussion 
it will be sufficient to insist that the kinematical Hilbert space 
$\cal H$ carries the 
Master Constraint Operator $\widehat{M}$ as a positive and self -- 
adjoint operator.
\item[IV.] {\it Physical Hilbert space}\\
The idea to solve the Master Constraint is to apply 
spectral theory to it \cite{Test1}. Suppose that the Hilbert $\cal H$ 
decomposes into separable $\widehat{M}-$invariant subspaces ${\cal 
H}_\theta$ where $\theta$ labels the corresponding sectors. Then it is 
well known that ${\cal H}_\theta$ is unitarily equivalent to a direct 
integral Hilbert space
\be \label{3.1}
{\cal H}_\theta \cong {\cal H}^\oplus_\theta:=\int_{{\rm 
spec}(\widehat{M})}^\oplus\;d\mu(\lambda)\;{\cal H}^\theta_\lambda
\ee
Here the measure class $\mu$ on the spectrum spec$(\widehat{M})$ of 
the Master Constraint is unique and the multiplicities 
dim$({\cal H}^\theta_\lambda)$ are unique up to $\mu-$measure zero sets. 
The unitary map $U: {\cal H}_\theta\to {\cal H}^\oplus_\theta;\;
\psi \mapsto (\tilde{\psi}(\lambda))_\lambda$ is a generalisation of the 
Fourier transform and is such that $U\widehat{M} \psi=(\lambda 
\tilde{\psi}(\lambda))_\lambda$, that is $U \widehat{M} U^{-1}$ is 
represented as multiplication by $\lambda$ on ${\cal H}^\theta_\lambda$.
We have 
\be \label{3.2}
<\psi,\psi'>_{{\cal H}_\theta}=<U\psi,U\psi'>_{{\cal 
H}^\oplus_\theta}=\int_{{\rm spec}(\widehat{M})}\; d\mu(\lambda)\;  
<\tilde{\psi}(\lambda),\tilde{\psi}'(\lambda)>_{{\cal H}^\theta_\lambda}
\ee
The physical Hilbert space is now defined as 
\be \label{3.3}
{\cal H}_{{\rm phys}}:=\oplus_\theta\; {\cal H}^\theta_{\lambda=0}
\ee
There are several remarks in order about (\ref{3.3}):\\
1.\\
In order that this works one must Lebesgue decompose every space 
${\cal H}_\theta$ into the $\widehat{M}-$invariant pure point, 
absolutely continious and continuous singular pieces 
and then decompose them as a direct integral.\\
2.\\
The spaces ${\cal H}^\theta_{\lambda=0}$ are uniquely determined in the 
pure point case but in the absolutely continuous case (and continuous 
singular case, which usually is absent in practice) further input is 
required because here the set $\{\lambda=0\}$ is of $\mu-$measure zero.
Roughly speaking, one requires that the space ${\cal H}^\theta_0$ 
carries a non -- trivial, irreducible representation of the algebra of 
(strong) observables. See \cite{Test1} for details.\\
3.\\
Due to a bad choice of factor ordering involved in the construction 
of $\widehat{M}$ it may happen that $0\not \in {\rm spec}(\widehat{M})$.
This typically happens when the quantum constraints $\widehat{C}_I$ that 
enter the definition of $\widehat{M}$ are anomalous, that is,
if they do not close as a quantum algebra. This can easily happen 
especially in the case that the classical constraint algebra involves 
non trivial structure functions rather than structure constants. Hence,
although the Master Constraint always trivially forms a non anomalous 
algebra, possible anomalies in the original algebra are detected by it,
so nothing is swept under the rug. In this case, following 
\cite{Klauder}, we propose to replace 
$\widehat{M}$ by $\widehat{M}'=\widehat{M}-\lambda_0$ where 
$\lambda_0=\min({\rm 
spec}(\widehat{M}))$. Here $\lambda_0$ should be finite and 
$\lim_{\hbar\to 0}\lambda_0=0$ in order that both 
$\widehat{M},\;\widehat{M}'$ have the same classcial limit. This has 
worked so far in all studied cases \cite{Test2} where $\lambda_0$ is 
related to a reordering or normal ordering of the constraints into a
non anomalous form. \\
4.\\
In case that the constraints can be exponentiated to a Lie group
$\mathfrak{G}$ one can avoid the construction of the Master Constraint 
and apply a more heuristic technique called group averaging \cite{GA}. 
This is at most possible if the constraints form an honest Lie algebra 
with structure constants rather than structure functions. Since we 
may assume without loss of generality that the constraints and the 
structure constants are real valued, we assume that we are given a unitary
representation $U$ of $\mathfrak{G}$ on $\cal H$. Assume also that there 
is a 
Haar measure $\nu$ on $\mathfrak{G}$, that is, a not necessarily 
normalised but bi -- invariant (with respect to group translations) 
positive measure on $\mathfrak{G}$. Fix a dense domain $\cal D$ and let 
${\cal D}^\ast$ be the algebraic dual of ${\cal D}$, i.e. linear 
functionals on $\cal D$ with the topology of pointwise convergence of 
nets. We define the rigging map
\be \label{3.4}
\eta:\;{\cal D}\to {\cal D}^\ast;\;f\mapsto \int_{\mathfrak{G}}\;
d\nu(\mathfrak{g})\; <U(\mathfrak{g})f,.>
\ee
The reason for restricting the domain of $\eta$ to a dense 
subset $\cal D$ of $\cal H$ is that in general only then (\ref{3.4}) 
defines an element of ${\cal D}^\ast$.
 
The image of $\eta$ are solutions to the constraints in the sense 
that\footnote{In general, given an operator $O$ which together with 
its adjoint $O^\dagger$ is densely defined on ${\cal D}\subset {\cal H}$ 
and 
preserves ${\cal D}$ we define the dual $O'$ on the algebraic dual 
${\cal D}^\ast$ by $[O' l](f):=l(O^\dagger f)$ for all $f\in {\cal D}$.} 
\be \label{3.4a}
[\eta(f)](U(\mathfrak{g})f')=[\eta(f)](f')
\ee
for all 
$\mathfrak{g}\in \mathfrak{G}$ and all $f'\in {\cal D}$. Notice that if 
we would identify the distribution $\eta(f)$ with the formal vector 
\be \label{3.5}
\eta'(f):=\int_{\mathfrak{G}}\;d\nu(\mathfrak{g})\; U(\mathfrak{g})f
\ee
then its norm diverges unless $\nu$ is normalisable, that is, unless 
$\mathfrak{G}$ is compact so that $\eta'(f)$ is not an element of ${\cal 
H}$ in general. However, formally we have $<\eta'(f),f'>=[\eta(f)](f')$ 
and thus 
\be \label{3.6}
<U(\mathfrak{g})\eta'(f),f'>=
<\eta'(f),U(\mathfrak{g}^{-1})f'>
=\eta(f)[U(\mathfrak{g}^{-1})f']
=<\eta'(f),f'>
\ee
for all $\mathfrak{g},\;f'$. Thus formally 
$U(\mathfrak{g})\eta'(f)=\eta'(f)$ which shows that $\eta'(f)$ is a 
(generalised, since not normalisable) eigenvector of all the 
$U(\mathfrak{g})$ with eigenvalue 
equal to one as appropriate for a solution to the constraints. Hence
(\ref{3.4a}) is the rigorous statement of the formal computation 
(\ref{3.6}).  

We define the physical inner product on the image of $\eta$ by
\be \label{3.7}
<\eta(f),\eta(f')>_{{\rm phys}}:=\eta(f')[f]
\ee
and the physical Hilbert space is the completion of $\eta({\cal D})$ 
in the corresponding norm\footnote{Provided that (\ref{3.7}) is 
positive semidefinite and with removal of zero norm vectors 
understood.}.\\
5.\\
The spectral decomposition solution of the constraint can be seen as a 
special case of group averaging in the sense that in case of a single 
self -- adjoint constraint $\widehat{M}$ we can indeed exponentiate it
to obtain a one parameter unitary, Abelean group 
$U(t)=\exp(it\widehat{M})$. The Haar measure in this case would seem
to be the Lebesgue measure $d\nu(t)=dt/(2\pi)$. We then formally have
(we drop the label $\theta$)
\ba \label{3.8}
<\eta(f),\eta(f')>_{{\rm phys}}
&=&\int_{\Rl} \; d\nu(t)\; <U(t)f',f>  
\nonumber\\
&=& \int_{\Rl} \; d\nu(t)\; \int_{{\rm spec}(\widehat{M})} 
\;d\mu(\lambda) \; 
e^{-it\lambda}<\tilde{f}'(\lambda),\tilde{f}(\lambda)>_{{\cal 
H}_\lambda}  
\nonumber\\
&=&  \int_{{\rm spec}(\widehat{M})} 
\;d\mu(\lambda) \; 
<\tilde{f}'(\lambda),\tilde{f}(\lambda)>_{{\cal 
H}_\lambda}
\int_{\Rl} \; d\nu(t)\; e^{-it\lambda}
\nonumber\\
&=& c 
<\tilde{f}'(0),\tilde{f}(0)>_{{\cal H}_0}
\ea
where $c=[\int_{{\rm spec}(\widehat{M})} 
\;d\mu(\lambda) \; \delta(\lambda)]$. This calculation is formal in 
the sense that we have interchanged the sequence of the integrations.
Also the constant $c$ can be vanishing or divergent which is one of the 
reasons why group averaging is only formal. For instance in the case 
of a pure point spectrum the appropriate measure is not the Lebesgue 
measure but rather the Haar measure on the Bohr compactification of the 
real line. See \cite{Test1} for the details. However, at least 
heuristically one sees how these methods are related.  
\end{itemize}
This ends the outline of the quantisation programme. We now apply it to 
General Relativity.
   
\section{Status of the quantisation programme for Loop Quantum Gravity 
(LQG)}
\label{s4}

In this section we describe to what extent the canonical quantisation 
programme has been implemented for General Relativity, that is,
we give the status of Loop Quantum Gravity. As an aside we sketch the 
historical development of the subject. We will mostly consider pure 
gravity, matter coupling works completely similar \cite{QSDII}. Also, in 
order to avoid technicalities about boundary terms (which can be dealt 
with \cite{QSDII}) consider compact $\sigma$ without boundary unless 
stated otherwise.

\subsection{Canonical quantum gravity before LQG}
\label{4.0}

The canonical quantisation of General Relativity in terms of the ADM 
variables \cite{Wald} culminated in the seminal work by DeWitt 
\cite{DeWitt} which {\it formally} carried out many of the steps 
outlined in the previous section. These crucial papers laid the 
foundations for a substantial amount of 
work on the canonical quantisation of General Relativity 
that followed. However, the stress is here on the word 
{\it formally}. We mention just some problems with these 
pioneering papers.
\begin{itemize}
\item[1.] {\it Kinematics:}\\
The Hilbert space representation used there was given 
in terms of a formal path integral which must be called ill -- defined 
by the standards of a mathematical physicist. For instance, the 
``measure'' was defined to be an infinite Lebesgue measure $[Dq]$ over a 
space of three metrics, an object 
that does not exist mathematically; the integration space, which should 
be given the appropriate structure of a measurable space was not 
specified etc. 

Nonetheless, if one defines the three metric operator as a 
multiplication operator and the conjugate momentum operator as a 
functional differential operator times $i\ell_P^2$ then one arrives at a 
formal representation of the canonical commutation relations such
that the canonical coordinates are represented as formally symmetric 
operators. Moreover, the formal Lebesgue measure is formally invariant 
under infinitesimal spatial diffeomorphisms.
\item[2.] {\it Dynamics}
\begin{itemize} 
\item[2a.] {\it Spatial Diffeomorphism Invariance:}\\
In order to solve the spatial diffeomorphism constraint one can assume 
that wave functions are normalisable functionals of 
spatially diffeomorphism invariant functions of the tree metric such 
as integrals over $\sigma$ of scalar densities of weight one constructed 
from the metric, the curvature tensor and all its covariant derivatives.  
In order that those derivatives make sense one must assume that the 
functional integral is over smooth three metrics. However, even if 
the wave function is say of the form $\exp(-\int_\sigma d^3x 
\sqrt{\det(q)})$ which is damped for large $q$ then the functional 
integral is ill defined: Due to spatial diffeomorphism invariance of the 
wave function and measure, the infinite volume of Diff$(\sigma)$ must be
factored out. But even after that, the space of smooth metrics is 
typically of measure zero with respect to the Gaussian type measure 
$[Dq]\;\exp(-2\int_\sigma d^3x \sqrt{\det(q)})$. Finally
the function $\det(q)$ can stay small while components of $q_{ab}$ 
can become large, hence the exponent has flat directions so that the 
integral also has divergent modes.
Hence the norm of these type of states are dangerously close to being 
either plain infinite or plain zero. 
\item[2b.] {\it Hamiltonian Constraints}\\
The infinite number of 
Hamiltonian constraints were formally given as a functional differential 
equation of second order, which goes by the famous name Wheeler -- 
DeWitt 
equations.
However, these ``operators'', which are really products of operator 
valued distributions multiplied at the same point in $\sigma$, are 
hopelessly divergent on the space of wave functions just specified
where the divergence really orginates from the product of operator 
valued distributions. There was no ``normal ordering'' or 
renormalisation  
possible because no exact vaccum state could be found with respect to 
which one should normal order.
\end{itemize}
\end{itemize}
It is therefore fair to say that canonical Quantum Gravity got stuck 
at the level of \cite{DeWitt} in the mid sixties of the past century.

\subsection{The new phase space}
\label{s4.2}

In a sense, in terms of the ADM variables one could never even find a 
proper, background independent representation of the canonical 
commutation relations. Thus, even leaving the dynamics aside, one could 
never even finish the kinematical part of the programme.

With the advent of the new variables \cite{New} there was new hope. 
Initially the new variables consisted, instead of a three metric and 
(essentially) the extrinsic curvature as a canonical pair, of an 
$SL(2,\Cl)$ connection $A^\Cl$ and an imaginary $sl(2,\Cl)$ valued, 
pseudo two 
-- form\footnote{A pseudo -- two form is dual, via the totally 
antisymmetric, metric independent symbol, to a vector density.}
$E^\Cl$. This was attractive because the Hamiltonian constraint,
after multiplying it with the non -- polynomial factor\footnote{The 
determinant of the three metric is required to be everywhere non vanishing 
classically, hence the modified constraint captures the same information 
about the reduced phase space as the original one.} 
$\sqrt{\det(q)}$, becomes a {\it fourth order polynomial} 
$\tilde{H}=\sqrt{\det(q)} H$ in terms of $A^\Cl,\; E^\Cl$ which is no 
worse than in Yang -- Mills theory. Hence the dynamics seemed to be 
drastically simplified as compared to the ADM formulation with its non 
-- polynomial Hamiltonian constraint.  

The catch, however, was in the reality conditions: Namely, in order to 
deal with real rather than complex General Relativity one had to 
impose the reality conditions 
\be \label{4.1}
\overline{A^\Cl}+A^\Cl=2\Gamma,\;\overline{E^\Cl}+E^\Cl=0
\ee
where $\Gamma$ is spin connection of the triad $e$ determined by the 
three metric. Since essentially $E^\Cl=-i \sqrt{\det(q)} e$ it follows 
that $\Gamma$ and thus (\ref{4.1}) take a highly non polynomial form.
In fact, $\Gamma$ is a fraction whose numerator and denominator 
are homogeneous polynomials of degree three in $E^\Cl$ and its first 
partial derivatives. It is clear that to find a representation of 
the formal $^\ast-$algebra $\mathfrak{A}$ with (\ref{4.1}) as 
$^\ast-$relations is hopeless and to date nobody was successful. 

Despite of this, in \cite{AI,AL} an honest representation for a canonical 
theory 
based on an $SU(2)$ connection $A$ and a real $su(2)$ valued pseudo 
two form $E$ was constructed\footnote{That in this representation the 
pseudo two form is indeed an essentially selfadjoint operator valued 
distribution was only shown later in \cite{ALMMT}.}. More precisely, 
\cite{AI}
constructs a measurable space of generalised (distributional) connections
$\overline{{\cal A}}$
which turns out to be the Gel'fand spectrum of an Abelean 
$C^\ast-$subalgebra of the corresponding kinematical algebra 
$\mathfrak{A}$. In \cite{AL} a (regular, Borel, probability) measure 
$\mu_0$ 
on $\overline{{\cal A}}$ was constructed. Thus, the corresponding Hilbert 
space ${\cal H}:=L_2(\overline{{\cal A}},d\mu_0)$ is a space of square 
integrable functions on $\overline{{\cal A}}$. As expected \cite{MTV}, the 
space of 
classical (smooth) connections $\cal A$ is contained in a measurable 
subset of $\overline{{\cal A}}$ of measure zero. Hence, any (formal) 
state
which requires to be restricted to smooth connections in order that, say 
the Hamiltonian constraint be defined on it, has zero norm and thus 
can be discarded from $\cal H$. For the first time, these and related 
questions could be answered with absolute precision.

However, what does this Hilbert space have to do with General Relativity 
if the true phase space is in terms of $SL(2,\Cl)$ plus complicated 
reality conditions rather than $SU(2)$ with simple reality conditions?
In \cite{Barbero} it was pointed out that the Hilbert space $\cal H$
can still be considered as a representation space for the quantum 
kinematics of General Relativity. In fact, the connection $A$ and pseudo
two form $E$ are related to triad $e$ and extrinsic curvature $K$
by\footnote{If the $SU(2)$ Gauss law holds.} 
\be \label{4.2}
A_a^j=\Gamma_a^j+\beta K_{ab} e^b_j,\;E^a_j=\sqrt{\det(q)} e^a_j/\beta
\ee
where $a,b,c,..=1,2,3$ are spatial tensor indices,
where $j,k,l,..=1,2,3$ are $su(2)$ Lie algebra indices and the 
real number $\beta$ is called the Immirzi parameter \cite{Immirzi}. For 
any (nonvanishing) value of $\beta$, the variables $(A,E)$ are canonically 
conjugate and thus can be used as a kinematical starting point for the 
quantisation programme.  

The price to pay is that, in order 
to keep it polynomial, one has to multiply the Hamiltonian 
constraint (which of course depends explicitly on $\beta$) by a 
sufficiently large power of 
$\det(q)=|\det(E)|$. This was considered to be rather unattractive because 
these very high degree polynomials would intuitively drastically worsen 
the UV singularity structure of the Hamiltonian constraint as compared to 
the ADM formulation. In fact, this UV problem was already noticed at a 
rather formal
level with the quantum version of $\tilde{H}$ in terms of the complex 
variables \cite{BP}: All the formal solutions to the Hamiltonian 
constraint were soltions at the regularised level only (in some ordering).
When taking the (point splitting) regulater away, the result would be of 
the type zero times infinity. These problems were expected to even worsen 
when increasing the polynomial degree of the Hamiltonian constraint.
Hence, the initial excitement that formally Wilson loop functions of
smooth and non intersecting loops were formally annihilated by the 
Hamiltonian constraint dropped significantly.\\
\\
Hence, a critic could have said at this point:\\
{\it You have made the theory more complicated and you have not gained
anything: You may have a rigorous kinematical framework but that framework 
does not support the quantum dynamics of the theory.}\\
\\ 
In \cite{PRL} it was demonstrated how these obstacles can be overcome:\\
One can show that General Relativity or any other 
background independent quantum field theory is {\it UV self -- regulating}
provided one equips the Hamiltonian constraint with its natural density 
weight equal to one as it automatically appears in the classical analysis.
Notice that the Hamiltonian of the standard model on Minkowski space has 
density weight two rather than one. This is the reason why in background 
dependent quantum field theories UV singularities appear. One can 
intuitively understand this as follows: In background dependent theories, 
Hamiltonians are spatial 
integrals over sums of products of operator valued distributions evaluated 
at the same point. Such products are therefore divergent. In background 
independent theories such polynomials $P$ also appear, however, they 
appear as 
numerators in a fraction $P/Q$. The denominator $Q$ of that fraction  
is an appropriate power of $\sqrt{\det(q)}$ such that $P/Q$ is a scalar 
density of weight one. As one can show, if the numerator has the 
singularity structure of the $(n+1)-$th power of the 
$\delta-$distribution\footnote{Notice that the $\delta-$distribution 
$\delta(x,y)$ transforms as a density of weight one in, say $x$ and as a 
scalar in, say $y$.} (and its spatial derivatives) then the denominator 
has the singularity structure of the $n-$th power. This must be the case 
in order that the oparator valued distribution has the correct density 
weight. Hence, in a proper (point splitting) regularisation of $P/Q$
one can, intuitively speaking, ``factor out'' out $n$ of those 
$\delta-$distributions and one is left with a well -- defined integral
after removing the regulator. 

In other words, it was wrong to 
assume that the Hamiltonian constraint should be polynomial. Rather,
{\it it must be non -- polynomial} in order that it is well defined.
Of course, the details are not as simple as that and we will explain the 
open issues in the next section. However, even at this stage one can 
say:\\
\\
{\it What has been gained is that not only a rigorous kinematical 
framework has been erected, that framework also supports the quantum 
dynamics. In particular, the original problem of the reality conditions 
is completely resolved.}\\
\\
One of the most important issues is whether that dynamics defined by the 
final, regulator free, Hamiltonian constraint operator, which underwent 
rather non trivial regularisation steps until one removed the 
regulator, 
reduces to the classical one in an appropriate classical limit.
We will have much to say about this point in subsequent sections.\\
\\
Before we close this section, let us comment on some criticism that one 
might have encountered \cite{Samuel,Alexandrov}: The complex connection 
is actually the pull back to $\sigma$ of the (anti) self -- dual part of 
the 4D spin connection. Hence it has a covariant interpretation. The real
valued connection is not related to a covariant action as simply as that.
The relation is as follows: Additional to the Palatini action one 
considers a term which is topological on shell which amounts to the
total action
\be \label{4.3}
S=\int_M F_{IJ}\wedge \ast(e^I\wedge e^J)+\frac{1}{\beta} \int_M 
F_{IJ}\wedge
(e^I\wedge e^J)
\ee
Here $I,J,K,..=0,1,2,3$ are Lorentz indices, $e^I$ is the cotetrad
one form and $F_{IJ}$ is the curvature for a Lorentz connection $A_{IJ}$.
The first term in (\ref{4.3}) is the Palatini action. The second term is a 
total 
derivative when substituting the equation of motion for the connection 
$A^{IJ}$. Now when performing the Legendre transform of (\ref{4.3}) one 
encounters second class constraints \cite{Henneaux}. These must be 
eliminated by using the Dirac bracket or by partially fixing the Lorentz 
gauge symmetry $SO(1,3)$ (or its universal cover $SL(2,\Cl)$) to $SO(3)$ 
(or 
$SU(2)$) respectively, called the time gauge. 

Using the Dirac bracket 
leads to a Poisson structure with respect to which connections are not 
Poisson commuting. Hence, while manifestly originating from a covariant 
action, the Lorentz connections cannot be used as a configuration space 
in the quantisation programme, that is, they cannot be represented as 
(commuting) multiplication operators \cite{Alexandrov}. In fact, to date 
there is no honest representation based on Lorentz connections available.
On the other hand, the time gauge 
immediately leads to the phase space description sketched above 
with Immirzi parameter $\beta$. The manifest covariant origin of the phase 
space is lost due to the gauge fixing of the Lorentz group \cite{Samuel}, 
however, one can show easily \cite{TB,reviews} that 
symplectic reduction with respect to the $SU(2)$ Gauss constraint results
in the manifestly covariant ADM phase space. Hence, both criticisms are of 
purely aesthetical nature and do not give rise to either an obstacle or an
insight concerning the quantisation.

\subsection{Quantum kinematics}
\label{s4.3}

\subsubsection{Elementary functions}
\label{s4.3.1}

Having convinced ourselves that the cotangent bundle 
${\cal M}:=T^\ast({\cal A})$
over the space 
of smooth $SU(2)$ connections is an appropriate kinematical phase space of 
General Relativity we are supposed to choose an appropriate Poisson
$^\ast-$subalgebra $\mathfrak{E}$ of elementary functions. Experience from 
lattice gauge theory \cite{Gambini} shows that it is convenient to work 
with $SU(2)$ valued magnetic holonomies 
\be \label{4.4}
A(e):={\cal P}\exp(\int_e A)
\ee
and real valued electric fluxes 
\be \label{4.5}
E_f(S):=\int_S \rm{Tr}(n\;\ast E)
\ee
Here $e$ is a path in $\sigma$, $S$ is a two surface in $\sigma$ and $n$ 
is a Lie algebra valued scalar\footnote{For simplicity we assume that the 
$SU(2)$ principal bundle is trivial which is always possible. The final 
quantum theory turns out 
not to be affected by this assumption. The paths and surfaces are 
piecewise analytic for technical reasons.}.  
These functions 
separate the points of $\cal M$ since $G=SU(2)$ is compact \cite{Giles}. 
Moreover, they satisfy the reality conditions 
\be \label{4.6}
\overline{A(e)}=[A(e^{-1})]^T,\;\;
\overline{E_n(S)}=E_n(S)
\ee
as well as the Poisson brackets
\be \label{4.7}
\{A(e),A(e')\}=0,\;\; \{E_f(S),A(e)\}=8\pi G_{{\rm Newton}}\beta 
A(e_1)f(S\cap e) A(e_2)
\ee
Here we have assumed that $e$ and $S$ intersect transversally in an 
interior point of both $S,e$ thus splitting the path $e$ at $S\cap e$ as 
$e=e_1\circ e_2$ and $G$ is Newton's constant. Similar 
formulae can be derived if $S,e$ intersect in a more 
complicated way. 

The algebra $\mathfrak{E}$ can now be described as follows: Consider 
the algebra Cyl of cylindrical functions, that is, those which depend 
on a finite number of holonomies only. Hence, a cylindircal function is of 
the form $f(A)=f_\gamma(\{A(e)\}_{e\in E(\gamma)})$ where $\gamma$ is an 
oriented graph (a collection of paths, called edges, which intersect in 
their endpoints only), $E(\gamma)$ denotes the set of edges of $\gamma$ 
and $f_\gamma$ is a complex valued function on $SU(2)^N$ where $N$ is the 
number of edges of $\gamma$. Next, consider the vector field $u_{S,n}$, 
considered as a derivation on Cyl, defined by
\be \label{4.8}
u_{S,n}[f]:=\{E_n(S),f\}=\sum_{e\in E(\gamma)} \{E_n(S),[A(e)]_{mn}\}
\frac{\partial f_\gamma}{\partial [A(e)]_{mn}}
\ee
The algebra $\mathfrak{E}$ is now defined as the Lie algebra generated 
by the pairs $(f,u)$ where $f\in$Cyl and $u$ is a derivation on Cyl 
which is either of the form of a finite linear combinations of the  
$u_{n,S}$ or which is generated from those by the Lie bracket 
$\{(f,u),(f',u')\}=(u[f']-u'[f],[u,u'])$ where $[u,u']$ denotes the 
commutator of vector fields.  

Notice that, in this sense, the Poisson bracket 
between the $E_n(S)$ is generically non vanishing. Its result is such 
that, formally, the Jacobi identity holds in $\mathfrak{E}$. The reason 
for this is that we do not smear the fields in three but in less 
dimensions. If we would smear in three dimensions as usual, then the 
smeared 
electric fields would Poisson commute. See \cite{ACZ} for more details on 
this point. The reason for why we do not smear the fields in three 
dimensions is due to the fact that it is natural in a background 
independent theory to smear one forms in one dimension and two forms in 
two dimensions. This way we do not need a backround metric in order raise 
or lower indices. Moreover our holonomies and fluxes transform in a simple 
way under the kinematical part of the gauge group, that is, $SU(2)$ gauge 
transformations and spatial diffeomorphisms. In fact, consider the smeared 
Gauss constraint  and spatial diffeomorphism constraint respectively given 
by 
\be \label{4.9}
G(\Lambda):=\int\sigma \; d^3x\;  \Lambda^j \; G_j,\;\;
D(v):=\int\sigma \; d^3x\;  v^a \; C_a
\ee
(where $\Lambda,\;v $ are test functions) 
where\footnote{${\cal D}$ and $F$ are respectively the covariant 
differential and curvature determined by $A$ and $\tau_j,\;j=1,2,3$ 
denotes a basis of $su(2)$.}
\be \label{4.9a}
G_j={\rm Tr}(\tau_j {\cal D}_a E^a),\;
D_a={\rm Tr}(F_{ab} E^b)
\ee
and the one parameter 
families of canonical transformations generated by them. These are 
explicitly given by, say on $f\in$ Cyl 
\ba \label{4.10}
\alpha_{\exp(t\Lambda)}(f) &:=& \sum_{n=0}^\infty \frac{t^n}{n!}
\{C(\Lambda),f\}_{(n)}
\nonumber\\
\alpha_{\varphi^t_v}(f) &:=& \sum_{n=0}^\infty \frac{t^n}{n!}
\{C(v),f\}_{(n)}
\ea
where $b(e),\;f(e)$ denote respectively the beginning and final point
of $e$.
Here $t\mapsto \varphi^t_v$ is the one parameter family of diffeomorphisms
generated by $v$. 
It is not difficult to see that (\ref{4.10}) is the restriction to 
local gauge transformations of the form $g=\exp(t\Lambda)$ and spatial 
diffeomorphisms of the form $\varphi=\varphi^t_v$ of the following
action of the semidirect product $\mathfrak{G}={\cal G} \rtimes {\rm 
Diff}(\sigma)$ on Cyl given by
\ba \label{4.10a}
[\alpha_g(f)](A) &=& f_\gamma(\{g(b(e)) A(e) 
g(f(e))^{-1}\}_{e\in E(\gamma)})
\nonumber\\
{[}\alpha_\varphi(f)](A) &=& f_\gamma(\{A(\varphi(e))\}_{e\in E(\gamma)})
\ea
There is a similar action on the vector fields $u_{n,S}$. As the 
notation suggests, the maps $\alpha_g,\;\alpha_\varphi$ are  
automorphisms of $\mathfrak{E}$, that is 
$\alpha_.(\{a,b\})=\{\alpha_.(a),\alpha_.(b)\}$ for any $a,b\in 
\mathfrak{E}$ as one can easily verify. Hence we have a representation 
of $\mathfrak{G}$ as automorphisms on $\mathfrak{E}$.

\subsubsection{Quantum $^\ast-$algebra}
\label{s4.3.2}

We follow the standard construction of section \ref{s3}:\\
Consider the free $^\ast-$algebra $\mathfrak{F}$ generated by 
$\mathfrak{E}$. That is, we consider finite linear combinations 
of ``words'' $w$ constructed from $\mathfrak{E}$. A word is simply 
a formal finite sequence $w=(a_1..a_N)$ of elements $a_k$ of 
$\mathfrak{E}$.
Multiplication of words is defined as $w\cdot w'=(a_1 .. a_N a'_1 .. 
a'_{N'})$ where $w=(a_1.. a_N),\;w'=(a'_1.. a'_{N'})$. The $^\ast$
operation on $\mathfrak{F}$ is $w^\ast=(\bar{a}_N..\bar{a}_1)$. 

Consider the two sided ideal $\mathfrak{I}$ in $\mathfrak{F}$ generated 
by elements of the form 
\be \label{4.12}
(a) \cdot (b)-(b)\cdot (a)-i\hbar (\{a,b\})
\ee
for all $a,b\in \mathfrak{E}$.
Then the quantum $^\ast-$algebra is defined as the quotient
\be \label{4.13}
\mathfrak{A}:=\mathfrak{F}/\mathfrak{I}
\ee
We can now simply lift the automorphisms labelled by $\mathfrak{G}$ from 
$\mathfrak{E}$ to $\mathfrak{A}$ by $\alpha_.(w)=(\alpha_.(a_1) .. 
\alpha_.(a_N)$. 

\subsubsection{Representations of $\mathfrak{A}$}
\label{s4.3.3}

In quantum field theory representations of $\mathfrak{A}$ are never 
unique in contrast to the situation in quantum mechanics where the 
Stone -- von Neumann theorem guarantees that irreducible and weakly 
continuous representations of the Weyl algebra generated 
by the unitary operators $U(x)=\exp(ixq),\;V(y)=\exp(iyp),\;x,y\in \Rl$
are automatically unitarily equivalent to the Schr\"odinger 
representation. For example, an appeal to Haag's theorem \cite{Haag}
reveals that Fock representations for free massive scalar fields with 
different masses are unitarily inequivalent representations of the 
corresponding Weyl algebra. Hence already in this simplest case we 
have an uncountably infinite number of unitarily inequivalent 
representations of the canonical commutation relations and all of them 
satisfy the Wightman axioms, e.g. Poincar\'e invariance. 
In order to 
single out preferred representations one must use additional criteria 
from physics. In the case of the scalar field, the representation is 
fixed if we insist on the Wightman axioms plus specifying the mass of the 
scalar field. Hence we need dynamical input as pointed out in 
\cite{Araki}.
 
In the case of our algebra $\mathfrak{A}$ the idea is to use dynamical 
input from the kinematical gauge algebra $\mathfrak{A}$. Namely, we 
want a unitary representation of $\mathfrak{G}$ on the Hilbert space.
To do this, recall that for any $^\ast-$algebra such as our 
$\mathfrak{A}$ it is true that any representation is a (possibly 
uncountably infinite) direct sum of cyclic representations. Hence it is 
sufficient to consider cyclic representations. Next, any cyclic 
representation is in one to one correspondence with a state $\omega$ on 
$\mathfrak{A}$ via the GNS construction \cite{Haag}. Here a {\it state} 
is defined as a positive linear functional 
on $\mathfrak{A}$, that is, $\omega(w^\ast w)\ge 0$ for all $w\in 
\mathfrak{A}$. It is not to be confused with {\it vectors}, that is,
elements of some Hilbert space. Hence, it suffices to consider states 
on $\mathfrak{A}$. 

The physical input to have a unitary representation 
of $\mathfrak{A}$ on the GNS Hilbert space ${\cal H}_\omega$ determined     
by $\omega$ now amounts to asking that the state $\omega$ be 
$\mathfrak{G}-$invariant. To see this we have to recall some elements of 
the GNS construction:\\
The GNS construction means that there is a one to one correspondence 
between states $\omega$ on a (unital) $^\ast-$algebra $\mathfrak{A}$ and 
GNS data 
$({\cal H}_\omega,\;\pi(\omega),\;\Omega_\omega)$. Here ${\cal 
H}_\omega$ is a Hilbert space, $\pi_\omega$ is a representation of 
$\mathfrak{A}$ by densely defined and closable operators on ${\cal 
H}_\omega$ and $\Omega_\omega$ is a cyclic vector in ${\cal H}_\omega$.
Cyclic means that $\pi_\omega(\mathfrak{A})\Omega_\omega$ is dense in 
${\cal H}_\omega$. This is done as follows: Consider the subspace 
of $\mathfrak{A}$ (considered as vector space) defined by 
$\mathfrak{J}:=\{w\in \mathfrak{A};\;\omega(w^\ast w)=0\}$. It is not 
difficult to show that this is a left ideal. Consider the equivalence 
classes $[w]:=\{w+w';w'\in \mathfrak{J}\}$. Then ${\cal H}_\omega$ is 
the closure of the vector space $\mathfrak{A}/\mathfrak{J}$ of 
euivalence classes, $\Omega_\omega:=[{\bf\rm 1}]$ and 
$\pi_\omega(w)[w']:=[ww']$. The scalar product is defined as 
$<[w],[w']>_{{\cal H}_\omega}:=\omega(w^\ast w')$. Now if 
$\omega$ is in addition 
$\mathfrak{G}$ invariant then by using the automorphism property it is
easy to see that $U_\omega(\mathfrak{g})[w]:=[\alpha_{\mathfrak{g}}(w)]$ 
is a unitary representation of $\mathfrak{G}$ with $\mathfrak{G}$ -- 
invariant cyclic vector $\Omega_\omega$.\\
\\
The surprising result is now the following structural theorem 
\cite{LOST}.
\begin{Theorem} \label{th4.1} ~\\
The only $\mathfrak{G}$ -- invariant state on the holonomy -- flux algebra 
$\mathfrak{A}$ is the Ashtekar -- Isham -- Lewandowski state 
$\omega_{AIL}$ whose GNS data coincide with the Ashtekar -- Isham - 
Lewandowski representation.
\end{Theorem}
The surprising aspect to this theorem is that not the full gauge 
symmetry of the theory associated with the Hamiltonian constraint had 
to be used. In fact it is actually sufficient to just use the spatial 
diffeomorphism invariance in order to prove the theorem\footnote{The 
careful statement of the theorem uses semianalytic rather than smooth 
structures on $\sigma$. For every smooth structure there is always a 
semianalytic structure and semianalytic charts are equivalent up to 
smooth diffeomorphisms. Semianalyticity is the rigorous formulation of 
the more intuitive notion of piecewise analyticity. See \cite{LOST}
for details.}. 

The assumptions of the theorem are fairly weak as one can 
see. A possible generalisation is as follows: We have implicitly assumed 
that the flux operators themselves exist as self -- adjoint operators on 
the Hilbert space. This is equivalent to asking that the state is 
regular, i.e. weakly continuous with respect to the one parameter 
unitary groups they generate. This needs not to be the case. In 
\cite{Fleischhack} it was shown that including non -- regular states 
into the analysis does not change the uniqueness result modulo a slight 
additional assumption that one has to make. This is to say that the 
uniqueness result is fairly robust. It is rather important in the 
following sense: Suppose we had found a multitude of representations 
which satisfy the physical criterion of $\mathfrak{G}-$invariance. 
Then each of them would be a bona fide kinematical starting point for 
the Dirac quantisation programme which would amount to a large amount 
of ambiguity. The uniqueness result excludes this possibility and we can 
thus be confident to use the Hilbert space ${\cal H}_{AIL}$.

\subsubsection{The kinematical Hilbert space and its properties}
\label{s4.3.4}

There are several complementary characterisations of the kinematical 
Hilbert space ${\cal H}:={\cal H}_\omega$ which are useful in different 
contexts. This section is for the mathematically inclined reader and 
can be skipped by readers interested only in the conceptual framework.
\begin{itemize}
\item[1.] {\it Positive linear functional characterisation}\\
We notice first of all that 
every word $w$ can be written, using the commutation relations 
(\ref{4.7}), (\ref{4.12})  as a 
finite linear combination of reduced words. A reduced word
is of the form $f u_{n_1 S_1} .. u_{n_N S_N}$ with $f\in$Cyl and 
arbitrary $n_k,\;S_k$ and $N=0,1,..$. Due to linearity it suffices to 
specify $\omega$
on reduced words. The definition is
\be \label{4.14}
\omega(w)=\left\{ \begin{array}{cc}
0 & \mbox{ if } N> 0 \\
\omega_0(f) & \mbox{ if } N=0 \end{array} \right.
\ee
Here $\omega_0$ is the so -- called Ashtekar -- Lewandowski positive 
linear 
functional on the $C^\ast-$algebra completion $\overline{{\rm Cyl}}$ 
of Cyl with respect to the sup norm. It can be explicitly written 
as
\be \label{4.15}
\omega_0(f)=\int_{SU(2)^n}\;d\mu_H(h_1)..d\mu_H(h_n)\;
f_\gamma(h_1,..,h_n)
\ee
for $f(A)=f_\gamma(A(e_1),..,A(e_n)$, i.e. $f$ cylindrical over a graph 
with $n$ edges. Here $\mu_H$ is the Haar measure on $SU(2)$. The Hilbert 
space $\cal H$ is the GNS Hilbert space derived from (\ref{4.14}).
\item[2.] {\it $C^\ast-$algebraic characterisation}\\
The completion $\overline{{\rm Cyl}}$ of Cyl with respect to the 
sup norm $||f||:=\sup_{A\in {\cal A}} |f(A)|$ defines an Abelean 
$C^\ast-$algebra \cite{Bratteli}. Define the space of generalised 
connections 
$\overline{{\cal A}}$ as its Gel'fand spectrum\footnote{That is, the set 
of all homomorphisms from the algebra into the complex numbers.} 
$\Delta(\overline{{\rm 
Cyl}})$ \cite{Bratteli}, also called the Ashtekar -- Isham space of 
generalised 
connections. By the Gel'fand isomorphism we can think of 
$\overline{{\rm Cyl}}$ as the space $C(\overline{{\cal A}})$ of 
continuous functions on the spectrum. The spectrum of an Abelean 
$C^\ast-$algebra is a compact Hausdorff space if equipped with the 
Gel'fand topology of pointwise convergence of nets. Hence, by the 
Riesz -- Markov theorem \cite{Rudin} the positive linear functional
$\omega_0$ in (\ref{4.15}) is in one to one correspondence with a 
(regular, Borel) measure $\mu_0$ on $\overline{{\cal A}}$ also called the 
Ashtekar -- Lewandowski measure. The Hilbert 
space ${\cal H}:=L_2(\overline{{\cal A}},d\mu_0)$ is the space of square
integrable functions on $\overline{{\cal A}}$ with respect to that  
measure.
\item[3.] {\it Projective limit characterisation}\\
The spectrum of $\overline{{\rm Cyl}}$ abstractly defined above can be 
given a concrete geometric interpretation. It can be identified set 
theoretically and topologically as the set of homomorphisms from the 
groupoid $\cal P$ of paths into $SU(2)$, that is, there is a homeomorphism  
$\overline{{\cal A}} \to {\rm Hom}({\cal P},SU(2))$ \cite{Velhinho}.
Here the groupoid of paths is defined, roughly speaking, as the set of  
(piecewise analytic) paths modulo retracings and reparametrisations 
together with the operations of 1. connecting paths with common beginning 
or end point and 2. inversion of orientation. Now recall that an element
$A\in \overline{{\cal A}}$ is a homomorphism from $\overline{{\rm Cyl}}$ 
into the complex numbers. Consider a function $f\in$Cyl cylindrical over 
some graph $\gamma$. Since $A$ is a homomorphism we have 
$A(f)=f_\gamma(\{A(h_e)\}_{e\in E(\gamma)})$ where for $A\in {\cal A}$,
$h_e(A)=A(e)$ is the holonomy map. Hence it suffices to 
consider the action of $A\in \overline{{\cal A}}$ on holonomy maps.  
Now since $h_e h_{e'}=h_{e\circ e'},\;h_e^{-1}=(h_e)^{-1}$ and $A$ is a 
homomorphism it follows that every point in the spectrum defines an 
element of Hom$({\cal P},SU(2))$. That also the converse is true is shown 
e.g. in \cite{AL2}, hence there is a bijection. 

To see that this 
bijection is a homeomorphism we must specify a topology on 
Hom$({\cal P},SU(2))$. To do this, we describe the space Hom$({\cal 
P},SU(2))$ as a projective limit: For every graph $\gamma$ we consider 
the space Hom$(\gamma,SU(2))$ of homomorphisms from the subgroupoid of 
paths within $\gamma$ (also denoted $\gamma$) into the gauge group. Since 
such homomorphisms
are completely specified by their action on the edges of the graph, the 
sets Hom$(\gamma,SU(2))$ are identified topologically with $SU(2)^n$ where 
$n$ is the number of edges of the graph. As such, Hom$(\gamma,SU(2))$ is a
compact Hausdorff space. The set of subgroupoids is partially 
ordered and directed with respect 
to the inclusion relation, that is, for any two $\gamma,\gamma'$ there
is $\tilde{\gamma}$ (e.g. $\gamma\cup \gamma'$) such that 
$\gamma,\gamma'\subset \tilde{\gamma}$. Given $\gamma\subset \gamma'$ we 
say that $A_{\gamma'}\in {\rm Hom}(\gamma',SU(2))$ is compatible with 
$A_{\gamma}\in {\rm Hom}(\gamma,SU(2))$ provided that the restriction 
of $A_{\gamma'}$ to $\gamma$ coincides with $A_\gamma$, that is,
$A_{\gamma'|\gamma}=A_\gamma$. The projective limit Hom$({\cal 
P},SU(2))$ (of the spaces 
Hom$(\gamma,SU(2)))$ is the (automatically closed) subset of the 
infinite direct product $\overline{X}$ of the Hom$(\gamma,SU(2))$ 
restricted to the compatible points. The space $\overline{X}$ carries the 
natural Tychonov topology \cite{Munkres} with respect to which it is 
compact and Hausdorff. This property is inherited by the closed subset 
Hom$({\cal P},SU(2))$ in the subspace topology. As one can show, the 
compact Hausdorff topologies on $\overline{{\cal A}}$ and 
Hom$({\cal P},SU(2))$ are identified by the above mentioned bijection
$A\mapsto (A_{|\gamma})_\gamma$ where $A_{|\gamma}$ is the restriction of 
$A$ to $\gamma$.

Also the measure $\mu_0$ abstractly defined via the Riesz -- Markov 
theorem can be given a nice projective description: On each subgroupoid 
$\gamma$ we consider the product Haar measure $\mu_{0,\gamma}$ as in 
(\ref{4.15}). Let $p_\gamma:\;{\rm Hom}({\cal P},SU(2))\to
{\rm Hom}(\gamma,SU(2));\;A\mapsto A_{|\gamma}$ be the restriction map.
The system of measures $\mu_{0,\gamma}$ satisfies the following 
compatibility condition: For every $\gamma\subset \gamma'$ we have 
$\int d\mu_{0,\gamma'} f_\gamma=\int d\mu_{0,\gamma} f_\gamma$
for every $f=f_\gamma \circ p_\gamma$ cylindrical over $\gamma$. 
This property qualifies the $\mu_{0,\gamma}$ as the cylindrical 
projections \cite{Yamasaki} $\mu_{0,\gamma}=\mu_0\circ p_\gamma^{-1}$ of a 
measure on the projective limit. Here the translation invariance and 
normalisation of the Haar measure are absolutely crucial to establish this 
property.
\item[4.] {\it Inductive limit characterisation}\\
We consider the Hilbert spaces ${\cal H}_\gamma({\rm 
Hom}(\gamma,SU(2)),d\mu_{0,\gamma})$. For every $\gamma\subset\gamma'$ 
there is an isometric embedding 
$U_{\gamma\gamma'}:\;{\cal H}_\gamma \to {\cal H}_{\gamma'}$. These
isometries satisfy $U_{\gamma\tilde{\gamma}}=U_{\gamma'\tilde{\gamma}}
U_{\gamma\gamma'}$ for all $\gamma\subset\gamma'\subset\tilde{\gamma}$.
This qualifies the ${\cal H}_\gamma$ as an inductive system of Hilbert 
spaces. The Hilbert space ${\cal H}$ is the corresponding inductive limit.
\end{itemize}
It is not difficult to show that this representation of $\mathfrak{A}$ is 
irreducible \cite{Irred}. Moreover,
it turns out that the Hilbert space $\cal H$ has an orthonormal 
basis over which one has complete control, the spin network basis
\cite{SNW}. These provide an indispensible tool in all analytical 
calculations in LQG. A {\it spin network} (SNW) is a quadruple\footnote{It 
is understood that at bivalent vertices such that the incident edges 
are at least $C^{(1)}$ continuations of each other, then in the 
intertwiner decomposition of the state (see below) no trivial 
representation occurs. Otherwise this leads to an overcounting problem.
Hence, if the intertwiner is trivial then such points are not counted as 
vertices.}
$s=(\gamma,j,m,n)$ consisting of a graph $\gamma$, a collection of
spin quantum numbers $j=\{j_e\}_{e\in E(\gamma)}$ and two collections of 
magnetic quantum numbers $m=\{m_e\}_{e\in E(\gamma)},\;
n=\{m_e\}_{e\in E(\gamma)}$ subject to the conditions $j_e=1/2,1,3/2,..$ 
and $m_e,\;n_e\in \{-j_e,-j_e+1,..,j_e\}$. The analytical expression
for a {\it spin network function} (SNWF) is given by (we write 
$A(e):=A(h_e)$ for $A\in \overline{{\cal A}}$)
\be \label{4.16}
T_s(A):=\prod_{e\in E(\gamma)}\;\sqrt{2j_e+1}\; 
[\pi_{j_e}(A(e))]_{m_e n_e}
\ee
Here $\pi_j$ is the spin $j$ irreducible representation of $SU(2)$. 
Its dimension is $2j+1$ and we label the entries of the corresponding 
matrices by $[\pi(h)]_{mn}$. \\
\\
Three important properties of ${\cal H}$ follow from the existence of the 
SNW basis:\\
1.\\ 
Since the set of finite graphs is an uncountably infinite 
set, the kinematical Hilbert space is therefore non separable since it 
does not have a countable basis.\\
2.\\
Consider a vector field $v$ on $\sigma$ and let $t\mapsto \varphi^v_t$ 
be the one parameter family of spatial diffeomorphisms generated by 
it\footnote{These are obtained by computing the integral curves $c^v_x(t)$
defined by $\dot{c}^v_x(t)=v(c^v_x(t)),\;c^v_x(0)=x$ and setting 
$\varphi^v_t(x):=c^v_x(t)$.}. Then the one parameter unitary group 
$t\mapsto U(\varphi^v_t)$ is not weakly continuous, that is, it does not 
hold that $\lim_{t\to 0} <\psi, U(\varphi^u_t)\psi'>=<\psi,\psi'>$ for all 
$\psi,\psi'\in {\cal H}$. To see this choose $\psi=\psi'=T_s$ such that 
the graph $\gamma$ underlying $s$ has support in the support of $v$. Then 
$<T_s,U(\varphi^v_t) T_s>=0$ for all $\epsilon>|t|>0$ for some $\epsilon$
because $U(\varphi) T_s=T_{\varphi\cdot s}$ where 
$\varphi\cdot s=(\varphi(\gamma),j,m,n)$ if 
$s=(\gamma,j,m,n)$.
By Stone's theorem \cite{Stone} this means that the infinitesimal 
generators of spatial diffeomorphisms do 
not exist as (self adjoint) operators on $\cal H$. \\
3.\\
On SNWF's the operators $A(e)$ act by multiplication while 
$E_{n,S}:=u_{n,S}$ 
becomes a linear combination of the right invariant vector fields 
$X^j_e={\rm Tr}([\tau_j A(e)]^T\partial/\partial A(e))$ on a copy of 
$SU(2)$ coordinatised by $A(e)$.

\subsection{Quantum dynamics}
\label{s4.4}

The quantum dynamics consists in two steps: 1. Reduction of the system
with respect to the gauge transformations generated by the constraints and 
2. Introduction of a notion of time with respect to which observables
(gauge invariant operators) evolve. 
It is convenient to subdivide the discussion of the reduction step 
into the gauge transformations corresponding to $\mathfrak{G}$ and 
those generated by the Hamiltonian constraint. We will also mention 
spin foam models which are the path integral formultion of LQG. Spin foam 
models were completely neglected in \cite{NPZ} although half of the 
current activity in LQG is devoted to them. This was partly corrected 
in \cite{NPZ1}. The presentation will be brief 
since our main focus is on the criticisms of \cite{NPZ} towards the 
canonical formulation.

\subsubsection{Reduction of Gauss -- and spatial diffeomorphism 
constraint}
\label{s4.4.1}

\paragraph{Gauss constraint}
\label{s4.4.1.1}

~\\
\\
The SNWF are not invariant under $\cal G$. It is easy to make them gauge 
invariant as follows: Pick a vertex $v\in V(\gamma)$ in the vertex set of 
$\gamma$ and consider the edges $e_1,..,e_N$ incident at it. Let us assume 
for simplicity that the edges are all outgoing from $v$, the general case 
is similar but requires more book keeping. It is easy to see that at
$v$ the state transforms in the tensor product representation 
$j_1\otimes .. \otimes j_n$ where $j_k:=j_{e_k}$. Hence in order to make 
the state gauge invariant, all we need to do is to couple the $N$ spins 
$j_1,..,j_N$ to resulting spin zero. This is familiar from the quantum 
mechanics of the angular momentum: We begin with 
\be \label{4.17}
|j_1 m_1> \otimes |j_2 m_2>=\sum_{j_12} <j_{12} m_1+m_2|j_1 m_1;j_2 m_2>
\;|j_{12} m_1+m_2>  
\ee
The recoupling quantum numbers take range in $j_{12}\in 
\{|j_1-j_2|,..,j_1+j_2\}$ and $<j_{12} m_1+m_2|j_1 m_1;j_2 m_2>$
is the familiar Clebsch -- Gordan coefficient (CGC) . Next we 
repeat (\ref{4.17}) with the 
substitutions $(j_1, m_1; j_2, m_2) \to (j_{12}, m_1+m_2;j_3, m_3)$.

The procedure is now iterated until all spins have been recoupled
to total
angular momentum $J=0$ and total magnetic quantum number $M=m_1+..+m_N=0$.
Consider the corresponding coefficients $<j_1 m_1;..;j_N m_N|J M>$
in the decomposition of $|j_1 m_1> \otimes .. \otimes |j_N m_N>$ into the 
$|J M>$. As we just showed, these can be written explicitly as polynomials 
of CGC's. We are interested only in those coefficients with $J=0$, called 
intertwiners $I_v$. This 
imposes some restriction on the range of the $j_k$ in 
order that this is possible at all. The number of those intertwiners 
does not depend on the sequence in which we couple those spins which is 
called a recoupling scheme. Different recoupling schemes are related by a 
unitary transformation. We now take one of those intertwiners and 
sum the SNWF times the intertwiner over all $m_k\in \{-j_k,..,j_k\}$.
The result is a state which is gauge invariant at $v$. Now repeat this 
for all vertices. 

The resulting states are gauge invariant and 
orthonormal with respect to the kinematical inner product by the 
properties of the CGC's and they define an orthonormal basis of the 
${\cal G}$ invariant Hilbert space. We will also denote them by $T_s$ 
where now $s=(\gamma,j,I)$ and $I=\{I_v\}_{v\in V(\gamma)}$.

\paragraph{Spatial diffeomorphism constraint}
\label{s4.4.1.2}

~\\
\\
While the solutions to the Gauss constraint were normalisable with respect 
to the kinematical inner product, this turns out to be no longer the case 
with respect to the spatial diffeomorphism constraint. Let $\cal D$ 
be the finite linear span of SNWF's which by construction is dense in 
$\cal H$. We will look for solutions to the spatial diffeomorphism 
constraint in the algebraic dual ${\cal D}^\ast$ of $\cal D$. The 
algebraic dual of $\cal D$ are simply linear functionals on $\cal D$ 
equipped with the topology of pointwise convergence of nets (weak 
$^\ast-$topology). It is clear that an element $l\in {\cal D}^\ast$ is 
completely specified by the numbers $l_s:=l(T_s)$. Hence we can write 
any element of ${\cal D}^\ast$ formally as the uncountable direct sum  
\be \label{4.18}
l=\sum_s \; l_s\; <T_s,.>
\ee
where the sum is over all gauge invariant spin network labels.

Following the group averaging technique described earlier,
we say that an element $l\in {\cal D}^\ast$ is spatially diffeomorphism 
invariant provided that 
\be \label{4.19}
l(U(\varphi) f)=l(f)
\ee
for all 
$\varphi \in$Diff$(\sigma)$ and all $f\in {\cal D}$. As we have 
seen, this definition 
is a direct generalisation from vectors $\psi\in {\cal H}$ to 
distributions of the equation 
$U(\varphi) \psi=\psi$ for all $\varphi\in$Diff$(\sigma)$. The latter 
equation has only one solution (up to a constant) in ${\cal H}$, namely 
$\psi=\Omega_\omega=1$, the trivial spin network state. 

In order to see what this requirement amounts to we notice that it is 
sufficient to restrict attention to the $f=T_s$. Let 
$[s]=\{\varphi \cdot s;\;\varphi\in{\rm Diff}(\sigma)\}$ be the orbit of 
$s$. Then it is not difficult to see that (\ref{4.19}) amounts to asking
that $l_s=l_{s'}$ whenever $[s]=[s']$. Thus $l_s=l_{[s]}$ just 
depends on the orbit and not on the representative. It is therefore clear 
that no non zero solution except for the vector $1$ is normalisable with 
respect to the kinematical inner product $<l,l'>:=\sum_s \overline{l_s} 
l'_s$. Interestingly, the solutions are labelled by {\it generalised knot 
classes} where generalised refers to the fact that we allow for knots with 
intersections. Any solution obviously is a linear combination of the 
elementary 
solutions $T_{[s]}:=\sum_{s'\in [s]} <T_{s'},.>$.

We therefore have to define a new inner product on the solution space 
${\cal D}^\ast_{{\rm Diff}}$. This can be systematically done using the 
group averaging technique described in section \ref{s3}. The only known 
Haar measure on Diff$(\sigma)$ is the counting measure. Indeed, it is 
almost true that $T_{[s]}$ coincides with the image of the rigging map
\be \label{4.19a}
\eta(T_s):=\sum_{\varphi\in {\rm Diff}(\sigma)} <U(\varphi) 
T_s,.>
\ee
if it was not for fact that Diff$(\sigma)$ contains an uncountably 
infinite number of elements which have trivial action on any given $s$.
These trivial action diffeomorphisms form a subgroup (but not an 
invariant one) but that subgroup 
evidently depends on $s$. Hence one cannot take a universal factor
group (rather: coset) for the averaging in order to get rid of the 
associated infinity. 
However, we notice that
formally $\eta(T_s)[T_s']=0$ whenever $[s]\not=[s']$. Hence it is 
justified to decompose the kinematical Hilbert space into the direct sum 
of Diff$(\sigma)$ invariant subspaces 
${\cal H}_{[\gamma]}$ consisting of the finite linear span of SNWF's 
over the graphs $\gamma'$ in the orbit $[\gamma]$ of $\gamma$. The group 
averaging can now be done on these subspaces seperately because in any 
case their images under (\ref{4.19a}) would be orthogonal. This is done 
by identifying a 
subset Diff$_{[\gamma]}(\sigma)$ which is in one to one 
correspondence\footnote{That is, fix a representative $\gamma_0$ in 
every orbit and select diffeomorphisms which map $\gamma_0$ to every 
point in the orbit.}
with the points in $[\gamma]$. When restricting (\ref{4.19a}) only to
those diffeomorphisms and a discrete set of additional graph 
symmetries\footnote{These are 
diffeomorphisms which leave the range of the representative 
$\gamma_0$ invariant
but permute the edges among each other.} then we indeed get
get $\eta(T_s)=k_{[s]} T_{[s]}$ where $k_{[s]}$ is a positive constant
which is of the form of a positive number $k_{[\gamma(s)]}$ times an  
integer which is the ratio of the orbit sizes of the least symmetric 
$[s']$ with $[\gamma(s')]=[\gamma(s)]$ and the orbit size of $[s]$. 
See 
\cite{ALMMT} for the details. It follows that
the spatially diffeomorphism invariant inner product is determined by 
\be \label{4.20}
<T_{[s]},T_{[s']}>_{{\rm Diff}}
=\frac{1}{k_{[s]} k_{[s']}} \;<\eta(T_s),\eta(T_{s'})>_{{\rm Diff}}
=\frac{1}{k_{[s]} k_{[s']}} \;\eta(T_{s'})[T_s]
=\frac{\delta_{[s],[s']}}{k_{[s]}} 
\ee
Notice, however, that the relative normalisation of the $T_{[s]}$ is 
only fixed for those $s$ which have diffeomorphic underlying graphs
because we applied the averaging to all those ``sectors'' separately. In 
order to fix the normalisations between the sectors one needs to 
consider diffeomorphism invariant operators which are classically real 
valued and map between these 
sectors and require that they be self -- adjoint (or at least 
symmetric).     

Finally we mention that ${\cal H}_{{\rm Diff}}$ just like $\cal H$
is still not separable because the set of singular knot classes 
$[\gamma]$ has uncountably infinite cardinality \cite{Grot}. This is 
easy to understand from the fact that the group of semianalytic 
diffeomorphism reduces to $GL(3,\Rl)$ at each vertex. Hence, for 
vertices of valence higher than nine one cannot arbitrarily change, in a 
coordinate chart, all 
the angles between the tangents of the adjacent edges. It turns out that
valence five is already sufficient, that is, there are diffeomorphism 
invariant ``angles'', called moduli $\theta$ in all vertices of valence
five or higher. There are several proposals for an enlargement of the 
group of diffeomorphisms \cite{Zapata,Fairbairn,VelhinhoDiff}, however, 
these groups do not interact well with certain crucial operators in the 
the theory such as the volume operator which depend on at least 
$C^{(1)}(\sigma)$ structures while those extensions basically replace 
diffeomorphisms by homeomorphisms or even mor general bijective maps
on $\sigma$. We will see, however, that the non separability of ${\cal 
H}_{{\rm Diff}}$ is immaterial when we pass to the physical Hilbert 
space ${\cal H}_{{\rm phys}}$.

\subsubsection{Reduction of the Hamiltonian 
constraint}
\label{s4.4.2}

The informed reader knows that the implementation of the Hamiltonian 
constraint is the most important technical problem in canonical quantum 
gravity ever since. The source of these technical problems within LQG
can be appreciated when recalling the Dirac algebra 
$\mathfrak{D}$ (\ref{2.2}):
\begin{itemize}
\item[1.] 
The first relation in (\ref{2.2}) means that diff$(\sigma)$ is a 
subalgebra. However, the second relation says that this subalgebra is 
not an ideal. In other words, the Hamiltonian constraints are not 
spatially diffeomorphism invariant. In particular, if there is a 
quantum operator $\widehat{H}(N)$ associated with $H(N)$ then it 
{\it cannot be defined} on ${\cal H}_{{\rm Diff}}$. We stress this simple 
observation here because one often hears statements saying the contrary
in the literature. Spatially diffeomorphism invariant states do play a 
role but a quite different one as we will see shortly. The constraint
operators $\widehat{H}(N)$ must be defined on the kinematical 
Hilbert space ${\cal H}$ and nowhere else. One could try, as suggested in 
\cite{Habitat,Marolf} to define the dual 
$\widehat{H}'(N)$ of the constraint operator on some subspace ${\cal 
D}^\ast_\star$ of ${\cal D}^\ast$ invariant under the $\widehat{H}'(N)$ 
via 
\be \label{4.20a}
[\widehat{H}'(N) l](f):=l(\widehat{H}(N)^\dagger f)     
\ee 
for all $f\in {\cal D}$. However, in order to solve all 
constraints, eventually one wants to restrict 
${\cal D}^\ast_\star$ to the space of spatially diffeomorphism invariant 
distributions on which $\widehat{H}'(N)$ is ill defined. Hence the 
definition on $\cal H$ is the only option.
\item[2.] 
We have seen that the kinematical Hilbert Hilbert space is, under rather 
mild assumptions, uniquely selected. In other words, there is no other 
choice. Unfortunately, as we have seen, in this representation the 
diffeomorphisms are not represented weakly continuously and there is no
way out of this fact. This poses a problem in representing (\ref{2.2}) 
on $\cal H$ because evidently (\ref{2.2}) involves the infinitesimal 
generators $D(\vec{N})$ of spatial diffeomorphisms which are obstructed 
to exist 
as quantum operators as we just have said. As far as the first two 
relations in (\ref{2.2}) are concerned, there is a substitute involving 
finite (exponentiated) diffeomorphisms only. It is given by
\ba \label{4.21}
U(\varphi) U(\varphi') U(\varphi)^{-1} &=& U(\varphi\circ \varphi'\circ 
\varphi^{-1})
\nonumber\\
U(\varphi) \widehat{H}(N) U(\varphi)^{-1} &=& 
\widehat{H}(N\circ\varphi)
\ea
Indeed, if $\widehat{D}(\vec{N})$ would exist then one parameter 
subgroups of spatial diffeomorphisms would be given by 
$U(\varphi^{\vec{N}}_t)=\exp(it \widehat{D}(\vec{N})/(\hbar 8 \pi 
G_{{\rm Newton}}))$ and then (\ref{4.21}) would be equivalent to  
\ba \label{4.22}
[\widehat{D}(\vec{N}),\widehat{D}(\vec{N}')] 
&=& i 8\pi G_{{\rm Newton}}\hbar \;\; \widehat{D}({\cal 
L}_{\vec{N}}\vec{N}')
\nonumber\\    
{[}\widehat{D}(\vec{N}),\widehat{H}(N')] 
&=& i 8\pi G_{{\rm Newton}}\hbar \;\; \widehat{H}({\cal L}_{\vec{N}}N')
\ea
upon taking the derivative at $t=0$.
\item[3.]
In other words there is a finite diffeomorphism reformulation of the
first two relations in (\ref{2.2}). 
However, this is no longer possible 
for the third relation in (\ref{2.2}). The problem is the structure 
function involved on the right hand side of this relation which prevents 
us from exponentiating the Hamiltonian constraints. The 
commutator algebra of the Hamiltonian 
constraints is simply so complicated that the Dirac algebra 
$\mathfrak{D}$ is no longer a Lie group. Therefore we cannot 
exponentiate the third relation in (\ref{2.2}) and there seems to be no 
chance to find a substitute involving finite diffeomorphisms only. 
\item[4.] 
Even if the problem just mentioned could be solved, we would still have 
no idea for how to find the physical inner product because group 
averaging only works for Lie algebra valued constraints.
\end{itemize}
These remarks sound like an obstruction to implement the operator 
version of the Hamiltonian constraint in LQG at all. In what follows
we describe the progress that has been made over the past ten years 
with regard to this task. There are two constructions: The first, 
surprisingly, indeed 
proposes a quantisation of the Hamiltonian constraints as operators on 
the kinematical Hilbert space. The algebra of these operators {\it is 
non -- Abelean and 
closes} in a precise sense as we will see. We again stress this 
because one sometimes reads that the Hamiltonian constraint algebra is 
Abelean \cite{Habitat,Marolf} which is simply wrong. However, no 
physical scalar
product using these operators has so far been constructed due to the 
non -- Lie algebra structure mentioned above. Also, so far the 
correctness of the  
semiclassical limit of these constraint operators has not been 
established, in particular it is unsettled in which sense the third 
relation in 
(\ref{2.2}) is implemented in the quantum theory.

To make progress on these two open problems, the construction of the 
physical scalar product and the establishment of the correct classical 
limit which are interlinked in a complicated way as we will see, the 
Master Constraint Programme was launched 
\cite{MCP,QSDIII,Test1,Test2}. We have outlined it already in section 
\ref{s3} for a general theory and will apply it to the Hamiltonian 
constraints below.\\
\\
This section is organised as follows:\\

The Master Constraint Programme 
overcomes many of the shortcomings of the Hamiltonian constraint and 
is the modern version of the implementation of the Hamiltonian 
constraint in LQG. We will still describe first the old Hamiltonian 
constraint 
programme \cite{QSDI,QSDII,QSDIII} in order to address the 
criticisms spelled out in \cite{NPZ} 
and because the quantisation technique in \cite{MCP,QSDIII} is still 
heavily based on the {\it key techniques} developed in \cite{QSDI}.
Indeed, without the techniques developed in \cite{QSDI} the recent 
intriguing results of Loop Quantum Cosmology (LQC)\footnote{LQC is 
the usual homogeneous (and isotropic) cosmological model quantised 
by LQG methods. It is not the cosmological sector of LQG because 
LQG is a quantum field theory (infinite number of degrees of 
freedom) while LQC is a quantum mechanical toy model (finite number of 
degrees 
of freedom) in which the inhomogeneous excitations are switched off by 
hand.} \cite{BMT} such as avoidance of the big bang singularity could 
never have been achieved. A large amount of the success of LQC is a 
direct consequence of \cite{QSDI}. 

Then we describe the Master Constraint Programme which was not mentioned 
at all in \cite{NPZ} although it removes many of the criticisms stated 
there. In particular we describe recent progress made in a particular 
version of the Master Constraint Programme called Algebraic Quantum
Gravity (AQG) \cite{AQG} which establishes that the Master Constraint 
Operator has the correct semiclassical limit. The work \cite{AQG} 
also removes the criticism of \cite{NPZ} that no calculations 
involving the volume operator can be carried out in LQG. This, together 
with 
the general direct integral construction of the physical inner product 
already described make the Master Constraint Programme a promising
step forward in LQG. 

Unfortunalely the subsequent discussion is rather complictaed because 
the problems of anomaly freeness, semiclassical limit, dense definition,
representation of the Dirac algebra etc. for the Hamiltonian constraints
are interlinked in a complex way. In order to appreciate these 
interdependencies we have to go into some detail about the actual 
constructions. We will try our best at keeping the discussion as non 
technical as possible.

\paragraph{Hamiltonian constraint}
\label{s4.4.2.1}

~\\
\\
The task is to quantise the Hamiltonian constraints which on the new 
phase space are given by\footnote{This is only the simplest piece
of the geometry part of the constraint, the remaining piece as well 
as matter contributions can be treated analogously \cite{QSDII} and 
will be neglected here for pedagogical reasons.}
\be \label{4.23}
H(N)=\int_\sigma \; d^3x\; N(x)\; \frac{{\rm Tr}(F_{ab} E^a 
E^b)}{\sqrt{|\det(E)|}}(x)
\ee
where $N$ is a test function.
The non polynomial character of (\ref{4.23}) is evident and it is hard 
to imagine that there is any way to tame (\ref{4.23}) when replacing 
$A,E$ by their operator equivalents.

We now sketch the {\it key tools} developed in \cite{QSDI}. Let $R_x$ be 
any region in $\sigma$ containing $x$ as an interior point then 
\be \label{4.24}
e_a^j(x)=-\{A_a^j(x),V(R_x)\}/\kappa
\ee
where $\kappa=8\pi G_{{\rm Newton}}$ and
\be \label{4.25}
V(R_x):=\int_{R_x} \; d^3y\; \sqrt{|\det(E)|}(y)
\ee
is the volume of the region $R_x$. Using the relation 
$e^a_j=E^a_j/\sqrt{|\det(E)|}$ we can rewrite (\ref{4.23}) as
\be \label{4.26}
H(N)=-\frac{1}{\kappa} \int_\sigma\;  N(x)\; {\rm Tr}(F(x)\wedge 
\{A(x),V(R_x)\}) 
\ee
where now all the dependence on $E$ resides in the volume function
$V(R_x)$. The point of doing this that $V(R_x)$ admits a well defined 
quantisation as a positive essentially self -- adjoint 
operator\footnote{Actually there are two 
inequivalent volume 
operators \cite{RSVol,ALVol} which result from using two different 
background independent 
regularisation techniques. However, in a recent mathematical self -- 
consistency analysis \cite{GT} the operator \cite{RSVol} could be 
ruled out.}
$\widehat{V}(R_x)$ on ${\cal H}$. Following the rules of canonical 
quantisation one would then like to replace the 
Poisson bracket between the functions appearing in (\ref{4.26}) by the
commutator between the corresponding operators divided by $i\hbar$.

The problem is that the connection operator $\widehat{A}(x)$ does not 
exist on the Hilbert space $\cal H$. To see this, note the classical 
identity $A_a(x) \dot{p}^a(0)=(d/dt)_{t=0} A(p_t)$ where $p:\;[0,1]\to 
\sigma;\;s\mapsto p(s)$ is a path, $p_t(s)=p(ts)$ for $t\in [0,1]$, 
$p(0)=x$ and 
$A(p_t)$ is the holonomy along $p_t$. By varying the path we can recover 
the connection from the holonomy. Hence we would like to define the 
connection operator by this formula from the holonomy operator.
However, this does not work since the family of 
operators $t\mapsto A(p_t)$ is not weakly continuous on $\cal H$. Hence 
the derivative at $t=0$ is ill defined. It follows that the UV 
singularity structure of the Hamiltonian constraints is not at all 
determined by the $E$ dependence but 
rather by the $A$ dependence. In particular, the ambiguities discussed 
below coming from the loop attachment purely stem from the $A$ 
dependence.

It is at this point where we must regularise (\ref{4.26}). We consider a 
triangulation $\tau$ of $\sigma$ by tetrahedra $\Delta$. For each 
$\Delta$, let us single out a corner $p(\Delta)$ and denote the 
edges of $\Delta$ outgoing from $p(\Delta)$ by $s_I(\Delta),\;I=1,2,3$.
Likewise, denote by $s_{IJ}(\Delta)$ the edges of $\Delta$ connecting 
the end points of $s_I(\Delta),\;s_J(\Delta)$ such that the loop
$\beta_{IJ}(\Delta)=s_I(\Delta)\circ s_{IJ}(\Delta) s_J(\Delta)^{-1}$    
is the boundary of a face of $\Delta$. In particular 
$s_{JI}(\Delta)=s_{IJ}(\Delta)^{-1}$. It is then not difficult to see 
that 
\be \label{4.27}   
H_\tau(N):=\frac{1}{\kappa}\sum_{\Delta\in \tau}\; N(p_\Delta)\; 
\sum_{IJK} \; \epsilon^{IJK}\; {\rm Tr}(A(\beta_{IJ}(\Delta)) 
A(s_K(\Delta))\{A(s_K(\Delta))^{-1},V(R_{p(\Delta)})\})
\ee
is a Riemann sum approximation to (\ref{4.26}), that is, it converges 
to (\ref{4.26}) as we refine the triangulation to the continuum. We will 
denote the refinement limit by $\tau\to \sigma$.

Since (\ref{4.27}) is now written in terms of quantities of which the 
quantisation is known we immediately get a regularised Hamiltonian 
constraint operator on $\cal H$ given by
\be \label{4.28}
\widehat{H}^\dagger_\tau(N):=\frac{1}{i\ell_P^2}\sum_{\Delta\in \tau}\; 
N(p_\Delta)\; 
\sum_{IJK} \; \epsilon^{IJK}\; {\rm Tr}(A(\beta_{IJ}(\Delta)) 
A(s_K(\Delta))[A(s_K(\Delta))^{-1},\widehat{V}(R_{p(\Delta)})])
\ee
The reason for the adjoint operation in (\ref{4.28}) is due to the 
definition of the dual action in the footnote before (\ref{3.4a}) 
on elements in ${\cal 
D}^\ast$ which in turn would coincide with the action of $\widehat{H}(N)$ 
if elements of ${\cal D}^\ast$ would be normalisable.
Notice that (\ref{4.27}) is real valued so that classically
$H_\tau(N)=\overline{H_\tau(N)}$ so we may denote its operator 
equivalent with or without adjoint operation. 
It is not difficult to see that in this ordering the operator 
is densely defined\footnote{This is basically due to the properties of 
the volume operator $\widehat{V}(R)$: If the graph of a spin network 
state does not 
contain a vertex inside the region $R$ then it is annihilated.}  on 
$\cal D$ and closable (its adjoint is also densely 
defined on $\cal D$). However, it is not even symmetric in this 
ordering.
This may seem strange at first, however it is not logically required 
because we are only interested in the zero point of its spectrum.
It is not even possible to have a symmetric ordering as pointed out 
in \cite{HK} where it is shown that for reasons of anomaly freeness
in constraint algebras with structure functions symmetric orderings are
ruled out.

What we are interested in is in which operator topology (if any) the 
limit $\tau\to \sigma$ exists. Since $\widehat{H}_\tau(N)$ is not 
bounded, convergence in the uniform topology is ruled out. For the same 
reason that connection operators are not defined, convergence in the 
weak (and thus also strong) operator topology is ruled out. Hence we 
are looking for a weaker topology. There is only one {\it natural} 
candidate available: The weak$^\ast$ topology with respect to the 
algebraic dual ${\cal D}^\ast$ or a suitable subspace thereof. The only 
{\it natural} subspace is the space of spatially diffeomorphism invariant
distributions ${\cal D}^\ast_{{\rm Diff}}$ (finite linear combinations 
of the $T_{[s]}$ defined in section \ref{s4.4.1}). 

Before we do this, we 
must tame the limit $\tau\to \sigma$ somewhat: Notice that classically
$\lim_{\tau\to \sigma} H_\tau(N)=H(N)$ no matter how we refine the 
triangulation. This observation suggests the following strategy: Given a 
graph $\gamma$ we consider a family $\epsilon\mapsto 
\tau^\epsilon_\gamma$ of triangulations adapted to $\gamma$ where 
$\epsilon$ denotes the fineness of the triangulation and 
$\epsilon\to 0$ corresponds to $\tau\to \sigma$. This family is 
equipped with the following properties: For each vertex 
$v\in V(\gamma)$ and each triple of edges $e_1,e_2,e_3\in E(\gamma)$ 
incident at $v$ consider a tetrahedron $\Delta^\epsilon_v(e_1,e_2,e_3)$ 
such that 
$p(\Delta^\epsilon_v(e_1,e_2,e_3))=v$,
such that $s_I(\Delta^\epsilon_v(e_1,e_2,e_3))$ is a proper segment of 
$e_I$, 
such that 
the $s_{IJ}(\Delta^\epsilon_v(e_1,e_2,e_3))$ do not intersect $\gamma$ 
except in 
their end points and such that the $\Delta^\epsilon_v(e_1,e_2,e_3)$
are diffeomorphic for different values of $\epsilon$. That such tetrahedra 
always exist is proved in \cite{QSDI}.

Consider seven additional tetrahedra 
$\Delta^\epsilon_{v,1}(e_1,e_2,e_3),..,\Delta^\epsilon_{v,7}(e_1,e_2,e_3)$ 
which are obtained by analytically continuing\footnote{For sufficiently 
fine triangulation the segments can be 
taken to be analytic.} the segments 
$s_I(\Delta^\epsilon_v(e_1,e_2,e_3))$ through the vertex so that we 
obtain altogether
eight tetrahedra of equal coordinate volume which are like the eight 
octants of a Cartesian coordinate system. Denote by 
$W^\epsilon_v(e_1,e_2,e_3)$ the neighbourhood of $v$ they fill. Let 
$W^\epsilon_v$ be the 
region occupied by the union of the $W^\epsilon_v(e_1,e_2,e_3)$ as we 
vary the 
unordered triples of edges incident at $v$. For sufficiently fine 
triangulation, the $W^\epsilon_v$ are mutually disjoint. Finally let 
$W^\epsilon_\gamma$ 
be the union of the $W^\epsilon_v$. We have the following identity for 
any 
classical integral
\be \label{4.29}
\int_\sigma=[\int_{\sigma-W^\epsilon_\gamma}]+
\sum_{v\in V(\gamma)} \frac{1}{{ n_v \choose 3}} 
\sum_{e_1\cap e_2\cap 
e_3=v}\{[\int_{W^\epsilon_v-W^\epsilon_v(e_1,e_2,e_3)}]+
\int_{W^\epsilon_v(e_1,e_2,e_3)}]\}
\ee
where $n_v$ is the valence of $v$. We now triangulate the regions 
$\sigma-W^\epsilon_\gamma,\;W^\epsilon_v-W^\epsilon_v(e_1,e_2,e_3)$ 
arbitrarily and use the 
classical approximation $\int_{W^\epsilon_v(e_1,e_2,e_3)}\approx 
8\int_{\Delta^\epsilon_v(e_1,e_2,e_3)}$. Then, the tetrahedra within
$\sigma-W^\epsilon_\gamma,\;W^\epsilon_v-W^\epsilon_v(e_1,e_2,e_3)$ 
can be shown not to 
contribute to the action of the operator 
$\widehat{H}_{\tau^\epsilon_\gamma}(N)$ on any SNWF $T_s$ over 
$\gamma$ so that we obtain 
\ba \label{4.30}
\widehat{H}^\dagger_{\tau^\epsilon_\gamma}(N) T_s &=&
\frac{1}{i\ell_P^2}\sum_{v\in V(\gamma)} \;N(v)\; 
\frac{8}{{ n_v \choose 3}} \sum_{e_1\cap e_2\cap e_3} 
\sum_{IJK} \; \epsilon^{IJK}\; \times
\\
&& \times 
{\rm 
Tr}(A(\beta_{IJ}(\Delta^\epsilon_v(e_1,e_2,e_3))) 
A(s_K(\Delta^\epsilon_v(e_1,e_2,e_3)))
[A(s_K(\Delta^\epsilon_v(e_1,e_2,e_3)))^{-1},\widehat{V}(R_{p(\Delta)})])
\;\;T_s
\nonumber
\ea
For each $\gamma$ choose\footnote{Use the axiom of choice for each 
diffeomorphism equivalence class of loop assignments.} some 
$\epsilon_\gamma$ once and for all such that
$\tau_\gamma:=\tau^{\epsilon_\gamma}_\gamma$ satisfies the required 
properties and define $\widehat{H}^\dagger(N) 
T_s:=\widehat{H}^\dagger_{\tau_\gamma} 
T_s$ and 
$\widehat{H}^\dagger_\epsilon(N)T_s:=
\widehat{H}^\dagger_{\tau^\epsilon_\gamma} 
T_s$ whenever $s=(\gamma,j,I)$. Then, {\it due to spatial diffeomorphism 
invariance} we have the following notion of convergence
\be \label{4.31}
\lim_{\epsilon\to 0} |l(\widehat{H}^\dagger(N) 
f)-l(\widehat{H}^\dagger_\epsilon(N) f)|=0
\ee
for all $l\in {\cal D}^\ast_{{\rm Diff}}$ and all $f\in {\cal D}$. 
It is quite remarkable that precisely the space of diffeomorphism
invariant distribtions which is selected by one of the gauge symmetries 
of the theory naturally allows us to define an appropriate operator 
topology with respect to which it is possible to remove the regulator of 
the Hamiltonian constraints. Notice that despite the fact that we have 
worked with triangulations adapted to a graph, the operator is a linear
operator on $\cal H$ where it is, together with its adjoint, densely 
defined on $\cal D$.

One of the most striking features is that the Hamiltonian constraint 
operators {\it do not suffer from UV singularities} as we anticipated
in a background independent theory.\\
\\
Several remarks are in order:
\begin{itemize}
\item[1.] {\it Quantum Spin Dynamics (QSD)}\\
Intuitively, the action of the Hamiltonian constraint operator on spin 
network functions over a graph $\gamma$ is by creating the new edges 
$s_{IJ}(\Delta^{\epsilon_\gamma}_v(e_1,e_2,e_3))$ coloured with the spin 
$1/2$ representation and by changing the spin on the segements 
$s_I(\Delta^{\epsilon_\gamma}_v(e_1,e_2,e_3))$ from $j$ to $j\pm 1/2$.
Hence in analogy to QCD one could LQG call {\it Quantum Spin Dynamics 
(QSD)}.
\item[2.] {\it Locality}\\
The action of the Hamiltonian constraint operator has been criticised 
to be too local \cite{Smolin} in the following sense: The modifications
that the Hamiltonian constraint operator performs at a given vertex do 
not propagate over the whole graph but are confined to a neighbourhood 
of the vertex. In fact, repeated action of the Hamiltonian generates 
more and more new edges ever closer to the vertex never 
intersecting each other thus producing a fractal structure. In 
particular there is no action at the new vertices created. This is not 
what happens in lattice gauge theory where no new edges are created. 

Notice, however, that there is a large conceptual difference between 
lattice gauge theory which is a background dependent and regulator 
dependent (discretised) theory while LQG is a background independent 
and regulator independent (continuum) theory.  Even the role of 
the {\it single} 
QCD Hamiltonian (generator of physical time evolution) and the {\it 
infinite number} of Hamiltonian constraints (generator of unphysical 
time reparametrisations) is totally different. Hence there is no 
logical reason why one should compare the lattice QCD Hamiltonian 
with the QSD (or LQG) Hamiltonian constraints. In particular,
by inspection the infinite number of constraints $H(x)=0$ have a more 
local structure than a Hamiltonian $H=\int_\sigma d^3x H(x)$.

Next, it is actually technically incorrect that the actions of the 
Hamiltonian constraints $\widehat{H}_v,\;\widehat{H}_{v'}$ at different 
vertices $v,v'$ do not influence each other: In fact, these two operators 
do not commute, for instance if $v,v'$ are next neighbour, because for any 
choice function $\gamma\mapsto 
\epsilon_\gamma$ what is required is that the loop attachments at $v,v'$ 
do not intersect which requires that the action at $v'$ after the action 
at $v$ attaches the loop at $v'$ closer to $v'$ than it would before the 
action at $v$ and vice versa.

Finally, the action of the Hamiltonian constraints on spin network 
states did not fall from the sky but was derived from a proper 
regularisation. In particular it is not difficult to see that the 
operator would become anomalous (see below) if it would act at the 
vertices that it creates. This would indeed happen if one used the 
volume operator \cite{RSVol} rather than \cite{ALVol}. Fortunately,
the volume operator \cite{RSVol} was shown to be inconsistent 
\cite{GT} for totally independent reasons.

In summary, there is no conclusive reason for why this locality property 
of the constraints is a bad feature. In fact, in 3D \cite{QSDII} the 
solution space 
of those constraints selects precisely the physical Hilbert space of 
\cite{Witten}.
\item[3.] {\it Ambiguities}\\
The final Hamiltonian constraint operators seem to be highly ambiguous. 
There are several qualitatively different sources of ambiguities:
\begin{itemize}
\item[3.a.] {\it Factor ordering ambiguities}\\ 
We decided to order the $E$ dependent terms in (\ref{4.27}) to the right 
of the $A$ dependent terms. Could we have reversed the order? The answer 
is negative \cite{QSDI}: Any other ordering results in an expression 
which is no longer densely defined because the operator would map any spin 
network state to a state which is a linear combination of SNWF's whose 
underlying graphs are all tetrahedra of the triangulation. The resulting 
``state'' is not normalisable in the infinite refinement limit. Thus the 
factor ordering chosen is in fact {\it unique}.
\item[3.b.] {\it Representation ambiguities}\\
When replacing connections by holonomies we have used the holonomy in the 
defining representation of $SU(2)$. However, as pointed out in \cite{Gaul}
we could also work with higher spin representations without affecting the 
limit of the Riemann sum approximation. Recently it was shown 
\cite{Perezj}
that higher spin leads to spurious solutions to the Hamiltonian 
constraints in 3D (where all solutions are known) and therefore very likey 
also in 4D. Hence this representation ambiguity is very likely to be {\it 
absent}.

Notice also that such kind of ambiguities are also present in ordinary 
QFT: Consider a $\lambda \phi^4$ QFT. Classically we could replace 
$\pi(x)$ by $\pi_f(x):=e^{i\phi(f)} \pi(x) e^{-i\phi(f)}$ in the 
Hamiltonian where $\phi(f)=\int_{\Rl^3} d^3x f(x)\phi(x)$ with some 
suitable test function $f$ and $\phi,\pi$ are canonically conjugate. 
One could even replace $\exp(i\phi(f))$ by some other invertible 
functional $F$ of $\phi$ and consider 
$[F \pi F^{-1}+\bar{F}^{-1} \pi \bar{F}]/2$.
In 
quantum theory the Hamiltonian does change when performing this 
substitution leading to a different spectrum. Of course in QFT one would 
never do that because this factor 
ordering ambiguity generically spoils polynomiality of the Hamiltonian,
so one is guided by some simplicity or naturalness principle.
In General Relativity the Hamiltonian constraint is non polynomial from 
the outset, however, still $j=1/2$ is the simplest choice.  
\item[3.c.] {\it Loop assignment ambiguities}\\
The largest source of ambiguities is in the choice of the family of 
triangulations $\epsilon\mapsto \tau^\epsilon_\gamma$ adapted to a graph.  
In particular, while it is natural to align the edges of the tetrahedra 
of the triangulations with the beginning segments of the edges of the 
graph\footnote{There is no ambiguity in the fact that the only 
contributions of the operator result from the vertices of the graph. This 
is a direct consequence of the properties of the volume operator 
\cite{ALVol} and the unique factor ordering mentioned above.} because 
there are no other natural terahedra available in the problem, it is not 
the only logically possible choice. For instance, one could slightly 
detach the loops $\beta_{IJ}(\Delta^\epsilon_v(e_1,e_2,e_3))$ from the 
beginning segments of $e_1,e_2,e_3$ as mentioned in the review by Ashtekar 
and Lewandowski in \cite{reviews} which found its way into \cite{NPZ}. 
Our statement here is as follows: First of all there is an additional, 
heuristic
argument in favour of the alignment. Secondly, even if one does not 
accept that argument, all of these uncountably 
infinite number of ambiguities at the level of $\cal H$ are reduced to a 
countable number at the level of ${\cal H}_{{\rm phys}}$ of which all but 
a few are rather pathological in the sense that one could also use them in 
lattice gauge theory but does not due to reasons of naturalness. 

Concerning the first claim, we want to point out that one of the reasons
for
why we have decided to work with the expression $\{A_a^j(x),V(R_x)\}$ 
rather 
than
$(\epsilon_{abc} E^b_k E^c_l\epsilon_{jkl}/\sqrt{|\det(E)|})(x)$ is that 
direct 
quantisation of the latter would formally result on a spin network state 
over a graph $\gamma$ in an expression of the form (before introducing the 
point splitting regulator) 
\ba \label{4.33}
&& \int_\sigma d^3x 
\frac{1}{\sum_{v'\in V(\gamma)} \delta(x,v') \widehat{V}_v}
\sum_{v\in V(\gamma)} \sum_{e_1\cap e_2=v}
\int_0^1\; dt\; \dot{e}_1^a(t)\; \delta(x,e_1(t))
\int_0^1 \; ds\; \dot{e}_2^b(s)\; \delta(x,e_2(s)) 
\times\nonumber\\
&& \times 
F^j_{ab}(\frac{e_1(t)+e_2(s)}{2}) \epsilon_{jkl} X^k_{e_1} X^l_{e_2}
\ea
where $X^j_e$ is a right invariant vector field on the copy of $SU(2)$ 
corresponding to $A(e)$. Likewise $\widehat{V}_v$ is a well defined 
operator (not an operator valued distribution) built from those vector 
fields. Clearly (\ref{4.33}) involves the holonomy of an 
infinitesimal loop whose tangents at the $v$ are pairs of edges incident 
at $v$. This motivates the alignment mentioned above\footnote{A careful . 
point splitting regularsation removes the $\delta-$distribution in both 
numerator and the denominator as well as the the integral over $\sigma$ 
leaving only an integral over $s,t$ with support in infinitesimal 
neighbourhoods of the vertices of the graph in question.}. 
The only reason why (\ref{4.33}) is not used in place of (\ref{4.27}) 
is that the the operator $\widehat{V}_v$ has zero modes so that its 
inverse is not even densely defined.
  
Concerning the second claim we notice that solutions to all constraints 
will be elements $l$ of ${\cal D}^\ast_{{\rm Diff}}$ which satisfy
$l(\widehat{H}(N)^\dagger f)=0$ for all $N$ and all $f\in {\cal D}$.
Now since $l$ is spatially diffeomorphism invariant, the space of 
solutions to all constraints {\it only depends on the spatially, piecewise 
analytic diffeomorphism invariant characteristics} of the loops 
$\beta_{IJ}(\Delta^{\epsilon_\gamma}_v(e_1,e_2,e_3))$. Hence it matters
whether or not the tetrahedra $\Delta$ are just continuous at their 
corners or of higher differentiability class, how the additional edges 
are routed or braided through the edges of the graph and whether they 
are aligned 
or not. Concerning the braiding, a natural choice is the one 
displayed in \cite{QSDI} which makes use of 
Puisseaux' theorem\footnote{
Basically one wants that the arcs intersect the graph only in their end
points which for sufficiently fine triangulations can only happen for
edges $e$ incident at the vertex in question. 
One first shows that there always exists an 
adapted frame, that is, a frame such 
that $s_I,s_J$ lie in the $x,y$ plane for sufficiently short $s_I,s_J$. 
Now one shows that for any other edge $e$ of the graph whose beginning 
segment is not aligned with either $s_I$ 
or $s_J$ there are only two possibilities: A. Either for all adapted 
frames
the beginning segment of $e$ lies above or below the $x,y$ plane and 
whether it is above or below is independent of the adapted frame. B. Or
there exists an adapted frame such that $e$ lies above the $x,y$ plane.
This can be achieved simultaneously for all edges incindent at the 
vertex in question.
The natural prescription is then to let the edge $s_{IJ}$ be the 
straight line 
in the selected frame connecting the end points of $s_I,s_J$ at which 
it intersects transversally.}. It 
follows that the seemingly uncountably infinite set of 
possible 
loop assignments is reduced to a {\it discrete number of choices}, of 
which all but a finite number is unnatural\footnote{Like winding the 
segments 
$s_I$ of the tetrahedra of the triangulation an arbitrary number of 
times around the edges 
$e_I$ of the graph. Such a ridiculous choice could also be made in 
lattice gauge theory but is not considerded there due to reasons of 
naturalness.}, once we construct solutions of all constraints.

In summary, the most natural proposal is such that the edges 
$s_{IJ}(\Delta^{\epsilon_\gamma}_v(e_1,e_2,e_3))$ intersect the graph 
$\gamma$ transversally with the braiding described in \cite{QSDI}. This 
defines a concrete and non ambiguous 
operator in the sense that it uniquely selects a subspace of 
${\cal D}^\ast_{{\rm Diff}}$ as the space of solutions to all 
constraints.   
\item[3.d.] {\it Habitat ambiguities}\\
In \cite{NPZ} we find an extensive discussion about ``habitats''
${\cal D}^\ast_\star$. A habitat is a subspace of ${\cal D}^\ast$ 
containing ${\cal D}^\ast_{{\rm Diff}}$ with the minimal requirement that 
it is preserved by the dual action of the Hamiltonian constraints. 
Habitats were introduced in \cite{Habitat,Marolf}. The idea was to take 
the limit $\epsilon\to 0$ for the duals of the Hamiltonian constraints
on such a habitat in the sense of pointwise convergence. The habitat
ambiguity is that there maybe zillions of habitats on which a limit of
this kind can be performed. As was shown in those papers, there exists at 
least one such habitat and it has the property that the limit dual 
operators are Abelian.

We now show that this habitat ambiguity is actually absent: Namely,
the habitat spaces must be genuine extensions of ${\cal D}^\ast_{{\rm 
Diff}}$. Hence these spaces are not in the kernel of the spatial 
diffeomorphism constraint and are therefore unphysical. Hence the only 
domain where to define the Hamiltonian constraints (rather than their 
duals) is on $\cal D$, i.e. on a dense subspace of 
the kinematical Hilbert space $\cal H$. 
This is the same domain as for the spatial 
diffeomorphism constraints which thus treats both types of constraints 
democratically. This fact is widely 
appreciated in the LQG community and not a matter of debate, the habitat
construction presented in \cite{NPZ} is outdated. {\it Habitats are 
unphysical and completely irrelevant in LQG.}

On the kinematical Hilbert space the 
Hamiltonian constraints are non commuting, see below. The apparent 
contradiction with 
the Abelean nature of the limits of the duals on the afore mentioned 
habitat is resolved by the fact that effectively the commutator of the 
limiting 
duals on the habitat is the dual of the commutator on $\cal D$. While the 
commutator on $\cal D$ is non vanishing, its dual annihilates ${\cal 
D}^\ast_{{\rm Diff}}$ and also the habitat ${\cal D}^\ast_\star$ chosen 
which is a sufficiently small extension of 
${\cal D}^\ast_{{\rm Diff}}$.
\end{itemize}
Hence we see that the amount of ambiguity is far less severe than 
\cite{NPZ} perhaps make it sound once we pass to ${\cal H}_{{\rm 
phys}}$.
In fact, there are only a hand full of natural proposals available.
\item[4.] {\it Anomalies}\\
As already mentioned, the constraint algebra can only be checked in the 
form that only involves finite diffeomorphisms. Indeed it is not difficult 
to see that the first two relations of the Dirac algebra (\ref{2.2})
really hold in the form (\ref{4.21}) on ${\cal H}$ up to a 
spatial diffeomorphism \cite{TB}. Likewise one can check that 
\be \label{4.34}
l([\widehat{H}^\dagger(N),\widehat{H}^\dagger(N')]f)=0
\ee
for all test functions $N,N'$, all $f\in {\cal D}$ and all $l\in {\cal 
D}^\ast_{Diff}$. This can be read as an implementation of the third 
relation in (\ref{2.2}) because that relation involves an infinitesimal 
spatial diffeomorphism constraint whose dual action should annihilate 
${\cal D}^\ast_{Diff}$. Of course, the commutator does not involve an
infinitesimal diffeomorphism which does not exist in our theory. Rather
what happens is the following: The 
commutator $[\widehat{H}^\dagger(N),\widehat{H}^\dagger(N')]$ is non 
vanishing on $\cal H$. However, it can be shown that on SNWF's it is a 
finite 
linear combination of terms of the form 
$[U(\varphi)-U(\varphi')]\widehat{O}$ where $\widehat{O}$ is some 
operator on $\cal H$. 
\item[5.] {\it On shell closure versus off shell closure}\\
As we just just saw, the quantum constraint algebra is consistent, i.e. 
non -- anomalous. More precisely, the first relation in 
(\ref{2.2}) holds, in exponentiated form, on $\cal H$ exactly, it is non 
anomalous in every sense. The second relation in (\ref{2.2}) also 
holds in exponentiated form on $\cal H$ but only modulo a spatial 
diffeomorphism. How about the third relation in (\ref{2.2})?
In \cite{QSDII} it is shown that an independent quantisation of the 
classical function $D(q^{-1}(N' dN-N dN'))$ appearing on the right 
hand side of (\ref{2.2}) can be given. There is no contradiction to the 
non existence of the operator corresponding to $D(\vec{N})$ because 
$D(q^{-1}(N' dN-N dN'))$ is not of the form $D(\vec{N})$ due to the 
structure function $q^{-1}$ which is responsible for the existence of 
the composite operator corresponding to $D(q^{-1}(N' dN-N dN'))$.
That operator is constructed in a way analogous to the Hamiltonian 
constraint operator and is formulated in terms of the operators 
$U(\varphi)$. The duals of both operators 
$[\widehat{H}^\dagger(N),\widehat{H}^\dagger(N')]$ and 
$\widehat{D(q^{-1}(N' dN-N dN'))}$ annihilate ${\cal D}^\ast_{{\rm 
Diff}}$. 

Hence what one can say is the following: We define two operators on
$\cal H$ as equivalent $\widehat{O}_1 \sim \widehat{O}_2$ provided
that the dual of $\widehat{O}_1-\widehat{O}_2$ annihilates ${\cal 
D}^\ast_{{\rm Diff}}$. Then the classical identities (\ref{2.2})  
holds on $\cal H$ in the sense of equivalence classes (the first 
relation even identically).

One could call this partly on -- shell closure (partly because we did 
not use the full ${\cal H}_{{\rm phys}}$ but only ${\cal H}_{{\rm 
Diff}}$ in the equivalence relation). While it would be more 
satisfactory to have full off -- shell closure, it is not logically 
required: At the end we are only interested in physical states and these 
are in particular spatially diffeomorphism invariant. Those states 
cannot distinguish between different representatives of the equivalence 
classes.
\item[6.] {\it Semiclassical limit}\\
The problem with demonstrating off -- shell closure is that, in contrast 
to the first two, the third relation in (\ref{2.2}) does not 
hold by inspection, not even modulo a diffeomorphism. This is not 
surprising because even classically one needs a full page of calculation 
in order to bring the Poisson bracket between two Hamiltonian constraints 
into the form of the right hand side of the third relation in (\ref{2.2}).
This calculation involves reordering of terms, differential geometric 
identitiesand integrations by parts etc. which are difficult to perform
at the operator level. In order to make progress on this issue one would 
therefore like to probe the 
Dirac algebra with semiclassical states, the idea being that in 
expectation values with respect to
semiclassical states the operators can be replaced by their 
corresponding classical functions and commutators by Poisson brackets, up 
to $\hbar$ corrections.

There are two immediate obstacles with this idea:\\
The first is that the volume 
operator involved is not analytically diagonisable. Recently, however,
\cite{AQG} it was shown that analytical calculations involving 
the volume operator can be performed precisely using coherent states on 
$\cal H$ \cite{GCS}, so this problem has been removed\footnote{Notice that 
it is not possible to probe $\mathfrak{D}$
with semiclassical spatially diffeomorphism invariant states because 
none of the operators involved preserves ${\cal H}_{{\rm Diff}}$.}
The second is that the existing semiclassical tools are only 
appropriate for {\it graph non -- changing operators} such as the volume 
operator. Namely, as we will see, in order to be normalisable, coherent 
states are 
(superpositions of) states defined on specific graphs. The 
Hamiltonian constraint operator, however, is graph 
changing. This means that it creates new modes on which the 
coherent state does not depend and whose fluctuations are therefore 
not suppressed. Therefore the existing semiclassical tools are 
insufficient for graph changing operators such as the Hamiltonian 
constraint. The development of improved tools is extremely difficult and 
currently out of reach. 
\item[7.] {\it Solutions and physical inner product}\\
Solutions to all constraints can be constructed 
algorithmically  
\cite{QSDII}. These are the full LQG analogs of the LQC solutions of the 
difference equation that results from the single Hamiltonian constraint 
of LQC. They are the first rigorous solutions ever constructed in 
canonical quantum gravity, have non zero volume and are labelled
by {\it fractal knot classes} because the iterated action of the 
Hamiltonian constraint creates a self -- similar structure 
(spiderweb) around each vertex. However, as in LQC these solutions are not 
systematically 
derived from a rigging map which is why a physical inner product is 
currently missing for those solutions.
\end{itemize}
This finishes the discussion of the properties of the Hamiltonian 
constraint operators. We want to stress that while evidently several 
issues need to be resolved, this is the first time in history that 
canonical quantum gravity was brought to a level such that\\
1. these and related questions could meaningfully be asked and 
analysed with 
mathematical precision.\\
2. a concrete, natural proposal for the Hamiltonian constraint 
operators can 
be derived which is consistent (anomaly free), namely the one where the 
segments $s_I$ and $s_{IJ}$ respectively are aligned and transversal to 
the graph re3spectively and where the resulting loops $\beta_{IJ}$ are 
in the $j=1/2$ reprsentation.  \\
\\
Nobody in the LQG community believes that this concrete model is the 
``right'' or ``final'' one, but it provides a concrete proposal which can 
be studied and further improved.\\
 
As discussed, the most important open issues are the semiclassical limit 
and the physical inner product. These issues are overcome to a large 
extent by the  
Master Constraint Programme.

\paragraph{Master Constraint Programme}
\label{s4.4.2.2}

~\\
\\
The idea of the Master Constraint is to sidestep the complications of 
the Hamiltonian constraints that have their origin in the non Lie 
algebra structure of the Dirac algebra $\mathfrak{D}$. Consider the 
Master constraint
\be \label{4.35}
M:=\int_\sigma \; d^3x \; \frac{H(x)^2}{\sqrt{\det(q)}(x)}
\ee
It is not difficult to see that (\ref{4.35}) has the following 
properties:\\
1. $M=0$ is equivalent with $H(N)=0$ for all test functions $N$.\\
2. $\{F,\{F,M\}\}_{M=0}=0$ is equivalent with 
$\{F,H(N)\}=0$ when $H(N')=0$ for all test functions $N,\; N'$. \\
3. $M$ is spatially diffeomorphism invariant.\\
The first property says that the single constraint $M=0$ encodes the 
same constraint surface as 
the infinite number of Hamiltonian constraints while the second
property says that the single double 
Poisson bracket with $M$ selects the same weak Dirac observables as the 
infinite number of single Poisson brackets with Hamiltonian constraints.
In other words the Master constraints defines the same reduced phase 
space as the infinite number of Hamiltonian constraints.

The third property means that the complicated Dirac algebra 
$\mathfrak{D}$ can be replaced by the comparatively trivial Master 
Algebra $\mathfrak{M}$
\ba \label{4.36}
\{D(\vec{N}),D(\vec{N}')\} &=& \kappa D({\cal L}_{\vec{N}}\vec{N}')
\nonumber\\
\{D(\vec{N}),M\} &=& 0
\nonumber\\
\{M,M\} &=& 0
\ea
which now is a {\it true Lie algebra}. This removes almost all obstacles 
that we encountered with the Hamilotonian constraints:
\begin{itemize}
\item[1.] {\it Role of ${{\cal H}}_{{\rm Diff}}$}\\
Since $M$ is spatially diffeomorphism invariant, its operator 
version $\widehat{M}$ can be defined directly on the spatially 
diffeomorphism invariant Hilbert space ${\cal H}_{{\rm Diff}}$. In fact, 
if 
$\widehat{M}$ is a knot class changing spatially diffeomorphism 
invariant operator, then it {\it must} be defined on ${\cal H}_{{\rm 
Diff}}$, it cannot be defined on $\cal H$ \cite{ALMMT}. In retrospect, 
this justifies the construction of ${\cal H}_{{\rm Diff}}$ because for 
the solution of the Hamiltonian constraints the Hilbert space ${\cal 
H}_{{\rm Diff}}$ is unsuitable as an intermediate step towards the 
physical Hilbert space as it is not left invariant by the Hamiltonian 
constraints.
\item[2.] {\it Regulator removal}\\
Remember the awkward role of spatially diffeomorphism invariant states 
in the removal of the regulator of the Hamiltonian constraint operators
using a special type of weak$^\ast$ operator topology.
It turns out \cite{QSDVIII} that a knot class changing operator can 
indeed be constructed on ${\cal H}_{{\rm Diff}}$ by directly 
implementing the techniques of \cite{QSDI} sketched above. In the 
construction of this operator, the removal of the regulator is now 
in the standard weak operator topology of ${\cal H}_{{\rm Diff}}$.  
\item[3.] {\it Physical Hilbert space}\\
Since the infinite number of Hamiltonian constraints was 
replaced by a single constraint, provided $\widehat{M}$ can be defined 
as a positive self -- adjoint operator on ${\cal H}_{{\rm Diff}}$ and 
provided that ${{\cal H}}_{{\rm Diff}}$ decomposes into a direct sum 
of $\widehat{M}-$invariant, separable Hilbert spaces, we know that
direct integral decomposition guarantees the existence of a physical 
Hilbert space with a positive definite inner product induced from 
${{\cal H}}_{{\rm Diff}}$. In order to construct it explicitly one needs 
to know the projection valued measure associated with $\widehat{M}$, 
that is, only standard spectral theory is required. While this is a 
difficult task to carry out explicitly due to the complexity of the 
operator $\widehat{M}$,
we therefore have {\it an existence proof} for ${{\cal H}}_{{\rm 
phys}}$.
\item[4.] {\it Anomalies: Ambiguities, locality and the semiclassical 
limit}\\
Since there is only one Master constraint operator, it is trivially 
anomaly free. This enables one to consider a wider class of loop 
attachments, in particular those that would lead to an anomaly in the 
algebra of the Hamiltonian constraints. As examples show \cite{Test2}, 
such quantisations of the Master constraint based on anomalous individual 
constraints lead to spectrum which does not include zero. However, the 
prescription to subtract the spectrum gap from the Master constraint as 
mentioned in section \ref{s3} works in all examples studied. One might 
worry that this spectrum gap, which in free field theories is related to 
a normal ordering constant, is infinite. However, this is not the case:
The master constraint is not just a plain sum of squares of the 
individual constraints, it is a {\it weighted} sum. The weight in the 
case of gravity is the factor $1/\sqrt{\det(q)}$ in (\ref{4.35}) which 
is the natural object to consider in order to make (\ref{4.35}) 
spatially diffeomorphism invariant. For the case of the Maxwell field
Gauss constraint studied in \cite{Test2} the associated weight had to be 
a certain trace class operator on the one particle subspace of the Fock 
space. The weight function (operator) thus makes the normal ordering 
constant finite. Hence the master constraint programme can handle 
anomalous constraints.

For gravity of particular interest are constraints which are not graph 
changing, although the corresponding Hamiltonian constraints would be 
anomalous, for three reasons:
\begin{itemize}
\item[4.a.] {\it Ambiguities}\\
Since the loop to be attached is already part of the graph, the 
situation becomes closer to the situation in lattice gauge theory.
This tremendously reduces the number of choices for the loop attachmment 
and makes it no worse than the choice of a fundamental Hamiltonian in 
Wilson's approach to the renormalisation group.
\item[4.b.] {\it Locality}\\
With this option we are free to consider for instance ``next neighbour''
loop attachments. This leads to a spreading of the influence of the 
action of the Hamiltonian constraint from one vertex to all others thus 
removing the criticism of \cite{Smolin}. 
\item[4.c.] {\it Semiclassical limit}\\
A non graph changing Master constraint can be defined on the kinematical 
Hilbert space. This has the advantage that the semiclassical states 
which so far in LQG are elements of $\cal H$ can be directly used to 
analyse the semiclassical properties of $\widehat{M}$. This has been 
done recently in \cite{AQG} with the expected result that the 
infinitesimal generators (the Hamiltonian constraints) do have the 
correct semiclassical limit. Since these constraints determine the 
physical Hilbert space, this is an important step towards showing that
gauge invariant operators commuting with $\widehat{M}$ have the correct 
semiclassical limit on the physical Hilbert space. The semiclassical 
limit is of course only reached on graphs which are sufficiently fine.
Graphs with huge holes would corespond to spacetimes with degenerate 
metrics in macroscopic regions which is not allowed in classical General 
Relativity.

Notice that semiclassical states have so far not been constructed on 
${\cal H}_{{\rm Diff}}$. This means that the semiclassical limit of the 
graph changing Master constraint is currently out of reach, thus 
favouring the graph non changing version.
\end{itemize}
\end{itemize}
In what follows we will sketch both the graph changing Master 
constraint operator on 
${\cal H}_{{\rm Diff}}$ and the graph non changing operator on 
${\cal H}$.

\subparagraph{Graph changing Master constraint}
\label{s4.4.2.2.1}

~\\
\\
We follow closely \cite{QSDVIII}. We notice that classically ($\tau$ is 
again a triangulation)
\be \label{4.37}
M=\lim_{\tau\to\sigma} \sum_{\Delta\in \tau} [\tilde{C}(\Delta)]^2
\ee
where $\tilde{C}(\Delta)$ coincides with (\ref{4.27}) for the smearing 
function $N=\chi_\Delta$ (the characteristic function for a tetrahedron) 
and with $V(\Delta)$ replaced by $\sqrt{V(\Delta)}$. Thus, the heuristic 
idea is to define the quadratic form on ${\cal D}^\ast_{{\rm Diff}}$ 
by
\be \label{4.38}
Q_M(l,l'):=\lim_{\tau\to \sigma} \sum_{\Delta\in \tau} 
<l,[\widehat{\tilde{C}}'(\Delta)]^\ast [\widehat{\tilde{C}}'(\Delta)] 
l'>_{{\rm 
Diff}}
\ee
where the prime denotes the operator dual as usual and $^\ast$ denotes the 
adjoint on ${\cal H}_{{\rm Diff}}$. Unfortunately (\ref{4.38}) is ill 
defined as it stands because the operators $\widehat{\tilde{C}}'(\Delta)$
do not preserve ${\cal D}^\ast_{{\rm Diff}}$. The cure is to extend 
$<.,.>_{{\rm Diff}}$ to an inner product $<.,.>_\ast$ on all of ${\cal 
D}^\ast$. The final 
result turns out to be insensitive to the details of the extension because 
in the limit $\tau\to \sigma$ the Riemann sum becomes well defined on 
${\cal D}^\ast_{{\rm Diff}}$. 

Rather than going through the rigorous 
argument which can be found in \cite{QSDVIII} we will present here the 
shortcut already sketched in \cite{MCP}: Pretending that (\ref{4.38}) is 
well defined we can insert a resolution of unity (we assume that
the normalisation constants $k_{[s]}$ have been absorbed into the 
$T_{[s]}$
so that the $T_{[s]}$ form an orthonormal basis)
\be \label{4.39}
Q_M(T_{[s_1]},T_{[s_2]}):=\lim_{\tau\to \sigma} \sum_{\Delta\in \tau} 
\sum_{[s]} \;
<T_{[s_1]},[\widehat{\tilde{C}}'(\Delta)]^\ast T_{[s]}>_{{\rm Diff}} \;\;
<T_{[s]}, [\widehat{\tilde{C}}'(\Delta)] l'>_{{\rm Diff}}
\ee
Using the definition of the rigging map $\eta(T_s)=T_{[s]}$, the 
definition of the scalar product\\ $<\eta(T_s),\eta(T_s')>_{{\rm 
Diff}}=\eta(T_{s'})[T_s]$ and the definition of the dual operator 
$[O'l](f)=l(O^\dagger f)$ we obtain the now well defined equation
\be \label{4.39a}
Q_M(T_{[s_1]},T_{[s_2]})=\lim_{\tau\to \sigma} \sum_{\Delta\in \tau} 
\sum_{[s]} 
\overline{T_{[s_1]}(\widehat{\tilde{C}}^\dagger(\Delta) T_{s_0([s])})}
\;
T_{[s_2]}(\widehat{\tilde{C}}^\dagger(\Delta) T_{s_0([s])})
\ee
where $s_0([s])$ is some representative of $[s]$. Now for fixed 
$[s_1],[s_2]$ the number of $[s]$ contributing to (\ref{4.39}) is easily 
seen to be finite. Hence we can interchange the two sums in (\ref{4.39}).
Furthermore, for sufficinetly fine $\tau$ we just need to consider those 
terahedra $\Delta_v$ containing a vertx of the graph underlying 
$s_0([s])$. One can then show that the limit $\tau\to \sigma$ 
becomes trivial due to the diffeomorphism invariance of 
$T_{[s_1]},\;T_{[s_2]}$. Denoting $\widehat{\tilde{C}}^\dagger(\Delta)$ for 
$v\in 
\Delta$ by $\widehat{\tilde{C}}^\dagger_v$, which is the same as the 
coefficient of $N(v)$ in (\ref{4.30}) just that $\widehat{V}_v:=V(R_v)$ is
replaced by $\sqrt{\widehat{V_v}}$, we therefore obtain the final 
formula
\be \label{4.40}
Q_M(T_[s_1],T_{[s_2]})= 
\sum_{[s]} \sum_{v\in V(\gamma(s_0([s]))}
\overline{T_{[s_1]}(\widehat{\tilde{C}}^\dagger_v T_{s_0([s])})}
\;
T_{[s_2]}(\widehat{\tilde{C}}^\dagger_v T_{s_0([s])})
\ee
It is easy to show that (\ref{4.40}) is independent of the representative 
$s_0([s])$ again due to spatial diffeomorphism invariance.

Expression (\ref{4.40}) defines a positive quadratic form. However, it is 
not obvious that it presents the matrix elements of a positive operator.
In \cite{QSDVIII} it is shown that (\ref{4.40}) is closable thus 
presenting the matrix elements of a positive, self adjoint operator 
$\widehat{M}$ on ${\cal H}_{{\rm Diff}}$. Moreover, the non separable 
Hilbert space ${\cal H}_{{\rm Diff}}$ decomposes into an uncountable 
direct\footnote{Modulo some subtleties which can be found in 
\cite{QSDVIII} and that can be dealt with.} sum ${\cal H}_{{\rm 
Diff}}=\oplus_\theta {\cal 
H}^\theta_{{\rm Diff}}$. Here the sectors ${\cal H}^\theta_{{\rm Diff}}$ 
are separable and are labelled by the angle moduli mentioned earlier. They 
are left invariant by $\widehat{M}$ basically because it only creates 
three valent vertices which do not have moduli. It follows that the direct 
integral method is applicable to $\widehat{M}$ thus resulting in 
the physical Hilbert space ${\cal H}_{{\rm phys}}$ induced from ${{\cal 
H}}_{{\rm Diff}}$.

It is easy to show that $\widehat{M}$ allows for an infinite number of 
zero eigenvectors (elements of ${\cal H}_{{\rm Diff}}$). This follows 
immediately from the properties of the Hamiltonian constraints. One just 
has to choose $\gamma(s)$ to be out of the range of the graphs 
underlying the SNW's generated by the 
Hmiltonian constraints. Hence zero is contained in the point spectrum of 
this operator which constructed using non anomalous constraints. 
However, due to the present lack of graph changing and even spatially 
diffeomorphism invariant coherent states, a verification of the correct
semiclassical limit of the graph changing $\widehat{M}$ is currently out 
of reach.

\subparagraph{Non graph changing (extended) Master constraint}
\label{s4.4.2.2.2}

~\\
\\
In order to have control on the semiclassical limit one must currently use 
a non 
graph changing operator and an operator which can be defined on $\cal H$.
This can only be done by using underlying Hamiltonian constraints which 
are anomalous in the naive discretisation displayed below. However, 
there are techniques known from lattice gauge theory \cite{Hasenfratz} 
which make use of the renormalisation group flow and which might enable 
one to work with non anomalous constraints. This amounts to considering 
more sophisticated discretisations. We see here that the issues of the 
semiclassical 
limit and the anomaly freeness are interlinked in a complicated way.
Fortunately, anomalies do not pose any obstacles to the Master
constraint programme.

In order to define such an operator we need the 
notion of a minimal loop: Given a vertex $v$ of a graph $\gamma$ and two 
edges $e,e'$ outgoing from $v$, a loop $\beta(\gamma,v,e,e')$ 
within $\gamma$ based at $v$, outgoing along $e$ and incoming along $e'$ 
is said to be {\it minimal} if there is no other loop within $\gamma$ with 
the same 
properties and fewer edges traversed. Let $L(\gamma,v,e,e')$ be the set of 
minimal loops with the data indicated. Notice that this set is always non 
empty but may consist of more than one element. We now define 
$\widehat{M} T_s:= \widehat{M}_\gamma T_s$ on spin 
network states $T_s$ over $\gamma$ where
\ba \label{4.41}
\widehat{M}_\gamma &:=& \sum_{v\in V(\gamma)} \;
\widehat{\tilde{C}}_v^\dagger  
\widehat{\tilde{C}}_v
\\
\widehat{\tilde{C}}_v &:=& \frac{1}{|T(\gamma,v)|}\sum_{e_1,e_2,e_3 \in 
T(\gamma,v)} 
\frac{\epsilon_v(e_1,e_2,e_3)}{|L(\gamma,v,e_1,e_2)|}
\sum_{\beta
\in L(\gamma,v,e_1,e_2)}
{\rm Tr}([A(\beta)-A(\beta)^{-1}] A(e_3) 
[A(e_3)^{-1},\sqrt{\widehat{V}_v}])
\nonumber
\ea
Here $T(\gamma,v)$ is the number of ordered triples of edges incident at
$v$ (taken with outgoing orientation) whose tangents are linearly 
independent\footnote{We set $\widehat{\tilde{C}}_v=0$ if 
$T(\gamma,v)=\emptyset$.} and $\epsilon_v(e_1,e_2,e_3)={\rm 
sgn}(\det(\dot{e}_1(0),\dot{e}_2(0),\dot{e}_3(0))$. The volume operator
is given explicitly by
\be \label{4.42}
\widehat{V}_v=\sqrt{|\frac{i}{48}\sum_{e_1,e_2,e_3\in T(\gamma,v)}
\epsilon_v(e_1,e_2,e_3) \epsilon_{jkl} X^j_{e_1} X^k_{e_2} X^l_{e_3}|}
\ee
where $X^j_e T_s={\rm Tr}([\tau_j A(e)]^T\partial/\partial A(e))$ is 
the right invariant vector field on the copy of $SU(2)$ determined by the 
holonomy $A(e)$ as introduced earlier.

It is easy to see that the definition (\ref{4.41}) is spatially 
diffeomorphism invariant. Moreover, the results of \cite{AQG} imply that 
expectation values with 
respect to the coherent states constructed in \cite{GCS} which 
were barely mentioned in \cite{NPZ}, defined on graphs 
which are sufficiently fine, the zeroth order in $\hbar$ of 
$\widehat{M}_\gamma$ coincides with the classical expression. In other 
words, the correctness of the classical limit of $\widehat{M}$ has been 
established recently. The results of \cite{AQG} also imply that 
the commutator between the $\sum_v N_v \widehat{C}_v$  
reproduces the third relation in (\ref{2.2}) in the sense of 
expectation values with respect to coherent states where 
$\widehat{C}_v$ is the same as 
$\widehat{\tilde{C}}_v$ in (\ref{4.41}) just that 
$\sqrt{\hat{V}_v}$ is replaced by $\hat{V}_v$. This removes a further 
criticism spelled out in \cite{NPZ}, namely we have off shell 
closure of the Hamiltonian constraints to zeroth order in $\hbar$.
Possible higher order corrections (anomalies) are no obstacle for the 
Master 
constraint programme as already said.

\paragraph{Brief note on the volume operator}
\label{s4.4.2.3}

~\\
\\
In order to show this one has to calculate the 
matrix elements of (\ref{4.42}) which is non trivial because the spectrum 
of that operator is not accessible exactly\footnote{The matrix elements of 
the argument of the square root are known in closed form \cite{BT}.}.
However, one can perform an error controlled $\hbar$ expansion within 
coherent state matrix elements and compute 
the matrix elements of every term in that expansion analytically 
\cite{AQG}. The idea is extremely simple and it will surprise nobody that 
this works: In applications we are interested in expressions of the form
$Q^r$ where $Q$ is a positive operator, $0<r\le1/4$ is a rational number 
and its relation to the volume operator is $V=\root 4 \of{Q}$. The matrix 
elements of $Q$ in coherent states can be computed in closed form. Now
use the Taylor expansion of the function $f(x)=(1+x)^r$ up to some 
order $N$ including the remainder with 
$x=Q/<Q>-1$ where $<Q>$ is the expectation value of $Q$ with respect to 
the coherent state of interest. The operators $x^n$ in that expansion can 
be explicitly evaluated in the coherent state basis while the remainder 
can be estimated from above and provides a higher $\hbar$ correction than 
any of the $x^n,\;0\le n\le N$. This completely removes the criticism of 
\cite{NPZ} that ``nothing can be computed''. 

In \cite{NPZ} we also find a lengthy discussion about the regularisation 
of the volume operator in terms of flux operators. Actually the discussion 
in \cite{NPZ} follows closely \cite{ALVol}, however, the additional 
avaraging step performed in \cite{ALVol} is left out in \cite{NPZ} for 
reasons unclear to the present author. However, even if one considers 
that averaging procedure unconvincing or unmotivated, there are completely 
independent
abstract reasons for why (\ref{4.42}) is the only possibility to define 
the volume operator which were spelled out in \cite{ALVol}: 
Namley, the argument of the volume operator, which classically is given as 
the integral of $\sqrt{|\det(E)|}$ must be a completely 
skew expression in the right invariant vector fields 
because the only 
way to regularise it is in terms of flux operators. Now the relative  
coefficients between the terms for each triple are fixed, up to an overall 
constant, by spatial diffeomorphism covariance and cylindrical 
consistency\footnote{That is, the expression (\ref{4.42}) for a given 
graph reduces to the one on any smaller graph when applied to spin network 
functions over the smaller graphs.}. The task to do was to show that a 
regularisation indeed exists which produces (\ref{4.42}) which was done in 
\cite{ALVol} and to fix the constant which was done in \cite{GT}. 
In addition, there is an alternative point splitting regularisation 
\cite{TVol} which does not use the averaging which also results in 
\cite{ALVol}. Hence 
there can be absolutely no debate \cite{NPZ} about the correctness of 
(\ref{4.42}) 
in particular that now we know from \cite{AQG} that its classical limit is 
correct.

Finally, \cite{NPZ} stresses that the final operator that one gets should 
be independent of the regularisation scheme and it is criticised that 
the regularisation scheme that one uses for the volume operator seems to 
depend on ad hoc choices so that different choices could give a 
different operator. Again we state that \cite{ALVol,GT} fix the volume 
operator uniquely. Apart from that we would like to stress that in 
ordinary QFT there are only a hand full of regularisation schemes that one 
tests: Pauli -- Villars, minimal subtraction, dimensional regularisation, 
point splitting. Here two different schemes were used 
\cite{ALVol,TVol} which resulted in the same operator and thus the test is 
of the same order of ``generality'' .

\paragraph{Algebraic Quantum Gravity (AQG)}
\label{s4.4.2.4} 

~\\
\\
Notice that the framework of Algebraic Quantum Gravity 
(AQG) proposed in \cite{AQG} in many ways supersedes LQG: In contrast to 
LQG, AQG is a purely combinatorial theory, that is, topology and 
differential structure of $\sigma$ are semiclassical notions and not 
elements of the combinatorial formulation. Next, there are not an 
uncountably infinite number of finite, embedded graphs, there is only one 
countably infinite algebraic (or abstract) graph \cite{Stauffer}. In 
particular, the theory loses its graph dependence, only in the 
semiclassical sectors (corresponding to different $\sigma$) do embedded 
graphs play a role. Hence AQG can possibly deal with topology change. 

The Hilbert space of AQG is still not separable, but for an entirely 
different reason than in LQG: Since the graph is infinite we have to deal 
with an infinite tensor product of Hilbert spaces \cite{GCS}. 
However, as von 
Neumann showed, these Hilbert spaces naturally decompose into separable
Hilbert spaces which in our case turn out to be invariant 
under the algebraic version of $\widehat{M}$ so that on each sector the 
physical inner product exists by direct integral decomposition. 
Hence, non separability poses absolutely no obstacle. 
Some of 
these sectors can presumably be identified as approximations to Quantum 
Field Theories on curved backgrounds (namely when the geometry 
fluctuations around that background are small). In some sense, all QFT's
on curved spacetimes are included which must be the case in order to 
have a background independent theory. The Hilbert space therefore {\it has 
to be 
non separable} for we do not expect QFT's on different backgrounds to be 
unitarily equivalent and there are certainly uncountably many non 
diffeomorphic backgrounds.

Finally, since the natural
representation $U(\varphi)$ of Diff$(\sigma)$ is not available in the 
combinatorial theory (there is no $\sigma$), spatial diffeomorphism 
invariance has to be dealt with in an algebraic way. This possible by 
using the {\it extended} Master constraint whose classical expression for 
given $\sigma$ is given by 
\be \label{4.43} 
M_E=M+\int_\sigma\;d^3x\; \frac{q^{ab} D_a D_b}{\sqrt{\det(q)}}
\ee
It turns out that the additional piece in (\ref{4.43}) just like $M$ 
itself can be lifted to the algebraic level thus abstracting from the 
given $\sigma$. Actually, the additional piece could also be defined in 
LQG\footnote{Despite the fact that $D_a$ does not exist as an operator 
valued distribution in LQG. The too singular $D_a$ are tamed by the 
additional operator $q^{ab}/\sqrt{\det(q)}$.} and the results of 
\cite{AQG} also imply that $M_E$ has the correct classical limit in 
both LQG and AQG. However, 
within LQG $M_E$ is somewhat unmotivated because one already has the 
representation $U(\varphi)$ of spatial diffeomorphisms. In AQG on the 
other 
hand there is no choice and the advantage of $M_E$ is that it treats the 
Hamiltonian constraint and the spatial diffeomorphism constraint on equal 
footing (rather than defining the infinitesimal generator for the 
Hamiltonian constraints but only exponentiated diffeomorphisms). 
  
We refrain from displaying more details about AQG here as this is a rather 
recent proposal and because this is a review about LQG. The interested 
reader is referred to \cite{AQG}.

\subsubsection{Dirac observables and physical Hamiltonian}
\label{s4.4.3}

As mentioned in section \ref{s2}, General Relativity is an already 
parametrised system and in order to extract gauge invariant information 
and a notion of physical time evolution among observables  
one must deparamterise it, e.g. using the relational framework sketched in 
section \ref{s2}. There are many ways to do this but a minimal 
requirement is that the physical Hamiltonian is close to the Hamiltonian 
of the standard model at least when spacetime is close to being flat.
In \cite{Phantom} a particularly simple way of deparametrisation which 
fulfills this requirement has been 
recently proposed using scalar phantom matter. In fact one can write the 
Hamiltonian constraints in the equivalent form $H(x)=\pi(x)+C(x)$ where 
$\pi$ is 
the momentum conjugate to the phantom field $\phi$ and $C$ is a positive 
function on phase space which depends on all remaining matter and geometry 
only. Let now for any real number $\tau$
\be \label{4.44}
h_\tau:=\int_\sigma d^3x \;(\tau-\phi)(x)\; C(x),\;\;h:=\int_\sigma 
\;d^3x\; C(x)
\ee
Given a spatially diffeomorphism invariant function $F$ we set
\be \label{4.45}
F(\tau):=\sum_{n=0}^\infty \frac{1}{n!} \{h_\tau,F\}_{(n)}
\ee
Then $F(\tau)$ is a one parameter family of Dirac observables and 
$dF(\tau)/d\tau=\{h,F(\tau)\}$. In particular, $h$ is itself a Dirac 
observable, namely the physical Hamiltonian that drives the physical time 
evolution of the Dirac observables. 

This holds for the classical theory. In quantum theory (\ref{4.45}) should 
be replaced by
\be \label{4.46}
\widehat{F}(\tau):=\exp(\hat{h}_\tau/(i\hbar))\; \widehat{F} 
\exp(-\hat{h}_\tau/(i\hbar))
\ee
provided we can make sense out of $\hat{h}_\tau$ as a self -- adjoint 
operator. This is work in progress.
 
\subsubsection{Brief note on spinfoam models}
\label{s4.4.4}

Spin foam models \cite{Perez2} are an attempt at a path integral 
definition of LQG. They were heuristically defined in the seminal work
\cite{Reisenberger} which attempted at the construction of the physical 
inner product via the formal exponentiation of the Hamiltonian 
constraints of \cite{QSDI}. The reason that this approach was formal is 
that the Hamiltonian constraints do not form a Lie algebra and they are 
not even 
self -- adjoint. Thus, there are mathematical (exponentiation of 
non normal operators) and physical (non Lie group structure of the 
constraints prohibiting the possibility that functional integration over 
$N$ of $\exp(i\widehat{H}(N))$ leads 
to a (generalised) projector) issues with this proposal. 

This is why 
spin foam models nowadays take a different starting point. Namely, one 
starts from the Palatini action and writes it as a topological BF theory
$S_{BF}=\int_M {\rm Tr}(B\wedge F)$ 
together with additional simplicity constraint action $S(\Lambda,B)
=\int_M {\rm Tr}(\Lambda \otimes B\wedge B)$ where $\Lambda$ is a Lagrange 
multiplier tensor field with certain symmetry properties. Extremisation 
with respect to $\Lambda$ imposes 
the condition  
that the $B$ field two form comes from the wedge product of two tetrads.
The advantage of this formulation is that a lot is known about the 
topological BF theory and one can regard the additional simplicity 
constraint as a kind
of ``interaction'' term in addition to the ``free'' BF term. In order 
to define the spinfoam model one has to regularise it as in the 
canonical theory by introducing a finite 4D trinagulation $\tau$ and a 
corresponding discretisation of the action like Wilson's action for Yang 
Mills theory. The  
connection $A$ underlying the curvature $F$ is located as a holonomy on 
the edges of the
dual triangulation $\tau^\ast$ while the $B$ field is located on the 
faces of $\tau$. One 
integrates $\exp(iS_{BF}+iS(\Lambda,B))$ over $A$ with respect to the Haar
measure and over $B$ and the Lagrange multiplier $\Lambda$ with 
respect to Lebesgue measure. The integral over $\Lambda$ results in a 
$\delta-$distribution in $B$. This can be heuristically replaced by 
a $\delta$ distribution in the right invariant vector fields corresponding 
to the holonomies of the connection. One can then perform the $B$ integral 
resulting in an 
additional $\delta$ distribution in the holonomies which then are written 
as a sum over representations using the Peter\&Weyl theorem. The 
simplicity constraints in terms of the right invariant vectoir fields then 
impose restrictions on the occurring representations on the edges and 
intertwiners at the vertices.

These steps are simplest 
illustrated by modelling the situation by a one dimensional 
system $S_{BF}=BF,\; S(\Lambda,B)=\Lambda B^2$. Then formally
\ba \label{4.47}
&& \int dF \int dB \int d\Lambda \exp(i[BF+\Lambda B^2])
=\int dF \int dB \int \delta(B^2) \exp(iBF)
\nonumber\\
&=&\int dF \int dB \int \delta(-(d/dF)^2) \exp(iBF)
=\int dF \int \delta(-(d/dF)^2) \delta(F)
\ea
This brief paragraph does not reflect at all the 
huge body of research performed on spin foam models, we have barely 
touched only those aspects directly connected with the canonical 
formulation. Please refer to \cite{Perez2} and references therein for a 
more complete picture describing the beautiful connection with state sum 
models, TQFT's, categorification, 4D manifold invariants (Donaldson 
theory), non -- commutative geometry, emergence of Feynman graph language
and renormalisation groups etc. \\
\\
From the canonical perspective, spin foam models are very important as 
they provide a manifestly spacetime diffeomorphism covariant formulation 
of LQG. In order to reach this goal, the following issues have to be 
overcome: 
\begin{itemize}
\item[1.] The relation with the canonical formulation is somehow lost. 
In fact, it is well known how to obtain a path integral formulation of 
a given canonical constrained theory \cite{Henneaux}. The integration 
measure cannot be the naive one as used above if there are second class 
constraints. That this is indeed the case has been shown in the important 
work \cite{Noui} which is, in the mind of the author, not sufficiently 
appreciated.
\item[2.] While the simplicity constraints expressed in terms of $B$
are mutually commuting as operators, their replacement in terms of 
right invariant vector fields do not and in fact they do not form a closed 
algebra. Hence, considered as quantum constraints they are anomalous and 
it is 
remarkable that there exists a unique non trivial solution to the 
simplicity 
constraints \cite{BC} at all. For the corresponding model one can show 
that the path integral 
is dominated by degenerate metrics \cite{Christensen}, hence it seems not 
to have the correct semiclassical behaviour which is then maybe not too 
surprising. There should be a way to implement the symplicity constraints 
in their non anomalous form. 
\item[3.] In contrast to pure BF theory these constrained BF theories are 
no longer topological and thus not independent of the triangulation. Thus,
in order to get rid of the triangulation dependence one could sum over 
triangulations and the 
weights with which this should be done are motivated by group field theory
\cite{Laurent}. The result is supposed to give a formula defining a 
rigging map. While there are attractive features such as an emergent
Feynman graph language, it is presently unclear whether the sum converges 
(as it should in a fundamental theory) or whether
it is maybe not more appropriate to perform a refinement limit as in the 
theory of dynamical triangulations \cite{Loll}.
\end{itemize}
%

\section{Physical applications}
\label{s5}

We have so far mostly reported about the status of the quantisation 
programme. Since LQG is a non perturbative approach, preferrably 
one would complete the quantisation programme before one studies 
physical applications. Since the programme reached its current 
degree of maturity only relatively recently, physical applications could 
so 
far not attract much attention. Certainly what is needed in the future 
is an approximation scheme with respect to which physical states, the 
physical inner product, Dirac observables and the physical Hamiltonian 
can be computed with sufficient detail. The semiclassical states 
\cite{GCS} provide a possible avenue especially with respect to 
applications for which the quantum geometry can be regarded as almost
classical. Namely we can consider kinematical semiclassical states which 
are peaked on the constraint surface and on the gauge cut defined by the 
clock variables. These states are then approximately annihilated by the 
Master constraint and the power series defining the Dirac Observables 
can be terminated after a few terms just like in perturbation theory.
This procedure could be called {\it quantum gauge fixing} because we do 
not fix a gauge classically but rather suppress the fluctuations off the 
constraint surface and off the gauge cut. 

Despite the fact that such an approximation scheme has so far not been 
worked out in sufficient detail\footnote{In particular one would like to 
know how close the approximate kinematical calculations are to the
actual calculations on the physical Hilbert space.} there are already
some physical applications of LQG which are insensitive to 
the details of such an approximation scheme. These are 1. matter coupling, 
2.kinematical geometrical operators,   
3. Quantum black hole physics, 4. semiclassical states, 5. Loop quantum 
cosmology and 6. 
LQG phenomenology. We will say only very little about these topics 
here because our main focus is the mathematical structure of LQG.
Hence we will restrict ourseleves to the salient results and ideas.

\subsection{Matter coupling}
\label{s5.a}

We have so far hardly mentioned matter. However, in LQG all 
(supersymmetric) standard 
matter can be straightforwardly coupled as well \cite{QSDII,QSDIII}. 
As far as the kinematics is concernded, the background independent 
representation for the gauge fields of the standard matter is the same as 
for the gravitational sector because all the constructions work for an 
arbitrary compact gauge group. For Higgs fields, which are located at the 
vertices of the graph and other scalar matter one has a similar 
construction just that now states are labelled by points rather than 
edges. Finally for fermionic matter one uses a standard Berezin integral 
kind of construction. As far as the dynamics is concerned, the 
key technique of section \ref{s4.4.2.1} applies. All negative powers of 
$\det(q)$ which appear in the matter terms and which are potentially 
singular can be replaces by commutators between fractional powers of the 
volume operator and gravitational holonomy operators. 

The corresponding contributions to the Hamiltonian constraint have to be 
added up and are then squared in the Master constraint again without 
picking up an UV divergence. It is often criticised that LQG therefore 
does not impose any restriction on the allowed matter coupling. While
that may turn out to be phenomenologically attractive for the reasons 
mentioned in the introduction, it may actually be technically incorrect:
For it could be that the answer to the question, whether the spectrum of 
the Master constraint contains zero, critically depends on which type
of matter we couple. This is due to the fact that the shift of the 
minimum of the spectrum of the Master constraint away from zero 
is typically due to a kind of normal ordering correction. Now 
intuition from ordinary QFT  
suggests that there must be a critical balance between 
bosonic and fermionic matter in order that positive bosonic corrections 
cancel negative fermionic ones. Hence, maybe after all the spectrum only 
contains zero if we allow for supersymmetric matter. In order to decide 
this a more detailed knowledge of the spectrum of the Master constraint is 
required.

\subsection{Kinematical geometric operators}
\label{s5.3}

One of the most cited results of LQG is the discreteness of the spectrum 
of kinematical geometric operators such as the volume operator, the area 
operator \cite{RSVol,ALArea} or the length operator \cite{length}. The 
origin of this pure 
point spectrum is that these operators are functions of right invariant 
vector fields on various copies of $SU(2)$ and thus they are 
diagonalised by linear combinations spin network states with fixed 
graph and edge 
spin but varying intertwiners. Since for fixed edge spin the space of 
intertwiners is finite dimensional, it follows that these operators 
reduce to finite dimensional Hermitean matrices on these fixed graph 
and spin subspaces. 

However one should stress that the discreteness of the spectrum is a 
kinematical feature: None of these operators commutes with the spatial
diffeomorphism or the Master constraint. Whether or not these operators 
retain this property after having them made true Dirac observables 
via the relational machinery depends on the clock matter that is used to 
deparametrise the theory. See \cite{DT} for a discussion.   

However, if the discreteness of the spectrum is retained then this could 
be interpreted as saying that in LQG the geometry is discontinous or 
distributive at Planck scale. At macroscopic scales there is a 
correspondence principle at work, that is, the difference between 
subsequent eigenvalues rapidly decays for large eigenvalues.

\subsection{Semiclassical states}
\label{s5.0}

As already mentioned,
the development of semiclassical tools represent an important area of 
research
in the development of LQG because they allow to test whether LQG is 
really a quantum theory of General Theory and not of some pathological
phase thereof. These developments were hardly mentioned in \cite{NPZ}. 
Semiclassical states for LQG \cite{GCS,Varadarajan,shadow} have so far 
been constructed only for 
the kinematical Hilbert space because the primary goal was so far to test 
the semiclassical limit of the constraint operators which by definition 
annihilate physical semiclassical states and thus cannot be tested by 
them. 
However, we will present 
some ideas of how spatially diffeomorphism invariant or even physical 
semiclassical states might be constructed.\\
\\
The kinematical semiclassical states are actually coherent states and 
are all based on the {\it 
complexifier technique} \cite{GCS} which we will briefly sketch now:\\
Suppose that we are given a phase space of cotangent bundle structure 
${\cal M}=T^\ast {\cal A}$ where $\cal A$ is the configuration space.
A {\it complexifier} is, roughly speaking, a positive function $C$  
on $\cal M$ with the dimension of an action which grows stronger than 
linearly as $E^I\to \infty$ where
$E^I$ denotes the momentum coordinates on $\cal M$ and $I\in {\cal I}$
is a labelling set. Denoting the points in 
$\cal A$ by $A_I$ we define complex configuration coordinates
\be \label{5.0a}
Z_I=\sum_{n=0}^\infty \;\frac{i^n}{n!} \{A_I,C\}
\ee
explaining the name complexifier.
Suppose that the theory can be canonically quantised such that $\hat{C}$ 
becomes a positive, self -- adjoint operator on a Hilbert space ${\cal 
H}=L_2(\overline{{\cal A}},d\mu)$ of square integrable functions on some 
distributional extension $\overline{{\cal A}}$ of $\cal A$ with respect to 
some measure $\mu$. 
The quantum analog of (\ref{5.0a}) becomes, upon replacing Poisson 
brackets by commutators divided by $i\hbar$, the {\it annihilation 
operator}
\be \label{5.0b}
\hat{Z}_I=e^{-\hat{C}/\hbar}\;\hat{A}_I\;e^{\hat{C}/\hbar}
\ee
which explains the dimensionality of $C$. The operators $\hat{Z}_I$ 
are mutually commuting. The exponentials are defined 
via the spectral theorem. Let $\delta_A$ be the $\delta-$distribution
with respect to $\mu$ with support at $A$ and consider the distribution
\be \label{5.0c}
\Psi_A:=e^{-\hat{C}/\hbar} \delta_A
\ee
Due to the positivity of $\hat{C}$ the operator $\exp(-\hat{C}/\hbar)$ 
is a smoothening operator, explaining the required positivity of $C$. In 
fact,
if $\cal H$ is separable, then (\ref{5.0c}) will be an element of $\cal 
H$ (normalisable) if $C$ is suitably chosen. Now the growth condition
in the definition of $C$ typically ensures that $\Psi_A$ is {\it 
analytic} in A and hence can be analytically continued, as an $L_2$ 
function, to $Z$. Denoting the analytically continued object by $\Psi_Z$
we obtain immediately the defining property of a {\it coherent state} 
to be a simultaneous eigenstate of the annihilation operators
\be \label{5.0d}
\hat{Z}_I \;\Psi_Z =Z_I \;\Psi_Z
\ee
Notice, however, that if $\cal H$ is not separable then $\Psi_Z$ is only 
a coherent distribution even if $C$ has all the required properties. 

This construction in fact covers all coherent states that have been 
considered for finite or infinite systems of uncoupled harmonic 
oscillators, in particular the ``classical'' coherent states for 
the Maxwell field (QED). For the Maxwell field the complexifier turns out 
to be 
\be \label{5.0e}
C=\frac{1}{2 e^2}\int_{\Rl^3}\;d^3x\; \delta_{ab}\; E^a (-\Delta)^{-1/2} 
E^b
\ee
where $e$ is the electric charge, $E^a$ the electric field and $\Delta$ 
the flat space Laplacian. 

In fact, quadratic expressions in the momentum 
operators always are good choices for $C$. However, for LQG we may not use 
background dependent objects such as $\Delta$. In \cite{GCS} quadratic 
expressions in the area operator (see below) were used and semiclassical 
properties 
such as peakedness in phase space, infinitesimal Ehrenfest property, 
overcompleteness, semiclassical limit and small fluctuations were 
established. Of course, since the kinematical LQG Hilbert space ${\cal H}$  
is not separable, one must restrict the complexifier construction to separable 
subspaces. Natural candidates are the Hilbert spaces ${\cal H}_\gamma$ 
(closure of the span of SNWF's over $\gamma$) and ${\cal H}'_\gamma$ 
(closure of the span of SNWF's over all subgraphs of $\gamma$). The 
resulting states 
$\Psi_{Z,\gamma}=\sum_{\gamma(s)=\gamma} \Psi_{Z,s} T_s$ and 
$\Psi'_{Z,\gamma}=\sum_{\gamma(s)\subset\gamma} \Psi_{Z,s} T_s$
are respectively the spin network or cylindrical projections of the 
distributions $\Psi_Z=\sum_s \Psi_{Z,s} T_s$ (the sum is over all 
SNW's) and are called shadows 
\cite{shadow} or cut -- offs \cite{GCS} of $\Psi_Z$ 
respectively\footnote{In order to avoid confusion which may arise form 
corresponding remarks in \cite{NPZ}:
These states are functions of distributional connections 
$A\in \overline{{\cal A}}$
labelled by smooth fields $Z$. This is even the case for Maxwell coherent 
states. Hence one can surely get back the smooth fields of the classical 
theory in the classical limit.}.

This graph dependence of the present semiclassical framework of LQG
is an unpleasant feature which so far has prevented one from establishing 
the 
semiclassical limit of graph changing operators such as the Hamiltonian 
constraint. This is because the Hamiltonian constraint creates new edges 
whose fluctuations are not controlled by these graph dependent states.
Hence the above mentioned semiclassical properties only hold for graph non 
changing operators and this is why the graph non changing Master 
constraint is under much better control than the Hamiltonian constraints.
In AQG \cite{AQG} even the graph dependence is lost because there is only 
one fundamental graph. 

Finally, let us address the question of spatially diffeomorphism invariant 
or physical states. These Hilbert spaces do not obviously have a 
representation as $L_2$ spaces and moreover it is not easy to find 
complexifiers with the required properties which are either spatially 
diffeomorphism invariant or Dirac observables. Hence the complexifier idea 
is not immediately applicable. However, we have shown that there are 
(anti) linear rigging maps 
$\eta_{{\rm Diff}}:\;{\cal D}\to {\cal H}_{{\rm Diff}}$ and 
$\eta_{{\rm phys}}:\;{\cal D}^\ast_{{\rm Diff}}\to {\cal H}_{{\rm phys}}$
respectively. 
Now, given a, say cutoff state $\Psi_{Z,\gamma}$, we obtain spatially 
diffeomorphism invariant states 
$\Psi^{{\rm Diff}}_{Z,\gamma}:=\eta_{{\rm Diff}}(\Psi_{Z,\gamma})$ and 
physical states
$\Psi^{{\rm phys}}_{Z,\gamma}:=\eta_{{\rm phys}}\circ \eta_{{\rm 
Diff}}(\Psi_{Z,\gamma})$ which can serve as {\it Ans\"atze} for 
semiclassical states in the corresponding Hilbert spaces. Whether they 
continue to have the desired semiclassical properties with respect to
 spatially diffeomorphism invariant or Dirac observables respectively is 
the subject of current research.

\subsection{Quantum black hole physics}
\label{s5.1}

The main achievement of LQG in this application is to provide a 
microscopic explanation of the Bekenstein Hawking entropy, see 
\cite{BHReview} and references therein. The classical starting point 
is the theory of isolated and dynamical horizons \cite{Krishnan} which is 
somehow a local\footnote{The usual definition of a black hole region as 
the complement of the past of future null infinity obviously requires the 
knowledge of the entire spacetime and is inappropriate to do local 
physics.} definition of an event horizon and captures the intuitive idea 
of a black hole in equilibrium. The notion of an isolated horizon uses, 
among other things, the classical field equations and therefore is a 
classical concept which is imported into the quantum theory by hand.
In other words, the presence of the black hole is put in classically 
leading to an inner boundary of spacetime. It would be more desirable to 
have entirely quantum criteria at one's disposal, see e.g. \cite{Husain}
for first steps, however the following partly semiclassical framework is 
completely consistent and satisfactory. 

The presence of the inner boundary leads to boundary conditions which, 
intuitively speaking, reduce the gauge freedom at the boundary and thus 
give rise to boundary degrees of freedom. Remarkably, their dynamics is 
described by a $U(1)$ quantum Chern Simons theory. On the other hand, the 
bulk is described by LQG. In order to compute the entropy of the black 
hole one counts the number of eigenstates of the area operator of the
$S^2$ cross sections $S$ of the  
horizon\footnote{Due to the boundary conditions this turns out to be a 
Dirac observable. In particular, different cross sections have the same 
area.} whose eigenvalues fit into the interval 
$[{\rm Ar}_0-\ell_P^2,{\rm Ar}_0+\ell_P^2]$ where 
${\rm Ar}_0$ is some macroscopic area. 

This number would be infinite
if $S$ would be an arbitrary surface. Namely
a bulk state is described, near the 
horizon, by the ordered sets of punctures of the bulk graph with the 
surface 
$S$ and at each such puncture $p$ by the total spin $j_p$ to which the 
edges running into $p$ couple. The area eigenvalue for such a 
configuration is given by \cite{RSVol,ALArea}
\be \label{5.1}
\lambda=\hbar \kappa \beta \sum_p \sqrt{j_p(j_p+1)}
\ee
For fixed $j_p$ there are an infinite number of spin network states which 
couple to total $j_p$ (for instance let two edges run into $p$ 
with spins $j_1,\;j_2=j_1+j_p$ where $j_1$ is arbitrary). Hence 
if we would count bulk states, the entropy would diverge.    

However, the physical reasoning is 
that what we must count are horizon states of the Chern Simons theory
because the horizon degrees of freedom are the intrinsic description of 
the black hole and not the 
bulk degrees of freedom. Due to the quantum boundary conditions, the 
surface and bulk degrees of freedom are 
connected in the following way in the quantum theory: Around each 
puncture, the holonomy along the loop of the $U(1)$ Chern Simons 
connection must be, 
roughly speaking, equal to the signed area (flux) of the surface bounded 
by the loop. Hence what matters to the surface theory is the number $j_p$ 
and not the detailed recoupling that created it. In other words, one 
ignores the multiplicities of the $j_p$.

With this in mind one can count now the number of eigenvalues. This would 
again be infinite if there would not be an area gap, i.e. a smallest non 
vanishing area eigenvalue which one can read off from (\ref{5.1}). The 
result is \cite{Krasnov,Domagala,Meissner}
\be \label{5.2}
\ln(N)=\frac{\beta}{\beta_0} \frac{{\rm Ar}_0}{4\ell_P^2} + O(\ln ({\rm 
Ar}_0/\ell_P^2))
\ee
where $\beta_0$ is a numerical constant.
This is the Bekenstein Hawking formula if we set $\beta=\beta_0$ which has 
been suggested to be one way to fix the Immirzi parameter. This would be 
inconsistent if $\beta_0$ would depend on the hair of the black hole. 
However, the constant $\beta_0$ is universal, all black holes of the 
Schwarzschild -- Reissner -- Nordstrom -- Newman -- Kerr family are 
allowed as well as Yang Mills and dilatonic hair. Notice that these black 
holes are of astrophysical interest, they are non supersymmetric and far 
from extremal, in contrast to the similar calculations in string theory
which heavily depend on extremality.\\
\\
In summary, there is an unexpected, consistent interplay between classical 
black hole physics, quantum Chern Simons theory and LQG. Future 
improvements should include the development of quantum horizons and 
Hawking radiation.

\subsection{Loop quantum cosmology (LQC)}
\label{s5.2}

Loop quantum cosmology (LQC) is not the cosmological sector of 
LQG\footnote{So far there is no satisfactory derivation of LQC from LQG, 
LQC is constructed ``by analogy''.}. 
Rather it is 
the usual homogeneous minisuperspace quantised by the methods of LQG. This 
has a kinematical and a dynamical side \cite{ABL}. As far as the 
kinematics is 
concerned, although LQC has only a finite number of degrees of freedom
one can circumvent the Stone -- von Neumann uniqueness theorem for the 
representations of the canonical commutation relations by dropping the 
assumption of weak continuity of the Weyl operators. This is in complete 
analogy to LQG where holonomies but no connections are well defined as 
operators for precisely the same reason. The corresponding Hilbert space 
is then not of the Schr\"odinger type $L_2(\Rl,dx)$ but rather of the 
Bohr type $L_2(\overline{\Rl},d\mu_0)$. Here $\overline{\Rl}$ is the 
Bohr compactification of the real line. It is the counterpart of 
$\overline{{\cal A}}$ and can be 
defined as the Gel'fand 
spectrum of the Abelean C$^\ast$ algebra generated by the functions 
$q\mapsto \exp(i\mu q)$. This algebra is called the algebra of 
quasiperiodic functions and is the counterpart of $\overline{{\rm Cyl}}$.
Finally $\mu_0$ is the Haar measure on $\overline{\Rl}$ which is in 
complete analogy to the Ashtekar -- Lewandowski measure of LQG.

On the dynamical side the situation in LQG is matched in the sense that 
in the Hamiltonian constraint one cannot work with connections but only 
with holonomies. Hence one has to modify the classical constraint by 
working with say $\sin(\mu_0 q)/\mu_0$ 
rather than $q$ where $\mu_0$ is an arbitrarily small but finite 
constant\footnote{See \cite{ABL} for arguments to fix the value of 
$\mu_0$. In LQG the analog of $\mu_0$ would be the regulator $\epsilon$ 
in the loop attachment, however, in LQG all values of $\epsilon$ are 
equivalent due to spatial diffeomorphism invariance. This does not happen 
in LQC because in LQC the spatial diffeomorphism group is gauge fixed
so that the appearance of $\mu_0$ could be considered as an artefact of 
the simplicity of the model.}. Since also inverse powers of the momentum 
$p$ conjugate to $q$ appear in the Hamiltonian constraint one uses the 
same key kind of key identities as in LQG such as 
\be \label{5.3}
ir\mu_0 \frac{{\rm sgn}(p)}{|p|^{1-r}}=e^{-i\mu_0 q}\{|p|^r,e^{i\mu_0 q}\}  
\ee
where $0<r<1$ is a rational number, in order to avoid negative powers of 
the $p$ operator in the quantum theory. 

The main advantage of this model is that one can carry out almost 
all steps of the quantisation programme and compare it with the 
conventional Schr\"odinger quantisation (Wheeler DeWitt theory). 
The predictions of the model are in fact quite intriguing: Avoidance of 
curvature singularities, deterministic quantum gauge flow through the 
would be singularity, inflation from quantum geometry, avoidance of chaos 
in Bianchi IX cosmologies, recovery of conventional cosmology at large 
scale factor  etc. See \cite{BMT} for a review. The most 
mathematically precise treatment can be found in \cite{APS}.

Of course one is never sure whether the simplifications that are made in 
toy models spoil its predictive power, that is, whether the results of 
the toy model continue to hold in the full theory. 
Partial confirmation of LQC singularity avoidance results within 
full LQG can be found in \cite{BT2}, although via a completly different 
mechanism. However, at least
the model tests important aspects of the full theory, in particular the 
key identities of the type (\ref{4.24}) without which these spectacular 
results of LQC would not have been possible.

Notice that LQG and in particular LQC can easily deal with de Sitter
space kind of situations while this seems to be harder in superstring 
theory whose effective low energy limit should be supergravity on de 
Sitter space. However, the de Sitter algebra does not admit a positive 
Hamiltonian indicating that supergravity on de Sitter space is unstable.
This is potentially alarming because recent observations indicate that 
the universe currently undergoes a de Sitter phase.

\subsection{LQG phenomenology}
\label{s5.4}

The field of LQG phenomenology has just started to develop, 
mostly because in the majority of cases there is no clear cut derivation 
of the phenomenological assumptions made from full LQG. See 
\cite{Hugo,Hossenfelder} 
for a review. One of the hottest topics in this fields are 
signatures of Lorentz invariance violation. A phenomenological 
description of this could be doubly special relativity (DSR) 
\cite{Kowalsky}, a theory in 
which not only the speed of light but also the Planck energy is 
(inertial) frame independent. In 3D it turns out to be possible to 
connect DSR \cite{FreidelDSR}, non commutative geometry 
\cite{Majid} and LQG in the spin foam formulation but 
in 4D this is still elusive.  

The intuitive idea behind Lorentz invariance violation in quantum gravity 
is the apparently Planck scale discreteness of LQG: If true, then 
quantum geometry looks more like a crystal than vacuum even if the 
gravitational vacuum state looks like Minkowski space on large scales. 
Hence there 
could be non trivial dispersion relations for light propagation leading 
to energy dependent time of arrival delays of photons of high energies
that have travelled a long distance. One possible source of such signals
are $\gamma$ ray bursts at cosmological distances \cite{Amelino,Biller}
which would be detectable by the GLAST detector provided that the effect 
is linear in $E/E_P$ where $E$ is the photon 
energy and $E_P$ the Planck energy. For first steps towards a systematic 
derivation from LQG see \cite{GP,Hanno}. 
Notice that for the five perturbative string theories on the Minkowski 
target space Lorentz invariance is built in axiomatically, hence Lorentz 
invariance violation could discriminate between LQG and string theory.  

Another hot topic concerns cosmology. To this realm belong the 
prediction of deviations from the scale invariance of the power spectrum 
of the cosmic microwave background radiation (CMBR) \cite{Hofmann,Singh}
using LQC (related) methods which might be detectable by the WMAP or  
PLANCK detectors. 

Suffice it to say that this field is largely unexplored and that it 
needs more input both from experiments and theory.

\section{Conclusions and outlook}
\label{s6}

We hope to have given a brief but self critical account of the status of 
LQG with special focus on the most important issue, the 
implementation of the quantum dynamics. In particular we hope to have 
addressed most if not all issues that have been brought up in 
\cite{NPZ}. We presented them from an ``inside'' point of view and 
showed why the mostly technically 
correct description in \cite{NPZ}, in our mind, is often   
unnecessarily sceptic, inconclusive or incomplete. Notice 
also that we only reported results well known in the LQG literature. We 
emphasise this because the unfamiliar reader may have the impression 
that only \cite{NPZ} unveiled the issues discussed there. 

We have indicated why non separable Hilbert spaces are no obstacle in 
LQG, they may even be welcome! There has been important progress 
recently on the frontiers of the semiclassical limit, the physical 
Hilbert space, physical (Dirac) observables, the physical 
Hamiltonian, the constraint algebra, the avoidance of anomalies and 
quantisation ambiguities, the covariant formulation \cite{NPZ1} as well 
as physical 
applications which 
were insufficiently appreciated in \cite{NPZ}. The report given in 
\cite{NPZ} in many ways displays the field of LQG as it was a decade ago 
and thus 
ignores the progress made since then during which the field quadrupled 
in size. We hope to have clarified in this report that important 
developments were left out in \cite{NPZ} thus hopefully reducing the 
negative flavour conveyed there.

Let us discuss further issues touched upon in \cite{NPZ} which were not 
yet mentioned in this article:
\begin{itemize}
\item[1.] A folklore statement that seems to have entered several 
physics blogs is that weakly discontinuous representations of the kind 
used in LQG do not work for the harmonic oscillator so why should they 
work for more complicated theories? This is the conclusion reached in 
\cite{Helling}. As we will now show, while \cite{Helling} is technically 
correct, its physical conclusion is {\it completely wrong}. In 
\cite{Helling} 
one used a representation discussed first for QED \cite{Narnhofer} in 
order to avoid the negative norm states of the Gupta -- Bleuler formulation. 
In this representation neither position $q$ nor momentum $p$ operators 
are well defined, only the Weyl operators 
$U(a)=\exp(iaq),\;V(b)=\exp(ibp)$ exist. Hence the usual harmonic 
oscillator Hamiltonian $H=q^2+p^2$ does not exist in this 
representation. Consider the substitute 
$H_\epsilon=[\sin^2(\epsilon q)+\sin^2(\epsilon p)]/\epsilon^2$. What is 
shown in 
\cite{Helling} is that this operator is ill defined as $\epsilon\to 0$. 
Is this a surprise? Of course not, we knew this without calculation 
because the representation is not weakly continuous. The divergence  
of $H_\epsilon$ as $\epsilon\to 0$ in discontinuous representations is 
therefore a trivial observation.
However, what is physically much more interesting is the following. 
Fix an energy level $E_0$ above which the harmonic oscillator becomes 
relativistic and thus becomes inappropriate to model the correct 
physics. Let\footnote{Notice that classically 
$H_\epsilon=|a_\epsilon|^2$.} 
$a_\epsilon^\dagger:=[\sin(q\epsilon)+i\sin(p\epsilon)]/\epsilon$. 
Consider the finite number of observables 
\be \label{6.1}
b_{\epsilon,n}:=\frac{1}{n!} (a_\epsilon)^n\;(a_\epsilon^\dagger 
a_\epsilon) 
(a_\epsilon^\dagger)^n,\;
n=0,..,N=E_0/\hbar
\ee
Let $\Omega_0$ be the Fock vacuum in the 
Schr\"odinger representation and $\omega$ the state underlying the 
discontinuous representation. Fix a finite measurement precision 
$\delta$. Since the Fock representation is weakly continuous we  
find $\epsilon_0(N,\delta)$ such that $|<\Omega_0, b_{\epsilon,n} 
\Omega_0>-n\hbar|<\delta/2$ for all $\epsilon\le\epsilon_0$. On the 
other 
hand, by Fell's theorem\footnote{The abstract statement is \cite{Haag}: 
The folium 
of a faithful state on a C$^\ast-$algebra is weakly dense in the set of 
all 
states. Here the folium of a states are all trace class operators on 
the corresponding GNS Hilbert space. The theorem applies to the 
unique C$^\ast$ algebra \cite{Slawny} generated by the Weyl operators 
$U(a),\;V(b)$ which we are considering here. The representations 
considered in 
\cite{Helling,Narnhofer} are faithful.} 
we 
find a trace class operator 
$\rho_{N,\delta}$ in the GNS 
representation determined by $\omega$ such that $|{\rm 
Tr}(\rho_{N,\delta} 
b_{\epsilon_0,n})-<\Omega_0,b_{\epsilon_0,n}\Omega_0>|<\delta/2$ for all 
$n=0,1,..,N$. It follows that with arbitrary, finite precision 
$\delta>0$ we find 
states in the Fock and discontinuous representations respectively whose 
energy 
expectation values are given with precision $\delta$ by the usual value 
$n\hbar$. This implies that the two states cannot be physically 
distinguished. 

In \cite{AFW,Fredenhagen2} even more was shown\footnote{In a 
representation 
which 
was continuous in one of $p$ or $q$ but discontinuous in the other. But 
similar results hold in this completely discontinuous representation 
considered here.}: There the spectrum of the 
operator $H_\epsilon$ was studied and the eigenvectors and 
eigenvectors were determined explicitly. One could show that by tuning
$\epsilon$ according to $N,\delta$ even the first $N$ eigenvalues do not
differ more than $\delta$ from $(n+1)\hbar$. Moreover, having fixed 
such an $\epsilon$, the non separable Hilbert space is a direct sum of 
separable $H_\epsilon$ invariant subspaces and if we just consider the 
algebra generated by $a_\epsilon$ each of them is superselected. Hence
we may restrict to any one of these irreducible subspaces and  
conclude that the physics of the discontinuous representation is 
indistinguishable from the physics of the Schr\"odinger representation 
within the error $\delta$. This should be compared with the statement
found in \cite{Helling} that in discontinuous representations the physics 
of the harmonic oscillator is not correctly reproduced.
\item[2.] Actually the paper \cite{Helling} was triggered by 
\cite{LQGString} where the following was shown:\\
Using discontinuous representations one can quantise the closed bosonic 
string in any spacetime dimension without encountering anomalies, 
ghosts (negative norm states) or a tachyon state (instabilities). The 
representation independent and purely algebraic no -- go theorem of 
\cite{Mack} that the Virasoro anomaly is unavoidable is circumvented by 
quantising the Witt group Diff$(S^1)\times$ Diff$(S^1)$ rather than its 
algebra diff$(S^1)\oplus$dif$(S^1)$. Since the representation of the 
Witt group is discontinuous, the infinitesimal generators do not 
exist and there is no Virasoro algebra in this discontinuous 
representation, exactly like in LQG. However, as in LQG, a unitary 
representation of the Witt group is sufficient in order to obtain the 
Hilbert space of physical states via group averaging techniques and even 
a representation of the invariant charges \cite{Pohlmeyer} of the closed 
bosonic string. 

Does this mean that the magical dimension $D=26$ cannot be seen in this 
representation? Of course it can: One way to detect it in the usual Fock 
representation of the string is by considering the Poincar\'e algebra
(in the lightcone gauge) and ask that it closes. For the LQG string 
\cite{LQGString} again the Poincar\'e group is represented unitarily but 
weakly discontinuously. However, we can approximate the generators as 
above in terms of the corresponding Weyl operators using some tiny but 
finite 
parameter $\epsilon$. Since these are a finite number of operators
in the corresponding C$^\ast$ algebra, an appeal to Fell's theorem and 
using continuity of the Weyl operators in the Fock representation
guarantees that we find a state in the folium of the LQG string with
respect to which the expectation values of the approximate Poincar\'e 
generators coincides with their vacuum (or higher excited 
state) expectation values in the Fock 
representation to arbitrary  precision $\delta$. 

Thus $D=26$ is also hidden in this discontinuous representation, it is 
just that for no $D$ there is a quantisation obstruction. Of course,
much still has to be studied for the LQG string, e.g. a formulation of 
scattering theory, however, the purpose of \cite{LQGString} was not 
to propose a phenomenologically interesting model but rather to indicate
that $D=26$ is not necessarily sacred but rather a feature of the specific 
Fock quantisation used.
\item[3.] In \cite{Haag} we find the statement that the 
instantaneous fields (smeared only in 3D rather than in 4D) are too 
singular in interacting quantum field 
theories. Indeed, in Wightman field theories one can read Haag's 
theorem as saying that the representation of the interacting field 
algebra (which contains dynamical information) is never unitarily 
equivalent to the representation of the canonical commutation 
relations of the free field algebra (which lacks the information about 
the interaction). This seems to imply an obstruction 
to canonical quantisation where one precisely starts from a purely 
kinematical representation of the Poincar\'e algebra of the 
instantaneous fields. The catch is in the assumptions of Haag's theorem. 
In LQG this no go theorem is circumvented because 
the quantum field theory that one constructs is not a Wightman field 
theory: It is a QFT of a new kind, namely a background independent QFT 
to which Haag's theorem as stated above does not apply because the 
Wightman axioms do not hold. In fact, in LQG the interaction is encoded 
in the self -- adjoint Master constraint which is well (densely) 
defined.
\item[4.] In \cite{NPZ} we find the question where in LQG does one 
find the counter terms \cite{Sagnotti} of perturbative quantum gravity?
More generally, how does one make contact with perturbative QFT and what 
role does the renormalisation group play in LQG, if any? 
These perturbative questions are hard to answer in a theory which is 
formulated non perturbatively, however, let us make a guess\footnote{Of 
course one could ask whether the question is meaningful if quantum 
gravity, which is believed to be non renormalisable, simply does not 
admit a perturbative formulation. However, it is believed that 
perturbative quantum gravity does make sense as an effective theory.}:

Once a physical Hamiltonian such as the one of section \ref{s4.4.4} has 
been successfully quantised one can in principle define scattering 
theory in the textbook way, that is, one would compute transition 
amplitudes between initial and final physical states. This may be hard 
to do technically but there is no obstruction in principle. In order 
to recover perturbation theory around Minkowski space one will consider
a physical state (vacuum) which is a minimal energy state with respect to 
that 
Hamiltonian and peaked on Minkowski space. The physical excitations of 
that state can be considered as the analogs of the graviton excitations 
of the perturbative formulation. Now by construction the transition 
amplitudes (n point functions) are finite, however, there will be 
of course screening effects, i.e. effectively a running of couplings
where the energy scale at which one measures is fed in by the physical 
state by which one probes a given operator. This is the way we expect to 
recover renormalisation effects.

As far as counterterms are concerned, as we have frequently stated, 
there are correction terms in all semiclassical computations done so far 
which depend on the Planck mass, see e.g. \cite{Hanno} where a finite 
but large quantum gravity correction to the cosmological 
constant\footnote{The cosmological ``constant'' therefore becomes 
dynamical.} is computed which results from photon field propagation on 
fluctuating spacetimes. Similar results are expected with respect to 
graviton propagation. These counter term operators are formulated in 
terms of the canonical fields but using the field equations one can 
presumably recast their classical limits into covariant counter term 
Lagrangeans.

Of course it is on the burden of LQG to show that this is really what 
happens, but it is not that there are no ideas for possible mechanisms.
\end{itemize}
We will now answer a number of frequently asked questions which one 
can find in \cite{NPZ}:
\begin{itemize}
\item[1.] {\it Is there only mathematical progress in LQG?}\\
A continuously updated and fairly complete list of all LQG publications 
to date can be found in \cite{CorichiList}. A brief look at this list 
will show that there are papers of all levels of rigour and that 
mathematically more sophisticated papers were motivated and driven by 
less rigorous papers which started from a physical idea. It is true that
in LQG one puts stress on mathematical rigour. The reason is that 
developing background independent QFT's means walking on terra incognita.
Hence, one does not have the luxury to be cavalier about mathematical 
details as in background dependent QFT's where well established theorems 
ensure that there are rigorous versions of formal calculations.

Section \ref{s5} should have indicated that current research is focussed
on hot research topics such as semiclassical quantum gravity 
(contact with QFT on curved spacetimes), quantum cosmology and quantum 
black hole physics. These results together with the huge body of work 
on spin foam models was hardly mentioned in \cite{NPZ}, see however 
\cite{NPZ1}. Ignoring this 
research performed over the past ten years means giving an out of date 
presentation of LQG which would be similar to writing a review on string 
theory without mentioning D -- branes, M -- theory and the landscape.

Notice also that being a much smaller and younger field 
than string theory or high energy physics\footnote{LQG is the focal point 
of only an order of $10^2$ reserachers worldwide.} which in addition 
cannot just use the techniques from 
ordinary particle physics but in fact must first develop its own 
mathematical framework from scratch, the amount of results obtained so far 
is naturally smaller due to lack of man power.
\item[2.] {\it Has there anything been gained as compared to the Wheeler 
--
DeWitt framework?}\\
Any serious theoretical physicist will confirm that it is almost a miracle 
that one can tame mathematical monsters such as the area, volume, 
Hamiltonian constraint or Master constraint operator {\it at all}. These 
operators are integrals over delicate, nonpolynomial functions of  
operator valued distributions evaluated at the same point which 
are hopelessely singular in usual background dependent Fock 
representations. Moreover, not only can one give mathematical 
sense to 
them, they are even free of UV divergences. This is the beauty of 
background independence and provides a precise implementation of the old 
physical idea that quantum gravity should provide the {\it natural 
regulator of ordinary QFT UV divergences.}

For the first time one can write down a concrete, mathematically well 
defined proposal for the Hamiltonian or Master constraint and study its 
physical properties. For the first time one can actually construct 
rigorous solutions thereof. For the first time one can precisely define a 
kinematical, spatially diffeomorphism invariant or physical Hilbert space. 
For the first time one could show that the semiclassical limit of at least 
the graph non changing Master constraint 
is the correct one with respect to rigorously defined, kinematical 
coherent states\footnote{Notice that it is meaningless the semiclassical 
limit of a constraint operator with physical coherent states which by 
definition are in its (generalised) kernel.}.

It is true that not all questions have been answered in connection with 
the quantum dynamics and research on it will continue to occupy many 
researchers during many years to come. However, what is asked for 
in \cite{NPZ} is too much: Nobody expects that one can 
completely solve the theory. We cannot even solve classical General 
Relativity completely and we will probably never be able to. General 
Relativity and even more so LQG are not integrable systems such as string 
theory on Minkowski background target space which is 
mathematically relatively trivial as a field theory. Even today people 
working 
in classical General Relativity struggle to get analytical results for 
the gravitational waves radiated by, say, a black hole merger. The 
problem was posed almost half a century ago but recent 
progress is mostly due to increasing computing power. Gravitational waves 
is just a tiny sector of classical General Relativity and in LQG we 
also must hope that we can at least analyse the theory in sufficient 
detail in those sectors. 

In \cite{NPZ} the authors ask for (approximately) physical semiclassical 
states that enable one to investigate the QFT on curved spacetimes limit 
of the theory. We claim that these exist: 
Kinematical coherent states have been introduced in \cite{GCS}. We can 
choose to have them peaked on the 
constraint surface and then those states solve the Master constraint 
approximately \cite{AQG}, that is, $||\widehat{M}\psi||\approx 0$. These  
states will enable us to perform semiclassical perturbation theory 
as described in \cite{AQG} and the non diagonisability of the volume 
operator poses absolutely no problem here.

Finally, again a glance at \cite{CorichiList} reveals that in LQG there is 
linear progress on the quantisation programme outlined in section 
\ref{s3} over the past twenty years. One never changed the rules of that 
programme which means that the velocity of progress is naturally 
decreasing as one faces the ever tougher steps of that programme. We just 
mention that because from \cite{NPZ} one could sometimes get the 
impression that what is criticised is that researchers in LQG did not 
identify the open problems mentioned in \cite{NPZ}. They did as one can 
see from the publications, but some of the problems simply have not yet 
been solved and are topics of current research. That does not mean that 
they cannot be solved at all and so far every hurdle in LQG was taken.
\item[3.] {\it Is general covariance broken in LQG?}\\
When reading \cite{NPZ} one may get the impression that spacetime 
diffeomorphism invariance is broken right from the beginning just because 
one performs a $3+1$ split of the action. This impression is wrong.
As we have tried to explain in section \ref{s2} the constraints require 
that physical observables are independent of the foliation that one 
introduces in the canonical formulation. This is the same in string theory 
where the Witt constraints require worldsheet diffeomorphism invariance
of physical observables. These constraints are implemented in the LQG 
Hilbert space and their kernel defines spacetime diffeomorphism 
invariant states. 
The question of off -- shell versus on -- shell closure is still open 
for the Hamiltonian constraint. This is why tha Master constraint 
programme was introduced as a possible alternative which seems to work
in the sense that for the Master constraint one could 
show that these constraints and their algebra have the correct classical 
limit \cite{AQG}. They are maybe implemented anomalously in the Master 
constraint but by 
subtracting from the Master constraint the minimum of the spectrum the 
anomaly can be cancelled which corresponds to some kind of normal 
ordering. Notice that this is an {\it off shell} closure of the 
constraints as asked for in \cite{NPZ}. 

In \cite{NPZ} the authors illustrate the importance of on -- shell
versus off -- shell closure by multiplying a given set of non anomalous 
quantum constraints with structure constants with a central operator. 
These modified constraints still
close on shell while the off shell algebra would now close only with 
structure functions. However, there are now possibly extra solutions, 
namely those in the kernel of the central operator, hence the physical 
Hilbert space would suffer from an infinite number of ambiguities. 
We find this example inconclusive for the following reason: In LQG one 
did not randomly multiply the Hamiltonian constraint by something else
but rather just used the classical expression and directly quantised 
it by reasonable regularisation techniques. Furthermore, when modifying 
the classical constraints by multiplying it with a Dirac observable, the 
modified constraints define the same constraint surface as the original 
ones only where the Dirac observable does not vanish. Hence the extra 
solutions in the kernel of the central operator are physically not 
allowed and therefore, in this example at least, the modified quantum 
constraints in 
fact define the same physical Hilbert space as the original constraints.
\item[4.] {\it Does non -- separability of the Hilbert space
prevent the emergence of the continuum in the semiclassical limit?}\\  
In \cite{NPZ} the authors point out the non separability of the 
kinematical Hilbert space which originates from the weak discontinuity 
of the holonomy operators. They call this the {\it pulverisation of the 
continuum} in the sense that all, even infinitesimally different, edges 
lead to orthogonal spin network states. The only topology on the set 
graphs with respect to which the scalar product is continuous is the 
{\it discrete topology} (every subset is open). They then ask 
whether the the continuum can be recovered in the semiclassical limit. 
The answer is in the affirmative: The approximately physical states
\cite{GCS} (kinematical coherent states which are peaked on the 
constraint surface of the phase space) are labelled by {\it smooth}
classical fields and the corresponding expectation values of physical 
operators such as the Master constraint depend even smootly on 
those fields, not only continuously, see e.g. \cite{AQG}.
\item[5.] {\it Is the ambiguity in the Hamiltonian constraint 
comparable to non renormlisability of perturbative quantum 
gravity?}\\
Definitely not:\\
First of all there is a crucial qualitative difference: Perturbative
quantum gravity cannot make sense as a fundamental theory because the 
perturbation series diverges for all possible choices of the 
renormalisation constants. LQG is a finite theory for any choice of the 
ambiguity parameters. Next, the countably infinite number of 
renormalisation constants in perturbative quantum gravity take 
continuous values so that the number of ambiguities is uncountably 
infinite while the physical Hilbert space of LQG depends only 
on a discrete number of ambiguities. Finally, in perturbative quantum 
gravity all values of the infinite number of renormalisation constants 
are, a priori, all equally natural while in LQG all of the discrete 
choices are pathological except for a finite number. As we have 
explained, without some notion of naturality, even ordinary QFT suffers 
from an infinite number of ambiguities (such as all possible 
discretisations of Yang Mills theories on all possible lattices). Hence
applying naturalness, the amount of ambiguity in LQG reduces to a finite 
number of ambiguities which is comparable to the degree of ambiguity of 
a renormalisable ordinary quantum field theory.

An interesting speculation is that the ambiguities in the definition of 
the Hamiltonian constraints are 
maybe not unlike the discretisation ambiguities in Wilson's 
approach to (Euclidean) quantum Yang -- Mills theories. A priori, 
there are infinitely many possible definitions of the microscopic 
Lagrangean. However, their flow under (some background independent 
version of) the renormalisation group may end 
up in a common UV fix point. All those Lagrangeans are then in the same 
universality class and something similar might happen in LQG to the 
Hamiltonian constraints.
%
\end{itemize}
In the appendix the interested reader can find an example where the 
necessity of mathematical machinery is illustrated in a concrete physical
question, namely whether the so called Kodama state is a physical state of 
LQG. This would not be possible without it and therefore exemplifies 
``what has been gained''. \\
\\
\\
\\
{\large Acknowledgements}\\
\\
We thank the members of the Algebraic Quantum Field Theory community, 
especially
Dorothea Bahns, Detlev Buchholz, Klaus Fredenhagen, Karl -- 
Henning Rehren, Robert Schrader and 
Rainer Verch
for their encouragement to write this article. Also the pressure on 
the author from the 
Editorial Board of {\it Classical and Quantum Gravity} to write an answer 
to 
\cite{NPZ} is gratefully acknowledged. The author has benefitted 
from fruitful discussions 
with Abhay Ashtekar, Carlo Rovelli and Lee Smolin and is grateful for 
many comments by Herman Nicolai, Kasper Peeters and Marija 
Zamaklar. Furthermore we want to 
thank Hans Kastrup very much for frequently pointing out the necessity 
to respond to \cite{Helling}. Finally, we are grateful to 
Ion Olimpiu Stamatescu who invited and urged the author to write this 
article for the forthcoming book 
{\it
An Assessment of Current Paradigms in the Physics of Fundamental 
Phenomena}, (Springer Verlag, Berlin, 2006). 

The part of the research performed at the Perimeter Institute for 
Theoretical Physics was supported in part by funds from the Government of  
Canada through NSERC and from the Province of Ontario through MEDT.

\begin{appendix}

\section{The Kodama state}
\label{sa}

We end this article by displaying a concrete
example which illustrates the necessity of all the 
mathematical machinery in order to reach a conclusive answer for precise 
questions. The example is the so called {\it Kodama state} 
\cite{Kodama} which 
is frequently claimed to be an exact solution to all constraints 
of LQG \cite{SmolinKodama}. In fact the Kodama state has attracted 
much attention in the early days of LQG (see \cite{Pullinbook} and 
references therein) because of its formal connection with the Jones 
polynomial \cite{WittenJones} which would therefore seem to be an exact 
solution (in the loop representation) of all the constraints of LQG.

We will now show that this does not hold for various reasons. To see 
what is going on, consider pure gravity with a cosmological constant. 
After multiplying the Hamiltonian constraint by the factor 
$\sqrt{\det(q)}$ it is given by 
\be \label{6.2}
\tilde{H}=\epsilon^{jkl}\epsilon_{abc} E^a_j E^b_k[B^{\Cl c}_l+\Lambda 
E^c_l]
\ee
where $\Lambda$ is the cosmological constant and 
$B^{\Cl c}_j=\epsilon^{abc} F^{\Cl j}_{ab}/2$ the magnetic field of the 
complex connection $A^\Cl$ which is the pull back to $\sigma$ of the 
self dual part of the spin connection (annihilating the tetrad).

The idea underlying the Kodama state is that the $SL(2,\Cl)$ Chern -- 
Simons action 
\be \label{6.3}
S_{CS}[A^\Cl]:=\int_\sigma \;{\rm Tr}(F^\Cl\wedge A^\Cl-\frac{1}{3}
A^\Cl \wedge A^\Cl \wedge A^\Cl)
\ee
is the generating functional of the magnetic field, that is, $\delta 
S_{CS}/\delta A^{\Cl j}_a(x)=B^{\Cl a}_j(x)$ where the functional 
derivative is in the sense of the space of smooth $SL(2,\Cl)$ 
connections ${\cal A}_\Cl$. Now the 
canonical brackets $\{E^a_j(x),A_b^{\Cl k}(y)\}=i\kappa \delta(x,y)$
suggest to formally define a Hilbert space ${\cal H}_\Cl=L_2({\cal 
A}_\Cl,[dA_{\Cl}\; d\overline{A^\Cl}])$ of square integrable, 
holomorphic functions on 
${\cal A}_\Cl$
with respect to formal Lebesgue measure and to represent $A^{\Cl 
j}_a(x)$ as a 
multiplication operator and $E^a_j(x)$ as $-\ell_P^2 \delta/\delta 
A^{\Cl j}_a(x)$. This formally satisfies the canonical commutation 
relations. 

As is well known, the Chern -- Simons action is invariant under 
infinitesimal gauge transformations and as an integral of a three form 
over all of $\sigma$ it is also spatially diffeomorphsim invariant.
Moreover, in the ordering (\ref{6.2}) the Kodama state 
\be \label{6.4}
\Psi_{{\rm Kodama}}=e^{\frac{1}{\Lambda \ell_p^2} S_{CS}}
\ee
is annihilated by the Hamiltonian constraint. This is exciting because 
the nine conditions $B=-\Lambda E$ satisfied by this state is easily 
seen to correspond to de Sitter space for the appropriate sign of 
$\Lambda$ \cite{SmolinKodama}. Hence, the Kodama state could be argued 
to correspond to the de Sitter vacuum of LQG.

There are several flaws with this formal calculation:
\begin{itemize}
\item[A.] {\it Adjointness relations}\\
The formal representation of the canonical commutation relations just     
outlined is not a representation of the $^\ast$algebra generated by 
$E,\;A^\Cl$, that is, the adjointness relations are not satisfied. These 
demand that $E$ is self adjoint and that 
$A^\Cl+\overline{A^\Cl}=2\Gamma(E)$ where $\Gamma$ is the spin 
connection of $E$. It is clear that the ``measure'' $[dA^\Cl 
d\overline{A^\Cl}]$ cannot implement these adjointness relations, hence 
we have to incorporate a formal kernel $K(A^\Cl, \overline{A^\Cl})$. 
A formal Fourier transform calculation \cite{TB} reveals that 
\be \label{6.5}
K(A^\Cl,\overline{A^\Cl})=\int \; [dE] \exp(i\int_\sigma \;d^3x\;
[\frac{A_a^{\Cl j}+\overline{A_a^{\Cl j}}}{2}-\Gamma_a^j] E^a_j
\ee
Even without specifying the details of this functional integral, with this 
kernel the inner product is no longer positive (semi) definite. 
\item[B.] {\it Euclidean gravity}\\
Suppose we replace $A^\Cl$ by a real connection. This formally corresponds 
to Euclidean gravity and now the formal Hilbert space would be 
${\cal H}=L_2({\cal A}, [dA])$ which does give rise to a formal 
representation of the algebra underlying $A,E$ if 
$E^a_j(x)=i\ell_P^2\delta/\delta A_a^j(x)$. Now the Kodama state 
becomes   
\be \label{6.4a}
\Psi_{{\rm Kodama}}=e^{\frac{i}{\Lambda \ell_p^2} S_{CS}}
\ee
Being a pure phase, it is not normalisable in that formal inner product.
\item[C.] {\it Measurability}\\ 
Now consider instead the rigorous Ashtekar -- Isham -- Lewandowski 
representation 
$L_2(\overline{{\cal A}},d\mu_0)$. Certainly the operator 
corresponding to (\ref{6.2}) does not exist in that representation but 
let us forget about the origin of $\Psi_{{\rm Kodama}}$ and just ask 
whether it defines an element of that Hilbert space. Being a pure phase it 
is formally normalisable because the measure $\mu_0$ is normalised.
However, the question is whether the Kodama state is a measurable 
function\footnote{A function is said to be measurable if the preimages of 
open subsets of $\Cl$ are measurable subsets. In our case, the measurable
stes are generated by the Borel sets of $\overline{{\cal A}}$.} in order 
that we can compute inner products between the Kodama state and, say, 
spin network functions. It is easy to see that this is not the case. 
For instance this follows from the fact that if we would triangulate 
the integral over the the Chern -- Simons action in order to replace the 
integral by a Riemann sum over certain holonomies (these are measurable 
functions) and consider the infinite refinement limit, then in this limit 
the Kodama state has non vanishing inner product with an uncountably 
infinite number of spin network functions. Thus, it is not normalisable
when viewed as a proper $L_2$ function. This can also be interpreted 
differently: Recall that $L_2$ 
functions are only defined up to sets of measure zero. The Chern Simons
action is a priori defined only on the measure zero subset of smooth 
connections. 
The extension to $\overline{{\cal A}}$ that we just tried by representing 
it as a linear combination of spin network functions is no longer a phase.
\item[D.] {\it Distributional solution}\\
One could interpret the last item as saying that the Chern Simons state 
defines a rigorous element of ${\cal D}^\ast$ and now the question is 
whether it is annihilated by the rigorously defined dual of the 
Hamiltonian constraint constructed insection \ref{s4}. It is easy to see 
that this is not the case because the Hamiltonian constraint with a 
cosmological constant term, although its 
dual formally acquires the ordering as in (\ref{6.2}), is not proportional 
to $B+\Lambda E$ because the volume operator that enters the 
cosmological constant term is not quantised in the form 
$E^a_j e_a^j$ but rather as $\sqrt{\det(|E|)}$. One could of course write 
the smeared constraint in the form
\be \label{6.5a}
H(N)=\int_\sigma\;N\; {\rm Tr}([F+\Lambda \ast E]\wedge \{A,V\})
\ee
where $V$ is the volume functional and proceed as in section \ref{s4}
although it would be somewhat awkward to define the volume operator in 
this way.
However, even if this would work, this would still only define a solution 
to the Euclidean constraint.
\end{itemize}
This discussion hopefully illustrates the physical importance of the 
mathematical notions introduced and shows that LQG has been brought to a 
level of mathematical rigour that allows to actually answer physical 
questions. Without it we could not have decided if and in which sense 
the Kodama state is a physical state. 

\end{appendix}

\end{document}